\documentclass[letterpaper,10pt]{article} 
\usepackage{opticameet3} %% use version 3 for proper copyright statement

\newcommand\authormark[1]{\textsuperscript{#1}}

\usepackage{amsmath,amssymb}
\usepackage{amsthm}
\usepackage[colorlinks=true,bookmarks=false,citecolor=blue,urlcolor=blue]{hyperref} %pdflatex

\usepackage{natbib}

\usepackage{graphicx} % Required for inserting images
\usepackage{adjustbox}
\usepackage{geometry}
\usepackage{multirow}
\usepackage[toc,page]{appendix}

\usepackage{lscape}
\usepackage{color}
\usepackage{undertilde}
\usepackage{accents}
\usepackage[utf8]{inputenc}
\usepackage{booktabs}
\usepackage{float}
\usepackage{setspace}
\usepackage{subcaption}
\usepackage{tcolorbox}
\usepackage{mathtools}
\usepackage{multicol}
\usepackage{csquotes}
\usepackage{etoolbox}
\usepackage{rotating}
\usepackage{enumitem}
\usepackage{amsbsy}
\usepackage[flushleft]{threeparttable}

\graphicspath{{./Images/}}

\DeclareMathOperator*{\argmax}{arg\,max}
\DeclareMathOperator*{\argmin}{arg\,min}

\hypersetup{ colorlinks = true, citecolor= blue}

\newcommand{\uiota}             {\mbox{\boldmath$\uiota$}}

\def\beq{\begin{equation}}
\def\eeq{\end{equation}}
\def\beqa{\begin{eqnarray}}
\def\eeqa{\end{eqnarray}}
\def\beqan{\begin{eqnarray*}}
\def\eeqan{\end{eqnarray*}}

\def\bc{\begin{center}}
\def\ec{\end{center}}
\def\btable{\begin{table}[htbp]}
\def\etable{\end{table}}
\def\bfig{\begin{figure}[htbp]}
\def\efig{\end{figure}}

\def\bi{\begin{itemize}}
\def\ei{\end{itemize}}

\newlength{\tempheight}
\newlength{\tempwidth}
\newcommand{\rowname}[1]% #1 = text
{\rotatebox{90}{\makebox[\tempheight][c]{\textbf{#1}}}}
\newcommand{\columnname}[1]% #1 = text
{\makebox[\tempwidth][c]{\textbf{#1}}}

\begin{document}

\title{Impact of methodological assumptions and covariates on the cutoff estimation in ROC analysis}

% \author{Author name(s)}
% \address{Author affiliation and full address}
% \email{e-mail address}
%%Uncomment the following line to override copyright year from the default current year.
%\copyrightyear{2022}

%\author{Soutik Ghosal,\authormark{1} Author Two,\authormark{2,*} and Author Three\authormark{2,3}}
\author{Soutik Ghosal}
\address{Division of Biostatistics, Department of Public Health Sciences, School of Medicine, University of Virginia, Charlottesville, VA 22903}
%\address{\authormark{1} Publications Department, Optica Publishing Group, 2010 Massachusetts Avenue NW, Washington, DC 20036\\
%\authormark{2}Publications Department, Optica Publishing Group, 2010 Massachusetts Avenue NW, Washington, DC 20036\\
%\authormark{3}Currently with the Department of Electronic Journals, Optica Publishing Group, 2010 Massachusetts Avenue NW, Washington, DC 20036}

\email{\authormark{*}soutik.ghosal@virginia.edu} %% email address is required

\begin{abstract}
The Receiver Operating Characteristic (ROC) curve stands as a cornerstone in assessing the efficacy of biomarkers for disease diagnosis. Beyond merely evaluating performance, it provides with an optimal cutoff for biomarker values, crucial for disease categorization. While diverse methodologies exist for threshold estimation, less attention has been paid to integrating covariate impact into this process. Covariates can strongly impact diagnostic summaries, leading to variations across different covariate levels. Therefore, a tailored covariate-based framework is imperative for outlining covariate-specific optimal cutoffs. Moreover, recent investigations into cutoff estimators have overlooked the influence of ROC curve estimation methodologies. This study endeavors to bridge this gap by addressing the research void. Extensive simulation studies are conducted to scrutinize the performance of ROC curve estimation models in estimating different cutoffs in varying scenarios, encompassing diverse data-generating mechanisms and covariate effects. Additionally, leveraging the Alzheimer's Disease Neuroimaging Initiative (ADNI) dataset, the research assesses the performance of different biomarkers in diagnosing Alzheimer's disease and determines the suitable optimal cutoffs. \\
%\keywords{Diagnostic accuracy, cutoff estimation, ROC curve, AUC, Alzheimer's disease.}
\end{abstract}

\section{Introduction} \label{Sec: Intro}

The receiver operating characteristic (ROC) curve is a graphical tool for evaluating the diagnostic accuracy of biomarkers for detecting disease with binary outcome, making it one of the most widely embraced tools in medical research. This graphical representation \citep{green1966signal} is crafted by plotting sensitivities (true positive rates) against $1-\text{specificities}$ (false positive rates) across various biomarker thresholds. Over decades, ROC curve analysis has been integral to biomarker assessment, with the area under the ROC curve (AUC) emerging as a pivotal metric for quantifying performance in disease discrimination. However, ROC curve analysis goes beyond mere summarization. It facilitates the identification of optimal biomarker cutoffs, enabling precise disease categorization for future biomarker assessments. This additional dimension of the ROC curve elevates its significance beyond its graphical representation, enhancing its practicality and relevance in clinical settings. Note that, throughout the article, in addition to the term \enquote{cutoff}, the terms \enquote{cutpoint} or \enquote{threshold} are also used interchangeably to refer to the same concept.

In the domain of optimal threshold estimation, a multitude of frameworks have emerged in the literature over the past 70 years. Among these, the Youden Index \citep{youden1950index} reigns as the oldest and most widely recognized. This method identifies the optimal cutoff by maximizing the sum of sensitivity and specificity, offering a foundational approach to threshold determination. Subsequently, \citet{perkins2006} introduced the closest-to-(0,1) criterion, which chooses the optimum cutoff by minimizing the distance of ROC curve points from the perfect classifier point (0,1), where both sensitivity and specificity are at their maximum. Building upon this, \citet{liu2012classification} proposed the concordance probability method, which seeks the optimal cutoff by maximizing the product of sensitivity and specificity, adding a nuanced perspective to threshold selection. In the latest decade, \citet{unal2017defining} introduced the index of union approach, which derives an optimal cutpoint by concurrently maximizing sensitivity and specificity values from the AUC value. %Moreover, \citet{unal2017defining} undertakes a comprehensive comparative analysis of these methodologies within a simulation framework, offering invaluable insights into their respective performances.

While these methodologies have been actively applied in the literature, their capacity to incorporate covariates remains largely unexplored. Biomarker performance seldom exhibits uniformity and its performance may vary across distinct subpopulations characterized by specific covariates. Consequently, covariates hold the potential to significantly influence the diagnostic efficacy of biomarkers, prompting the need for methodological frameworks to appropriately accommodate them \citep{tosteson1988general, toledano1996ordinal, pepe1997regression,ishwaran2000general, de2013bayesian}. It is logical to presume that if diagnostic summaries differ across various levels of covariates, optimal cutoffs will correspondingly vary. For example, in the literature, various cerebrospinal fluid biomarkers are recognized for screening Alzheimer’s disease, and their overall performance has been evaluated for this purpose. Research indicates significant differences in biomarker levels between sexes \citep{sundermann2020sex, mielke2020consideration}, suggesting potential variations in diagnostic capacity at different sex groups. Consequently, it's essential to employ a framework that facilitates the estimation of sex-specific cutoffs. So far in the literature,  \citet{inacio2017nonparametric} proposed a covariate-adjusted framework to estimate only the Youden's index and \citet{to2022receiver} extended some of the aforementioned optimal threshold estimators for ROC surface. However, all of the aforementioned methodologies have yet to be formally extended in the context of ROC curve to incorporate covariate considerations, leaving a notable gap in current research efforts. 

%In the ROC regression setting, where the performance of the biomarker could be impacted by various covariates, numerous methodologies of ROC regression have been proposed to estimate covariate-specific diagnostic accuracy estimates. When the diagnostic accuracy can vary with covariates, the corresponding cutoff estimates should vary accordingly.

Over the years, numerous classes of ROC curves have emerged across various methodological frameworks, spanning empirical \citep{delong1988comparing}, parametric \citep{dorfman1968maximum,metz1998maximum}, semiparametric \citep{pepe2000interpretation}, and nonparametric \citep{hsieh1996nonparametric,zou1997smooth,lloyd1998using} domains. While the empirical ROC curve remains popular among researchers for its simplicity and lack of distributional assumptions, model-based ROC curves offer distinct advantages, including the ability to generate smooth estimates. In recent years, an alternative modeling framework for ROC curves gained traction, centered around placement value (PV), a standardization of diseased biomarker scores relative to healthy biomarker distributions \citep{pepe2003statistical}. PV-based models have proven valuable due to their direct link to the ROC curve, as well as their capability to accommodate covariate effects \citep{sullivan2004analysis,cai2004semi,alonzo2002distribution,stanley2018beta} and constraints \citep{ghosal2019discriminatory,ghosal2022estimation}. Additionally, the literature has seen the emergence of shape-constrained ROC models, ensuring strictly concave ROC curves to avoid \enquote{improper} curves with hooks at extreme specificity levels. Notable examples include bibeta \citep{mossman2016using}, bigamma \citep{dorfman1996proper}, and bichi-squared \citep{hillis2016equivalence} ROC curves. \citet{gonccalves2014roc} has aptly summarized many of these methodologies. It is important to note that methodological assumptions of ROC curve modeling could play a significant role in optimal cutoff estimation. Given that the ROC curve is dependent on the distributions of healthy and diseased biomarkers, estimating cutoffs involves optimizing the functions involving sensitivity and specificity, both of which are functions of these distributions. Any variation in these distributions within the ROC framework can naturally influence the estimation of cutoff points. However, the comparative evaluation of different cutoff estimators has overlooked the impact of various ROC estimation models \citep{unal2017defining,hajian2018choice,rota2014finding}, creating a gap in research.

%Researchers often use one methodology over another for their own purposes and often different ROC curves result in similar diagnostic accuracy estimates, especially the commonly used AUC. Although areas from different ROC curves could be similar and might infer the similar performance of different biomarkers, the corresponding cutoff estimates could be different. Even, the distributional choice of modeling the ROC curve can impact both the AUC and the cutoff estimate.

%The ROC curve is essentially a distribution function and hence non-decreasing. However, the ROC curve is not necessarily concave and hence, often a non-concave ROC curve can show hooks at low or high specificity levels resulting in an 'improper' ROC curve. Sometimes, the area under a non-concave ROC curve can fall below 0.5 making the summary measure uninterpretable. In the literature, these issues have been addressed and some concave ROC methodologies have been proposed (REF bibeta \citep{mossman2016using}, bigamma \citep{dorfman1996proper}) that can fix these issues and ensure we obtain meaningful sensitivity estimates at all specificity levels and the resultant summary measures are always interpretable. Regarding the choice of concave or non-concave ROC, the estimation of the cutoffs can vary significantly. However, in the literature, the impact of the concave ROC curve on the estimation of the cutoff has not been tested.

The novelty of our work lies in addressing these gaps in diagnostic accuracy research. Primarily, we aim to shed light on the performance of optimal threshold methods across various ROC estimating models. Additionally, we seek to enhance current optimal cutoff frameworks by integrating covariate considerations. This comprehensive approach promises to provide valuable insights into both threshold estimation methodologies and the impact of covariates on diagnostic accuracy assessment. The rest of the article is organized as follows. Section \ref{Sec: Method} provides detailed descriptions of different ROC curve models, optimal threshold frameworks, covariate adjustment, and estimation mechanisms. We demonstrate the performance of the methodology through extensive simulation in Section \ref{Sec: Sim} and present application with the Alzheimer's disease data in Section \ref{Sec: Application}. We conclude with a brief discussion in Section \ref{Sec: Discussion}.

%Various types of ROC regression
%Role of ROC methodologies – discrimination and future data classification (often the 2nd role is not considered)
%The cutoff can be impacted by how ROC is estimated
%Covariate-wise different cutoff, when the performance of the biomarker can vary with covariate
%Also the cutoff could be different whether the ROC is concave or non-concave
%Additionally the AUC of different ROC curves could be similar even though the cutoffs are different
%This paper intends to find the robustness in the estimation of AUC and the cutoff.

\section{Methodology} \label{Sec: Method}

\subsection{General framework} \label{Sec: Structure}

Let $Y_0$ and $Y_1$ be the healthy and diseased biomarkers respectively discriminating the disease $D$ of binary classification ($D = 0$ or $D=1$ respectively denoting whether the corresponding group is healthy or diseased). Conventionally, we assume that,
\begin{equation}
Y_0 \sim F_0(\cdot) \text{ and } Y_1 \sim F_1(\cdot), \nonumber
\end{equation}
\noindent
where $F_0$ and $F_1$ denote the healthy and diseased biomarker distribution respectively. Then the corresponding ROC curve and its summary AUC can be written as:
\begin{align}
ROC(t) &= 1-F_1\left( F_0^{-1}(1-t)\right), \text{ } t \in (0,1). \label{Eq: ROCdef} \\
AUC &= \int_0^1 ROC(t)dt. \label{Eq: AUCdef}
\end{align}
Without loss of generality, let's assume that higher values of the biomarker indicate a diseased group, i.e. $Y_0 < Y_1$. Based on the distributional assumption of $Y_0$ and $Y_1$, we can write the sensitivity ($se$) and specificity ($sp$) at a certain cutpoint $c$ as
\begin{align} \label{Eq: se_sp}
se(c) &= P[Y_1 > c] = 1-F_1(c), \text{ and} \\ \nonumber
sp(c) &= P[Y_0 < c] = F_0(c). \nonumber
\end{align}

\subsubsection{Special ROC frameworks} \label{Sec: specialcase}
%\paragraph{Special cases}: 

Drawing from the general definition of the ROC curve in (\ref{Eq: ROCdef}), various types of ROC frameworks have emerged in the literature, including empirical, parametric, and nonparametric approaches. In this study, we will compare several of these ROC models, including the empirical (Emp) model, the Kernel-based nonparametric model (NonPar), and the widely used binormal (BN) model. For a comprehensive overview of these frameworks and others, please refer to Section~\ref{AppSec: Noted_general_ROC_models} in the Appendix.

\subsubsection{Optimal cutoffs} \label{Sec: Cutoff}
In this section, we introduce the mathematical definitions of the most renowned optimal cutoff methods employed in diagnostic accuracy research. \citet{unal2017defining} provided a thorough summary of various cutoff estimators. However, for the sake of completeness, we will provide a brief overview of them here as well.
 %By examining their theoretical foundations and mathematical formulations, we aim to elucidate their mechanisms for determining the optimal threshold in biomarker-based disease diagnosis.
\begin{enumerate}
    \item \textbf{Youden's index} ($J$): Youden's index \citep{youden1950index} stands out as one of the oldest and most widely used optimal cutoff frameworks. It determines the threshold by maximizing the sum of sensitivities and specificities across various cutoff points.
    \begin{align}\label{Eq: def_J}
    %c_{J} &= max_c J(c) = max_c \{ se(c) + sp(c) - 1\} 
    J(c) &= se(c) + sp(c) - 1, \\ \nonumber
    c_{J} &= \argmax_{c \in \Re} J(c) \nonumber
    \end{align}
    \item \textbf{Closest to $(0, 1)$ criteria} (ER): In this criterion \citep{perkins2006}, the optimal threshold is derived by minimizing the Euclidean distance between the ROC curve and the perfect classifier point, which is located at coordinates $(0,1)$. The optimal cutoff is thus determined as the threshold for which the sensitivity and $1-\text{specificity}$ pair on the ROC curve is closest to $(0,1)$.
        \begin{align}\label{Eq: def_ER}
        %c_{ER} &= min_c ER(c) = min_c \{ \sqrt{(1-se(c))^2 + (1-sp(c))^2} \} 
        ER(c) &= \sqrt{(1-se(c))^2 + (1-sp(c))^2}, \\ \nonumber
        c_{ER} &= \argmin_{c \in \Re} ER(c)
        \end{align}
    \item \textbf{Concordance probability method criteria} (CZ): This criterion \citep{liu2012classification} operates by maximizing the product of sensitivity and specificity across different thresholds. The optimal cutoff is identified as the threshold that achieves the maximum product.
    \begin{align} \label{Eq: def_CZ}
        %c_{CZ} &= max_c CZ(c) = max_c \{ se(c) \times sp(c) \} 
        CZ(c) &= se(c) \times sp(c), \\ \nonumber
        c_{CZ} &= \argmax_{c \in \Re} CZ(c) \nonumber
    \end{align}
    \item \textbf{Index of union criteria} (IU): The index of union, as outlined by \citet{unal2017defining}, represents one of the latest criteria. It operates on the premise that the optimal threshold occurs where sensitivity and specificity are simultaneously close to the AUC, while also minimizing the difference between sensitivity and specificity.
        \begin{align} \label{Eq: def_IU}
        %c_{IU} &= min_c IU(c) = min_c \{ |se(c) - AUC| + |sp(c)-AUC| \} 
        IU(c) &= |se(c) - AUC| + |sp(c)-AUC|, \\ \nonumber
        c_{IU} &= \argmin_{c \in \Re} IU(c) \nonumber
        \end{align}
\end{enumerate}

As previously mentioned, sensitivity and specificity are functions of $F_0$ and $F_1$ (from equation (\ref{Eq: se_sp})). By substituting equation (\ref{Eq: se_sp}) into equations (\ref{Eq: def_J}) - (\ref{Eq: def_IU}), we can express the optimal cutoffs in terms of $F_0$ and $F_1$ as follows:
\begin{align} \label{Eq: def_cutoff_F}
c_{J} &= \argmax_{c \in \Re} J(c) = \argmax_{c \in \Re} \{ F_0(c)-F_1(c)\} \\ \nonumber
c_{ER} &= \argmin_{c \in \Re} ER(c) = \argmin_{c \in \Re} \{ \sqrt{(F_1(c))^2 + (1-F_0(c))^2} \} \\ \nonumber
c_{CZ} &= \argmax_{c \in \Re} CZ(c) = \argmax_{c \in \Re} \{ (1-F_1(c)) \times F_0(c) \} \\ \nonumber
c_{IU} &= \argmin_{c \in \Re} IU(c) = \argmin_{c \in \Re} \{ |1-F_1(c) - AUC| + |F_0(c)-AUC| \} \nonumber
\end{align}

\subsection{Alternative ROC framework: Placement value-based model} \label{Sec: PV}

Unlike the conventional approach of individually modeling $F_0$ and $F_1$ within the general ROC curve framework, this alternative framework utilizes placement value (PV). PV can be conceptualized as a standardization of the diseased biomarkers ($Y_1$) with respect to the distribution of healthy biomarkers ($F_0$). It quantifies the separation between healthy and diseased biomarkers \citep{pepe2003statistical}. Given $Y_0$ and $Y_1$, PV can be calculated as:
\begin{equation}
Z = 1-F_0(Y_1).\label{Eq:PV} 
\end{equation}
The adoption of PV proves beneficial, as it can be shown that the distribution function of $Z$ corresponds to the ROC curve. This alternative framework leveraging PV enables the direct modeling of the ROC curve. Furthermore, this approach streamlines the incorporation of covariates into the model as needed.
 
\subsubsection{Special PV-based ROC frameworks} \label{Sec: specialPVcase}

Assuming $F$ represents the CDF of the PV random variable $Z$, representing the ROC curve, the PV-based framework requires the modeling of $F_0$ and $F$.
% Then, in the PV-based framework, we have
%\begin{equation}
%Y_0 \sim F_0(\cdot) \text{ and } Z = 1-F_0(Y_1)  \sim F (\cdot), \nonumber
%\end{equation}
%\noindent
%where $F$ is the CDF of $Z$ i.e. the ROC curve.
Hence, the overall structure of modeling PV-based ROC is:
\begin{align}\label{Eq: PVmodelStructure}
    Y_0 &\sim F_0(\cdot), \\ \nonumber
    Z &= 1 - F_0(Y_1), \\ \nonumber
    \eta^{-1}(Z) &\sim F(\cdot) 
\end{align}
\noindent
where, $\eta$ is a suitable transformation on the PV. Based on how we specify $F$, $\eta$ can be chosen accordingly. Based on how $F_0$ and $F$ are specified, different types of PV-based models can be proposed. For example, \citet{chen2019} showed the performances of the PV-based models by choosing Gaussian distributions for both $F_0$ and $F$, and using both logit and probit links as $\eta$. Later, \citet{ghosal2019discriminatory} proposed a transformed normal PV-regression model by accounting for covariates and using similar distributional assumptions. \citet{stanley2018beta} used a quantile regression approach for estimating the covariate-adjusted conditional distribution of $F_0$ and assumed Beta distribution for $F$ and used an identity link for $\eta$. There are also examples of semiparametric and nonparametric considerations for $F_0$ and $F$ in the literature \citep{ghosal2022estimation, inacio2022covariate, de2018bayesian}.

In this article, we employ several PV-based frameworks, including a parametric PV (PV) model and a semiparametric PV (Semi.PV) model, to assess their efficacy in estimating the cutoffs. These models are delineated in Section~\ref{AppSec: Noted_PV_ROC_models} in the Appendix.

\subsubsection{Optimal cutoffs} \label{Sec: Cutoff_PV}

In this section, we introduce the process of estimating cutoffs within the PV-based framework, which depends on estimating $F_0$ and $F$. Now, based on equation (\ref{Eq:PV}), we can write
\begin{align*}
F_0(Y_1) &= 1-Z, \\
\Rightarrow Y_1 &= F_0^{-1}(1-Z). 
\end{align*}
Then, analogous to equation (\ref{Eq: se_sp}), the sensitivity and specificity for the PV-based setup can be rewritten in terms of $F_0$ and $F$ as
\begin{align} \label{Eq: se_sp_PV}
sp(c) &= P[Y_0 < c] = F_0(c),\\ \nonumber
se(c) & = P[Y_1 > c] = P[F_0^{-1}(1-Z) > c] \nonumber \\
& = 1-P[F_0^{-1}(1-Z) < c] \nonumber \\
& = 1-P[1-z < F_0(c)] \nonumber \\
& = 1-P[z > 1-F_0(c)] \nonumber \\
& = P[z < 1-F_0(c)] \nonumber \\
&= F\left( 1-F_0(c) \right) \nonumber
\end{align}
Hence, substituting the above into equations (\ref{Eq: def_J}) - (\ref{Eq: def_IU}), we can express the optimal cutoffs for a PV-based framework in terms of $F_0$ and $F$ as:
\begin{align} \label{Eq: def_cutoff_PV}
c_{J} &= \argmax_{c \in \Re} J(c) = \argmax_{c \in \Re} \{ F_0(c) + F\left( 1-F_0(c)\right) – 1 \} \\ \nonumber
c_{ER} &= \argmin_{c \in \Re} ER(c) = \argmin_{c \in \Re} \{ \sqrt{(1-F\left( 1-F_0(c) \right))^2 + (1-F_0(c))^2} \} \\ \nonumber
c_{CZ} &= \argmax_{c \in \Re} CZ(c) = \argmax_{c \in \Re} \{ F\left( 1-F_0(c) \right) \times F_0(c) \} \\ \nonumber
 c_{IU} &= \argmin_{c \in \Re} IU(c) = \argmin_{c \in \Re} \{ |F\left( 1-F_0(c) \right) - AUC| + |F_0(c)-AUC| \} \nonumber
\end{align}

\subsection{Covariate adjustment} \label{Sec: Cov}
In this section, we extend the optimal threshold estimating frameworks to allow for covariates in the estimation of the thresholds. For the sake of simplicity, we illustrate the adjustment for one covariate, however, it can easily be extended for multiple covariates. Assume that $X_0$ and $X_1$ are respectively the covariates for the healthy and diseased biomarker groups. In the next subsections, we will introduce the covariates in the general and the PV-based framework, and lay out the forms of optimal cutoffs for corresponding frameworks.

\subsubsection{Covariates in general ROC framework} \label{Sec: Cov_gen_ROC}

For the general framework of the ROC curve, we can model $Y_0$ and $Y_1$ on the corresponding covariates as: 
\begin{align*}
Y_0 &\sim F_0(\cdot|X_0) \text{ and } Y_1 \sim F_1(\cdot|X_1).
\end{align*}
A covariate $x$-specific ROC and AUC estimates will have the forms given as:
\begin{align} \label{Eq: ROC_AUC_x_def}
ROC_x(t) &= 1-F_1\left( F_0^{-1}(1-t|X_0 = x)|X_1 = x\right), \text{ } t \in (0,1),\\ \nonumber
AUC_x &= \int_0^1 ROC_x(t)dt. \nonumber
\end{align}
Furthermore, the covariate $x$-specific sensitivity and specificity can be written as
\begin{align*}
se(c|x) &= P[Y_1 > c|X_1 = x] = 1-F_1(c|x), \\
sp(c|x) &= P[Y_0 < c|X_0 = x] = F_0(c|x). 
\end{align*}
Following equation (\ref{Eq: def_cutoff_F}), we can introduce covariates in the threshold estimators as:
\begin{align} \label{Eq: def_cutoff_Fx}
c_{J,x} &= \argmax_{c \in \Re} J_x(c) = \argmax_{c \in \Re} \{ F_0(c|x)-F_1(c|x)\} \\ \nonumber
c_{ER,x} &= \argmin_{c \in \Re} ER_x(c) = \argmin_{c \in \Re} \{ \sqrt{(F_1(c|x))^2 + (1-F_0(c|x))^2} \} \\ \nonumber
c_{CZ,x} &= \argmax_{c \in \Re} CZ_x(c) = \argmax_{c \in \Re} \{ (1-F_1(c|x)) \times F_0(c|x) \} \\ \nonumber
c_{IU,x} &= \argmin_{c \in \Re} IU_x(c) = \argmin_{c \in \Re} \{ |1-F_1(c|x) - AUC_x| + |F_0(c|x)-AUC_x| \} \nonumber
\end{align}
\noindent
where, $AUC_x$ is the covariate $x$-specific estimate of AUC. 

Given this structure, covariates can be incorporated into various frameworks of ROC curve modeling. For instance, when considering covariates within the widely popular BN framework, we anticipate employing separate linear regression models to characterize the healthy and diseased biomarkers, as illustrated below:
\begin{align*}
y_{0i} &= \beta_{00} + \beta_{10} X_{0i} + \epsilon_{0i}, \text{ } \epsilon_{0i} \sim N(0, \sigma_0^2), \text{ } i=1,2,\ldots,n_i,  \\
y_{1j} &= \beta_{01} + \beta_{11} X_{1j} + \epsilon_{1j}, \text{ } \epsilon_{1j} \sim N(0, \sigma_1^2), \text{ } j=1,2,\ldots,n_j,
\end{align*}
\noindent
where $i$ and $j$ respectively corresponds to healthy and diseased subjects, $n_k$ corresponds to the sample size, $\beta_{0k}$ and $\beta_{1k}$-s are different intercept and slope parameters, and $\epsilon_{k}$-s are the errors corresponding to $k^{\text{th}}$ groups with $k=0,1$. Then, following the conventional BN framework in Section~\ref{AppSec: Noted_general_ROC_models}, we have the covariate $x$-specific BN similar to (\ref{Eq: BN_ab}) as:
    \begin{align*} 
        a_x = \frac{(\beta_{01}-\beta_{00}) + (\beta_{11}-\beta_{10})\cdot x}{\sigma_1}, \text{ } b = \frac{\sigma_0}{\sigma_1},
    \end{align*}
\noindent
and the covariate $x$-specific ROC and AUC estimates:
    \begin{align*}
        ROC_x(t) &= \Phi(a_x + b \cdot\Phi^{-1}(t)), \text{ } t \in (0,1), \\  
        AUC_x &= \Phi \left( \frac{a_x}{\sqrt{1+b^2}} \right).
    \end{align*}
We can specify $F_0(\cdot |X_0 = x) = \Phi(\cdot; \beta_{00}+\beta_{10}x, \sigma_0)$ and $F_1(\cdot |X_1 = x) = \Phi(\cdot; \beta_{01}+\beta_{11}x, \sigma_1)$ in (\ref{Eq: def_cutoff_Fx}) to estimate BN version of different cutoffs, where $\Phi(y)$ corresponds to a standard normal CDF obtained at $y$ and $\Phi(y; \mu, \sigma)$ denotes a normal CDF with mean $\mu$ and standard deviation $\sigma$ obtained at $y$.

\subsubsection{Covariates in PV-based framework} \label{Sec: Cov_PV_ROC}

In the PV-based framework, PV is estimated by incorporating the diseased biomarker $Y_1$ and corresponding covariate $X_1$ in the healthy biomarker CDF $F_0$ as 
\begin{align*}
z|(X_1 = x_1) &= 1-F_0(y_1|X_0 = x_1). 
\end{align*}
Then, the covariate $x$-specific ROC curve and AUC have the form in terms of $Z$ as
\begin{align} \label{Eq: ROC_AUC_PVx_def}
ROC_x(t) &= F(t|X_1 = x) = P[Z\leq t|X_1 = x], \text{ } t \in (0,1),\\ \nonumber
AUC_x &= \int_0^1 ROC_x(t)dt. \nonumber
\end{align}

Then, the covariate $x$-specific sensitivity and specificity (following (\ref{Eq: se_sp_PV})) can be written as
\begin{align} \label{Eq: se_sp_PVx}
sp(c|x) &= P[Y_0 < c|X_0 = x] = F_0(c|x), \\ \nonumber
se(c|x) &= F\left( 1-F_0(c|X_0 = x)|X_1 = x \right) \\ \nonumber
\end{align}

Then, based on equation (\ref{Eq: def_cutoff_PV}), we have the following forms of covariate-specific optimal thresholds for the PV-based setup:
\begin{align} \label{Eq: def_cutoff_PVx}
c_{J,x} &= \argmax_{c \in \Re} J_x(c) = \argmax_{c \in \Re} \{ F_0(c|x) + F\left( 1-F_0(c|x)|x\right) – 1 \} \\ \nonumber
c_{ER,x} &= \argmin_{c \in \Re} ER_x(c) = \argmin_{c \in \Re} \{ \sqrt{(1-F\left( 1-F_0(c|x)|x \right))^2 + (1-F_0(c|x))^2} \} \\ \nonumber
c_{CZ,x} &= \argmax_{c \in \Re} CZ_x(c) = \argmax_{c \in \Re} \{ F\left( 1-F_0(c|x)|x \right) \times F_0(c|x) \} \\ \nonumber
 c_{IU,x} &= \argmin_{c \in \Re} IU_x(c) = \argmin_{c \in \Re} \{ |F\left( 1-F_0(c|x)|x \right) - AUC_x| + |F_0(c|x)-AUC_x| \} \nonumber
\end{align}

For a parametric PV regression model (similar to one in Section~\ref{AppSec: Noted_PV_ROC_models}), we can accommodate covariates and estimate covariate $x$-specific ROC and AUC in the following way:
\begin{align*}
y_{0i} &= \beta_{00} + \beta_{10} X_{0i} + \epsilon_{0i}, \text{ } \epsilon_{0i} \sim N(0, \sigma_0^2), \\
z|(X_1=x_1)  &= 1 - \Phi \left( y_1; \beta_{00}+\beta_{10}x_1, \sigma_0\right),\\
%\Phi^{-1}(z_j) &= \beta_0 + \beta_{1}X_{1j} + \epsilon_{j}, \text{ } \epsilon_{j} \sim N(0, \sigma^2), \text{ } j=1,2,\ldots,n_j, \\
ROC_x(t) &= \Phi \left(\Phi^{-1}(t); \beta_{0}+\beta_{1}x, \sigma \right), \text{ } t \in (0,1),\\
AUC_x &= \int_0^1 ROC_x(t)dt.
\end{align*}

Finally, we can specify $F_0(\cdot |X_0 = x) = \Phi(\cdot; \beta_{00}+\beta_{10}x, \sigma_0)$ and $F(\cdot |X_1 = x) = \Phi(\cdot; \beta_{0}+\beta_{1}x, \sigma)$ in (\ref{Eq: def_cutoff_PVx}) to estimate parametric PV version of different cutoffs.

The inclusion of covariates in a Semi.PV model proceeds in a similar manner. For further details, please refer to equation (\ref{Eq: semiparametricPV}) and consult \citet{ghosal2022estimation} and \citet{inacio2017nonparametric}.

\subsection{Bayesian computational aspects} 
\label{Sec: Likelihood}

We take a Bayesian approach for the inference purposes discussed so far. We use proper objective prior for all the parameters. Specifically, each of the mean parameters $\mu$-s along with regression coefficients such as intercept ($\beta_0$-s) and slope ($\beta_1$-s) parameters follow $N(0,100)$ priors and variance parameter $\sigma^2$ follows $IG(0.01, 0.01)$.

We use \texttt{RJAGS} to implement Monte Carlo Markov chain (MCMC) algorithms to generate samples from the posterior distribution of the model parameters given the data. Both visual inspection of the trace plots and diagnostic tools \citep{gelman1992inference} are used to ensure convergence of the MCMC chains. After convergence, we thin the iterations to produce a sample of 5000 to produce posterior means, standard deviations and 95\% credible intervals. The algorithm is implemented in \texttt{R}. %R code of implementing data analysis will be made available online.

\section{Simulation} \label{Sec: Sim}
%Previous research has thoroughly investigated the efficacy of diverse ROC estimating frameworks and various methods for determining cut points using the same ROC estimation technique. However, the effectiveness of an optimal cut-point estimator may be influenced by how we estimate the ROC curve itself. In a study by Unal et al. (2017), the authors assessed the performance of different cut-point estimation methods, employing a unified framework for ROC curve estimation.

%Beyond the impact of ROC estimating methodology on optimal cut-point algorithms, there is a notable absence of a covariate-specific optimal cut-point selection framework. 

To assess the efficacy of various optimal threshold estimators across diverse ROC estimation methodologies, we have conducted extensive simulations. These simulations encompass scenarios both with and without the inclusion of covariates, which will be detailed in the following two subsections.
%In this section, we conduct comprehensive simulations to examine how existing cut-point selection algorithms perform across different ROC estimating frameworks, both with and without the inclusion of covariates.

\subsection{Simulation without covariate} \label{Sec: Sim_noCov}

A list of simulation scenarios is considered in this subsection to understand the performances of different cutoff methodologies for different ROC curve estimation frameworks. The simulation settings vary concerning different data-generating mechanisms, varying AUC levels (low, medium, and high), and sample sizes (small, medium, and high). The different data-generating mechanisms mimic that of \citet{hajian1997comparison} and \citet{faraggi2002estimation} to ensure that the biomarker values are generated from varying distributions (normal, skewed, and mixture distributions). The details of the scenario are tabulated in Table \ref{Tab:Sim_nocov_mechanism}.

%\begin{table}[!ht]
%\begin{center}
%\caption{\label{Tab:Sim_nocov_mechanism} Details of simulation mechanism without covariate.}
%\begin{adjustbox}{width=\textwidth}
%\begin{tabular}{llll}
%\hline
%\textbf{Data}       & \textbf{Fitting}                              & \textbf{AUC}   & \textbf{Sample}    \\
%\textbf{generating} & \textbf{models}                               & \textbf{level} & \textbf{size}      \\
%\textbf{mechanisms}   & \textbf{}                                     & \textbf{}      & \textbf{}          \\ \hline
%BN equal            & Empirical (Emp)                               & Low            & Small ($N = 50$)   \\
%BN unequal          & Binormal (BN)                                 & Medium         & Medium ($N = 100$) \\
%Skewed I            & Bigamma (BG)                                  & High           & High ($N = 500$)   \\
%Skewed II           & Proper Binormal (BiChi)                       &                &                    \\
%Skewed III          & Kernel-based nonparametric (NonPar)           &                &                    \\
%Mixed I             & Placement value-based parametric (PV)         &                &                    \\
%Mixed II            & Placement value-based semiparametric (Semi.PV) &                &                    \\ \hline
%\end{tabular}
%\end{adjustbox}
%\end{center}
%\end{table}

\begin{table}[!ht]
\begin{center}
\caption{\label{Tab:Sim_nocov_mechanism} Details of simulation mechanism without covariate.}
%\begin{adjustbox}{width=\textwidth}
\begin{tabular}{llll}
\hline
\textbf{Data}       & \textbf{Fitting}                              & \textbf{AUC}   & \textbf{Sample}    \\
\textbf{generating} & \textbf{models}                               & \textbf{level} & \textbf{size}      \\
\textbf{mechanisms}   & \textbf{}                                     & \textbf{}      & \textbf{}          \\ \hline
BN equal            & Empirical (Emp)                               & Low            & Small ($N = 50$)   \\
BN unequal          & Binormal (BN)                                 & Medium         & Medium ($N = 100$) \\
Skewed I            & Kernel-based nonparametric (NonPar)                                   & High           & High ($N = 500$)   \\
Skewed II           & Placement value-based parametric (PV)                       &                &                    \\
Skewed III          & Placement value-based semiparametric (Semi.PV)           &                &                    \\
Mixed I             &          &                &                    \\
Mixed II            & 		 &                &                    \\ \hline
\end{tabular}
%\end{adjustbox}
\end{center}
\end{table}

By varying the types of data generating mechanisms and AUC levels, for each sample size we have $7 \times 3 = 21$ distinct simulation settings. For each of the simulation settings, a series of ROC curve estimation methodologies have been used including several parametric, nonparametric, semiparametric methods mentioned in Table~\ref{Tab:Sim_nocov_mechanism} to estimate AUC, ROC, and the four different optimal cutpoints discussed in Section \ref{Sec: Cutoff}. We create 1000 data replicates and report  median and IQR (inter quartile range) of biases obtained by different ROC estimation methods to estimate AUC and four different optimal cutpoints. We also plot the biases incurred by different ROC methods for different simulation mechanism. We opted for median and IQR instead of mean and standard deviation (SD) due to occasional skewness in the distribution of cutoff estimates obtained from different data replicates. Additionally, this choice aligns with the presentation in the figures for consistency. The details of the data-generating mechanism and corresponding true parameter values are tabulated in Table \ref{Tab:Sim_nocov_generation}. Subsequently, the true AUC and cutoff estimates are tabulated in Table A.1 of the supplementary material. In this paper, we will illustrate the simulation results with medium sample size, i.e., $N=100$ for both healthy and diseased biomarker samples. The simulation results with the rest two sample size categories will be described in the supplementary material.

\begin{table}[!ht]
\begin{center}
\caption{\label{Tab:Sim_nocov_generation} Data generation mechanism details with true parameters, no covariate framework.}
\begin{adjustbox}{width=\textwidth}
\begin{tabular}{ccccc}
\hline
\textbf{Data}       & \textbf{Data}                                          & \multicolumn{3}{c}{\textbf{True}}                                                                            \\
\textbf{generating} & \textbf{generation}                                    & \multicolumn{3}{c}{\textbf{parameter}}                                                                       \\ \cline{3-5} 
\textbf{scenario}   &                                                        & \textbf{Low AUC}                   & \textbf{Medium AUC}                & \textbf{High AUC}                  \\ \hline
BN equal            & $Y_0 \sim N(\mu_0, \text{ } \sigma^2)$                 & $\mu_0 = 0$, $\mu_1 = 0.2$,        & $\mu_0 = 0$, $\mu_1 = 1$,          & $\mu_0 = 0$, $\mu_1 = 2.5$,        \\
                    & $Y_1 \sim N(\mu_1, \text{ } \sigma^2)$                 & $\sigma = 1$                       & $\sigma = 1$                       & $\sigma = 1$                       \\ \hline
BN unequal          & $Y_0 \sim N(\mu_0, \text{ } \sigma_0^2)$               & $\mu_0 = 0$, $\mu_1 = 0.2$,        & $\mu_0 = 0$, $\mu_1 = 1$,          & $\mu_0 = 1$, $\mu_1 = 2.9$,        \\
                    & $Y_1 \sim N(\mu_1, \text{ } \sigma_1^2)$               & $\sigma_0 = 1.2$, $\sigma_1 = 0.8$ & $\sigma_0 = 1.2$, $\sigma_1 = 0.5$ & $\sigma_0 = 0.5$, $\sigma_1 = 1.2$ \\ \hline
Skewed I            & $Y_0^{\frac{1}{2}} \sim N(\mu_0, \text{ } \sigma_0^2)$ & $\mu_0 = 0$, $\mu_1 = 0.2$,        & $\mu_0 = 0$, $\mu_1 = 1$,          & $\mu_0 = 1$, $\mu_1 = 2.5$,        \\
                    & $Y_1^{\frac{1}{2}} \sim N(\mu_1, \text{ } \sigma_1^2)$ & $\sigma_0 = 1.2$, $\sigma_1 = 1$   & $\sigma_0 = 1$, $\sigma_1 = 0.7$   & $\sigma_0 = 1$, $\sigma_1 = 0.5$   \\ \hline
Skewed II           & $log(Y_0) \sim N(\mu_0, \text{ } \sigma_0^2)$          & $\mu_0 = 0$, $\mu_1 = 0.2$,        & $\mu_0 = 0$, $\mu_1 = 1$,          & $\mu_0 = 1$, $\mu_1 = 2.5$,        \\
                    & $log(Y_1) \sim N(\mu_1, \text{ } \sigma_1^2)$          & $\sigma_0 = 1$, $\sigma_1 = 1$     & $\sigma_0 = 1$, $\sigma_1 = 0.7$   & $\sigma_0 = 1$, $\sigma_1 = 0.5$   \\ \hline
Skewed III          & $Y_0 \sim Gamma(k, \text{ } \theta_0)$                 & $k = 0.5$, $\theta_0 = 0.1$        & $k = 0.5$, $\theta_0 = 0.1$,       & $k = 0.5$, $\theta_0 = 0.1$,       \\
                    & $Y_0 \sim Gamma(k, \text{ } \theta_1)$                 & $\theta_1 = 0.15$                  & $\theta_1 = 0.6$                   & $\theta_1 = 7$                     \\
                    & $k$: shape, $\theta_j$: scale                          &                                    &                                    &                                    \\ \hline
Mixed I             & $Y_0 \sim N(\mu_0, \text{ } \sigma_0^2)$               & $\mu_0=0$, $\sigma_0=1$,           & $\mu_0=0$, $\sigma_0=1$,           & $\mu_0=0$, $\sigma_0=1$,           \\
                    & $Y_1 \sim N(\pi \mu_{11} + (1-\pi) \mu_{12},$          & $\pi = 0.5$,                       & $\pi = 0.5$,                       & $\pi = 0.5$,                       \\
                    & $\pi^2 \sigma_{11}^2 + (1-\pi)^2\sigma_{12}^2)$        & $\mu_{11}=0$, $\sigma_{11}=1$      & $\mu_{11}=0$, $\sigma_{11}=1$      & $\mu_{11}=0$, $\sigma_{11}=1$      \\
                    &                                                        & $\mu_{12}=1$, $\sigma_{12}=5$      & $\mu_{12}=4$, $\sigma_{12}=5$      & $\mu_{12}=8$, $\sigma_{12}=5$      \\ \hline
Mixed II            & $Y_0 \sim N(\pi_0 \mu_{01} + (1-\pi_0) \mu_{02},$      & $\mu_{01}=0$, $\sigma_{01}=1$,     & $\mu_{01}=0$, $\sigma_{01}=1$,     & $\mu_{01}=0$, $\sigma_{01}=1$,     \\
                    & $\pi_0^2 \sigma_{01}^2 + (1-\pi_0)^2\sigma_{02}^2)$    & $\mu_{02}=1$, $\sigma_{02}=2$,     & $\mu_{02}=1$, $\sigma_{02}=5$,     & $\mu_{02}=1$, $\sigma_{02}=5$,     \\
                    & $Y_1 \sim N(\pi_1 \mu_{11} + (1-\pi_1) \mu_{12},$      & $\pi_0=0.5$                        & $\pi_0=0.5$                        & $\pi_0=0.5$                        \\
                    & $\pi_1^2 \sigma_{11}^2 + (1-\pi_1)^2\sigma_{12}^2)$    & $\mu_{11}=0$, $\sigma_{11}=1$,     & $\mu_{11}=0$, $\sigma_{11}=1$,     & $\mu_{11}=0$, $\sigma_{11}=1$,     \\
                    &                                                        & $\mu_{12}=1.5$, $\sigma_{12}=2.5$, & $\mu_{12}=2.5$, $\sigma_{12}=2.5$, & $\mu_{12}=5$, $\sigma_{12}=2.5$,   \\
                    &                                                        & $\pi_1=0.4$                        & $\pi_1=0.4$                        & $\pi_1=0.4$                        \\ \hline
\end{tabular}
\end{adjustbox}
\end{center}
\end{table}

Table~\ref{Tab: bias_nocov_medSamp} presents the median and IQR of biases in estimating AUC and four optimal threshold estimates across various ROC fitting models under different data generating mechanisms, specifically focusing on a medium sample size of $N = 100$. When the data are generated from \enquote{BN equal}, all models exhibit similar performance across different AUC levels. However, at low AUC level, the estimation of Youden's index shows higher variability. Conversely, for data generated from \enquote{BN unequal}, the empirical (Emp) and kernel-based (NonPar) models perform less effectively compared to the binormal (BN) and PV-based parametric (BN) and semiparametric (Semi.PV) models. For the \enquote{Skewed I} data generating mechanism, Emp and NonPar demonstrate minimal biases in estimating the cutoffs at low and medium AUC levels, with BN and PV models also performing well. In the case of \enquote{Skewed II} data generating mechanism, Emp and NonPar consistently perform satisfactorily across all AUC levels, while Semi.PV occasionally outperforms them, especially at low AUC levels. Here, the performance of the Semi.PV model occasionally emerges as the superior choice, notably excelling in certain scenarios such as estimating $J$ and $IU$ at low AUC level, and estimating $J$ at medium AUC level. Furthermore, in the majority of cases, it closely rivals the performance of the Emp and NonPar models. Conversely, while the performance of the BN and PV models is not subpar, they seldom achieve top rankings, with only occasional instances of outperforming others, such as in estimating $ER$ at high AUC levels. In the context of \enquote{Skewed III} mechanism, Emp and NonPar maintain consistent performance at low AUC levels, while Semi.PV and BN models display the lowest bias at medium and high AUC levels, respectively. For the \enquote{Mixed I} and \enquote{Mixed II} mechanisms, BN and PV models exhibit the least bias, with Semi.PV closely following. In the case of these final two mechanisms, the performances of the Emp and NonPar models were notably suboptimal, attributed to the data generating mechanism involving a mixture of normals. Interestingly, these mechanisms mirrored the \enquote{BN unequal} mechanism, yielding comparable findings.

\begin{table}[!htbp]
%\begin{landscape}
\begin{center}
\caption{\label{Tab: bias_nocov_medSamp} Biases of estimating AUC and optimal thresholds for different fitting models and different AUC levels for medium sample size, no covariate framework.}
\begin{adjustbox}{angle=90, width=0.5\textwidth}
\begin{tabular}{ccccccccccccccccc}
\hline
\textbf{Data}               & \textbf{Fitting} & \multicolumn{5}{c}{\textbf{Low AUC}}                                                                   & \multicolumn{5}{c}{\textbf{Medium AUC}}                                                                & \multicolumn{5}{c}{\textbf{High AUC}}                                                                  \\ \cline{3-17} 
\textbf{generating}         & \textbf{model}   & \multicolumn{5}{c}{\textbf{Median  $\pm$    IQR}}                                                      & \multicolumn{5}{c}{\textbf{Median  $\pm$  IQR}}                                                        & \multicolumn{5}{c}{\textbf{Median  $\pm$  IQR}}                                                        \\ \cline{3-17} 
\textbf{mechanism}          &                  & AUC                & J                  & ER                 & CZ                 & IU                 & AUC                & J                  & ER                 & CZ                 & IU                 & AUC                & J                  & ER                 & CZ                 & IU                 \\ \hline
\multirow{5}{*}{BN equal}   & Emp              & 0.001 $\pm$ 0.053  & 0.001 $\pm$ 0.099  & 0.002 $\pm$ 0.097  & 0.001 $\pm$ 0.096  & 0.001 $\pm$ 0.092  & 0.003 $\pm$ 0.042  & 0.003 $\pm$ 0.096  & 0.002 $\pm$ 0.095  & 0.003 $\pm$ 0.095  & 0.002 $\pm$ 0.095  & 0.003 $\pm$ 0.015  & 0.001 $\pm$ 0.095  & 0.002 $\pm$ 0.095  & 0.002 $\pm$ 0.095  & 0.001 $\pm$ 0.095  \\
                            & BN               & 0 $\pm$ 0.05       & 0.008 $\pm$ 0.646  & -0.001 $\pm$ 0.113 & -0.002 $\pm$ 0.124 & -0.001 $\pm$ 0.107 & 0 $\pm$ 0.041      & 0 $\pm$ 0.142      & 0 $\pm$ 0.089      & 0.002 $\pm$ 0.095  & -0.001 $\pm$ 0.139 & -0.001 $\pm$ 0.014 & 0.003 $\pm$ 0.1    & 0.004 $\pm$ 0.115  & 0.004 $\pm$ 0.103  & 0.003 $\pm$ 0.1    \\
                            & NonPar           & 0 $\pm$ 0.052      & 0.001 $\pm$ 0.13   & 0.001 $\pm$ 0.111  & 0.001 $\pm$ 0.11   & 0.001 $\pm$ 0.098  & 0.001 $\pm$ 0.042  & 0.004 $\pm$ 0.111  & 0.002 $\pm$ 0.104  & 0.002 $\pm$ 0.106  & 0.003 $\pm$ 0.104  & 0.001 $\pm$ 0.016  & 0.004 $\pm$ 0.107  & 0.002 $\pm$ 0.103  & 0.002 $\pm$ 0.107  & 0.004 $\pm$ 0.106  \\
                            & PV               & 0.001 $\pm$ 0.05   & 0.003 $\pm$ 0.645  & 0 $\pm$ 0.113      & -0.001 $\pm$ 0.124 & -0.001 $\pm$ 0.107 & -0.001 $\pm$ 0.041 & 0 $\pm$ 0.141      & 0 $\pm$ 0.089      & 0.002 $\pm$ 0.095  & 0 $\pm$ 0.139      & -0.003 $\pm$ 0.014 & 0.003 $\pm$ 0.1    & 0.003 $\pm$ 0.114  & 0.004 $\pm$ 0.103  & 0.003 $\pm$ 0.1    \\
                            & Semi.PV          & 0 $\pm$ 0.049      & 0.009 $\pm$ 0.651  & 0.001 $\pm$ 0.114  & 0.001 $\pm$ 0.123  & 0 $\pm$ 0.108      & -0.004 $\pm$ 0.043 & -0.001 $\pm$ 0.142 & -0.004 $\pm$ 0.092 & 0 $\pm$ 0.098      & -0.001 $\pm$ 0.137 & -0.006 $\pm$ 0.016 & 0.001 $\pm$ 0.104  & -0.001 $\pm$ 0.117 & 0.001 $\pm$ 0.107  & 0.001 $\pm$ 0.104  \\ \hline
\multirow{5}{*}{BN unequal} & Emp              & 0.001 $\pm$ 0.052  & 0.726 $\pm$ 0.109  & 0.163 $\pm$ 0.105  & 0.2 $\pm$ 0.105    & 0.013 $\pm$ 0.095  & 0.003 $\pm$ 0.045  & 0.17 $\pm$ 0.092   & -0.049 $\pm$ 0.085 & 0.061 $\pm$ 0.088  & -0.106 $\pm$ 0.098 & 0.001 $\pm$ 0.026  & 0.147 $\pm$ 0.086  & 0.282 $\pm$ 0.088  & 0.179 $\pm$ 0.087  & 0.217 $\pm$ 0.089  \\
                            & BN               & 0.001 $\pm$ 0.051  & -0.001 $\pm$ 0.19  & -0.001 $\pm$ 0.103 & 0 $\pm$ 0.114      & 0.001 $\pm$ 0.101  & -0.001 $\pm$ 0.043 & -0.001 $\pm$ 0.074 & -0.001 $\pm$ 0.062 & -0.001 $\pm$ 0.064 & 0.004 $\pm$ 0.089  & 0 $\pm$ 0.024      & 0.003 $\pm$ 0.072  & 0.004 $\pm$ 0.079  & 0.004 $\pm$ 0.076  & 0.002 $\pm$ 0.114  \\
                            & NonPar           & 0.001 $\pm$ 0.052  & 0.715 $\pm$ 0.176  & 0.156 $\pm$ 0.122  & 0.194 $\pm$ 0.128  & 0.014 $\pm$ 0.103  & 0 $\pm$ 0.045      & 0.165 $\pm$ 0.106  & -0.045 $\pm$ 0.09  & 0.058 $\pm$ 0.096  & -0.08 $\pm$ 0.133  & 0.001 $\pm$ 0.026  & 0.137 $\pm$ 0.097  & 0.269 $\pm$ 0.111  & 0.168 $\pm$ 0.101  & 0.199 $\pm$ 0.125  \\
                            & PV               & 0.001 $\pm$ 0.051  & -0.001 $\pm$ 0.19  & -0.001 $\pm$ 0.103 & 0 $\pm$ 0.114      & 0.001 $\pm$ 0.101  & -0.001 $\pm$ 0.043 & -0.001 $\pm$ 0.075 & -0.001 $\pm$ 0.061 & -0.001 $\pm$ 0.064 & 0.004 $\pm$ 0.089  & 0.001 $\pm$ 0.024  & -0.001 $\pm$ 0.077 & 0.007 $\pm$ 0.08   & 0.001 $\pm$ 0.077  & 0.005 $\pm$ 0.108  \\
                            & Semi.PV          & -0.001 $\pm$ 0.049 & 0.001 $\pm$ 0.195  & 0.001 $\pm$ 0.106  & 0.001 $\pm$ 0.114  & -0.006 $\pm$ 0.1   & -0.003 $\pm$ 0.043 & -0.003 $\pm$ 0.075 & -0.003 $\pm$ 0.065 & -0.001 $\pm$ 0.067 & 0.001 $\pm$ 0.09   & -0.014 $\pm$ 0.035 & -0.075 $\pm$ 0.178 & -0.056 $\pm$ 0.142 & -0.079 $\pm$ 0.172 & -0.067 $\pm$ 0.14  \\ \hline
\multirow{5}{*}{Skewed I}   & Emp              & -0.049 $\pm$ 0.058 & 1.014 $\pm$ 0.152  & 1.007 $\pm$ 0.15   & 1.008 $\pm$ 0.151  & 0.995 $\pm$ 0.172  & -0.169 $\pm$ 0.054 & 1.118 $\pm$ 0.157  & 0.982 $\pm$ 0.155  & 1.038 $\pm$ 0.154  & 1.048 $\pm$ 0.173  & 0.004 $\pm$ 0.029  & -5.615 $\pm$ 0.257 & -4.709 $\pm$ 0.255 & -4.834 $\pm$ 0.256 & -3.383 $\pm$ 0.253 \\
                            & BN               & -0.048 $\pm$ 0.056 & 1.603 $\pm$ 6.038  & 1.021 $\pm$ 0.302  & 1.026 $\pm$ 0.313  & 1.021 $\pm$ 0.145  & -0.2 $\pm$ 0.056   & 1.498 $\pm$ 0.718  & 1.036 $\pm$ 0.191  & 1.116 $\pm$ 0.219  & 1.095 $\pm$ 0.16   & -0.01 $\pm$ 0.039  & -5.626 $\pm$ 0.24  & -4.731 $\pm$ 0.324 & -4.84 $\pm$ 0.259  & -3.392 $\pm$ 0.238 \\
                            & NonPar           & -0.048 $\pm$ 0.057 & 1.016 $\pm$ 0.187  & 1 $\pm$ 0.177      & 1.003 $\pm$ 0.169  & 0.956 $\pm$ 0.259  & -0.17 $\pm$ 0.053  & 1.102 $\pm$ 0.193  & 0.96 $\pm$ 0.202   & 1.016 $\pm$ 0.206  & 0.996 $\pm$ 0.329  & 0 $\pm$ 0.028      & -5.647 $\pm$ 0.325 & -4.74 $\pm$ 0.303  & -4.866 $\pm$ 0.307 & -3.414 $\pm$ 0.308 \\
                            & PV               & -0.048 $\pm$ 0.055 & 1.595 $\pm$ 5.97   & 1.02 $\pm$ 0.301   & 1.025 $\pm$ 0.31   & 1.021 $\pm$ 0.145  & -0.2 $\pm$ 0.056   & 1.497 $\pm$ 0.72   & 1.035 $\pm$ 0.19   & 1.115 $\pm$ 0.218  & 1.095 $\pm$ 0.16   & -0.011 $\pm$ 0.039 & -5.626 $\pm$ 0.24  & -4.731 $\pm$ 0.324 & -4.84 $\pm$ 0.257  & -3.393 $\pm$ 0.238 \\
                            & Semi.PV          & 0.085 $\pm$ 0.149  & 1.367 $\pm$ 3.105  & 1.253 $\pm$ 1.204  & 1.33 $\pm$ 1.087   & 1.005 $\pm$ 1.062  & -0.166 $\pm$ 0.133 & 1.688 $\pm$ 2.817  & 1.166 $\pm$ 1.125  & 1.23 $\pm$ 1.099   & 1.151 $\pm$ 0.998  & -0.012 $\pm$ 0.092 & -5.089 $\pm$ 1.587 & -4.124 $\pm$ 1.296 & -4.294 $\pm$ 1.439 & -2.843 $\pm$ 1.37  \\ \hline
\multirow{5}{*}{Skewed II}  & Emp              & 0.001 $\pm$ 0.053  & 0.714 $\pm$ 0.259  & 0.697 $\pm$ 0.25   & 0.697 $\pm$ 0.255  & 0.672 $\pm$ 0.292  & 0.005 $\pm$ 0.041  & 1.098 $\pm$ 0.263  & 0.859 $\pm$ 0.268  & 0.95 $\pm$ 0.267   & 0.981 $\pm$ 0.29   & 0.005 $\pm$ 0.029  & 3.044 $\pm$ 0.735  & 2.268 $\pm$ 0.74   & 2.813 $\pm$ 0.736  & 2.809 $\pm$ 0.735  \\
                            & BN               & -0.011 $\pm$ 0.052 & 1.769 $\pm$ 2.853  & 0.843 $\pm$ 0.516  & 0.883 $\pm$ 0.624  & 0.719 $\pm$ 0.226  & -0.086 $\pm$ 0.058 & 1.59 $\pm$ 0.629   & 0.921 $\pm$ 0.291  & 1.14 $\pm$ 0.339   & 1.061 $\pm$ 0.313  & -0.06 $\pm$ 0.056  & 3.334 $\pm$ 0.932  & 2.007 $\pm$ 0.885  & 2.855 $\pm$ 0.785  & 2.972 $\pm$ 0.917  \\
                            & NonPar           & 0 $\pm$ 0.052      & 0.713 $\pm$ 0.299  & 0.675 $\pm$ 0.292  & 0.676 $\pm$ 0.288  & 0.596 $\pm$ 0.488  & 0.002 $\pm$ 0.041  & 1.052 $\pm$ 0.37   & 0.808 $\pm$ 0.402  & 0.904 $\pm$ 0.385  & 0.911 $\pm$ 0.479  & 0.001 $\pm$ 0.028  & 2.889 $\pm$ 1.258  & 2.116 $\pm$ 1.08   & 2.662 $\pm$ 1.112  & 2.656 $\pm$ 1.037  \\
                            & PV               & -0.012 $\pm$ 0.053 & 1.684 $\pm$ 2.689  & 0.818 $\pm$ 0.407  & 0.851 $\pm$ 0.496  & 0.718 $\pm$ 0.224  & -0.083 $\pm$ 0.059 & 1.548 $\pm$ 0.552  & 0.901 $\pm$ 0.277  & 1.107 $\pm$ 0.307  & 1.076 $\pm$ 0.3    & -0.058 $\pm$ 0.056 & 3.3 $\pm$ 0.914    & 2.003 $\pm$ 0.878  & 2.828 $\pm$ 0.77   & 3 $\pm$ 0.857      \\
                            & Semi.PV          & 0.068 $\pm$ 0.151  & 0.459 $\pm$ 4.221  & 0.733 $\pm$ 1.537  & 0.932 $\pm$ 1.491  & 0.565 $\pm$ 1.33   & -0.054 $\pm$ 0.212 & 0.922 $\pm$ 2.416  & 0.925 $\pm$ 1.585  & 1.01 $\pm$ 1.665   & 0.972 $\pm$ 1.447  & -0.022 $\pm$ 0.12  & 3.209 $\pm$ 4.981  & 2.979 $\pm$ 4.202  & 3.256 $\pm$ 4.472  & 3.486 $\pm$ 4.105  \\ \hline
\multirow{5}{*}{Skewed III} & Emp              & -0.002 $\pm$ 0.059 & 0.001 $\pm$ 0.009  & 0.031 $\pm$ 0.01   & 0.03 $\pm$ 0.009   & 0.031 $\pm$ 0.01   & -0.001 $\pm$ 0.047 & 0.064 $\pm$ 0.032  & 0.105 $\pm$ 0.034  & 0.093 $\pm$ 0.034  & 0.102 $\pm$ 0.037  & 0 $\pm$ 0.027      & 1.512 $\pm$ 0.381  & 1.578 $\pm$ 0.381  & 1.528 $\pm$ 0.381  & 1.57 $\pm$ 0.381   \\
                            & BN               & 0.012 $\pm$ 0.051  & 0.056 $\pm$ 0.023  & 0.043 $\pm$ 0.013  & 0.047 $\pm$ 0.016  & 0.033 $\pm$ 0.01   & -0.032 $\pm$ 0.032 & 0.075 $\pm$ 0.031  & 0.069 $\pm$ 0.022  & 0.089 $\pm$ 0.026  & 0.024 $\pm$ 0.017  & -0.164 $\pm$ 0.032 & 0.093 $\pm$ 0.048  & 0.076 $\pm$ 0.036  & 0.101 $\pm$ 0.047  & -0.044 $\pm$ 0.02  \\
                            & NonPar           & -0.003 $\pm$ 0.058 & 0.001 $\pm$ 0.011  & 0.03 $\pm$ 0.011   & 0.03 $\pm$ 0.011   & 0.029 $\pm$ 0.017  & -0.002 $\pm$ 0.047 & 0.06 $\pm$ 0.037   & 0.099 $\pm$ 0.043  & 0.088 $\pm$ 0.041  & 0.095 $\pm$ 0.054  & 0 $\pm$ 0.027      & 1.438 $\pm$ 0.514  & 1.505 $\pm$ 0.514  & 1.455 $\pm$ 0.514  & 1.497 $\pm$ 0.514  \\
                            & PV               & 0.012 $\pm$ 0.052  & 0.053 $\pm$ 0.022  & 0.042 $\pm$ 0.012  & 0.045 $\pm$ 0.015  & 0.033 $\pm$ 0.01   & 0.015 $\pm$ 0.038  & 0.05 $\pm$ 0.03    & 0.055 $\pm$ 0.021  & 0.064 $\pm$ 0.025  & 0.034 $\pm$ 0.018  & 0.037 $\pm$ 0.021  & 1.542 $\pm$ 0.349  & 1.609 $\pm$ 0.349  & 1.559 $\pm$ 0.349  & 1.6 $\pm$ 0.349    \\
                            & Semi.PV          & 0.061 $\pm$ 0.136  & 0.039 $\pm$ 0.131  & 0.048 $\pm$ 0.056  & 0.05 $\pm$ 0.059   & 0.033 $\pm$ 0.047  & 0.034 $\pm$ 0.158  & 0.024 $\pm$ 0.097  & 0.043 $\pm$ 0.066  & 0.038 $\pm$ 0.074  & 0.038 $\pm$ 0.063  & -0.005 $\pm$ 0.079 & 0.154 $\pm$ 1.265  & 0.161 $\pm$ 1.075  & 0.141 $\pm$ 1.179  & 0.155 $\pm$ 0.743  \\ \hline
\multirow{5}{*}{Mixed I}    & Emp              & 0.002 $\pm$ 0.056  & -1.142 $\pm$ 0.205 & -0.389 $\pm$ 0.197 & -0.577 $\pm$ 0.205 & 0.065 $\pm$ 0.18   & 0.003 $\pm$ 0.046  & -0.376 $\pm$ 0.193 & 0.094 $\pm$ 0.184  & -0.141 $\pm$ 0.183 & 0.249 $\pm$ 0.224  & 0.001 $\pm$ 0.027  & 0.349 $\pm$ 0.184  & 0.637 $\pm$ 0.185  & 0.413 $\pm$ 0.184  & 0.529 $\pm$ 0.194  \\
                            & BN               & 0 $\pm$ 0.051      & 0.004 $\pm$ 0.172  & -0.001 $\pm$ 0.126 & 0.001 $\pm$ 0.144  & 0.001 $\pm$ 0.144  & -0.001 $\pm$ 0.044 & 0.003 $\pm$ 0.147  & 0.001 $\pm$ 0.135  & 0.002 $\pm$ 0.136  & 0.007 $\pm$ 0.174  & -0.001 $\pm$ 0.025 & 0.008 $\pm$ 0.152  & 0.009 $\pm$ 0.162  & 0.008 $\pm$ 0.156  & 0.006 $\pm$ 0.237  \\
                            & NonPar           & 0.002 $\pm$ 0.055  & -1.089 $\pm$ 0.685 & -0.358 $\pm$ 0.258 & -0.543 $\pm$ 0.287 & 0.033 $\pm$ 0.188  & 0.002 $\pm$ 0.046  & -0.364 $\pm$ 0.237 & 0.085 $\pm$ 0.199  & -0.14 $\pm$ 0.215  & 0.18 $\pm$ 0.324   & 0.001 $\pm$ 0.027  & 0.328 $\pm$ 0.199  & 0.608 $\pm$ 0.231  & 0.392 $\pm$ 0.206  & 0.494 $\pm$ 0.266  \\
                            & PV               & 0 $\pm$ 0.051      & 0.003 $\pm$ 0.171  & -0.003 $\pm$ 0.127 & 0 $\pm$ 0.144      & -0.002 $\pm$ 0.145 & -0.001 $\pm$ 0.045 & -0.005 $\pm$ 0.147 & 0 $\pm$ 0.136      & -0.003 $\pm$ 0.14  & 0.005 $\pm$ 0.173  & 0.002 $\pm$ 0.024  & -0.007 $\pm$ 0.16  & 0.015 $\pm$ 0.165  & -0.002 $\pm$ 0.16  & 0.024 $\pm$ 0.222  \\
                            & Semi.PV          & -0.002 $\pm$ 0.052 & -0.029 $\pm$ 0.179 & -0.03 $\pm$ 0.13   & -0.043 $\pm$ 0.141 & -0.004 $\pm$ 0.148 & -0.005 $\pm$ 0.053 & -0.096 $\pm$ 0.174 & -0.059 $\pm$ 0.158 & -0.099 $\pm$ 0.173 & -0.003 $\pm$ 0.203 & -0.016 $\pm$ 0.04  & -0.186 $\pm$ 0.398 & -0.129 $\pm$ 0.304 & -0.187 $\pm$ 0.382 & -0.149 $\pm$ 0.303 \\ \hline
\multirow{5}{*}{Mixed II}   & Emp              & -0.016 $\pm$ 0.053 & -0.798 $\pm$ 0.136 & -0.159 $\pm$ 0.129 & -0.208 $\pm$ 0.133 & -0.083 $\pm$ 0.121 & -0.036 $\pm$ 0.047 & -0.4 $\pm$ 0.136   & -0.059 $\pm$ 0.13  & -0.134 $\pm$ 0.134 & -0.206 $\pm$ 0.127 & -0.036 $\pm$ 0.026 & -0.016 $\pm$ 0.126 & 0.113 $\pm$ 0.128  & 0.034 $\pm$ 0.127  & -0.018 $\pm$ 0.126 \\
                            & BN               & -0.017 $\pm$ 0.05  & -0.006 $\pm$ 0.28  & 0.003 $\pm$ 0.13   & -0.001 $\pm$ 0.143 & -0.048 $\pm$ 0.132 & -0.038 $\pm$ 0.045 & -0.002 $\pm$ 0.192 & -0.001 $\pm$ 0.121 & 0 $\pm$ 0.129      & -0.124 $\pm$ 0.139 & -0.041 $\pm$ 0.027 & 0.001 $\pm$ 0.122  & 0.006 $\pm$ 0.138  & 0.002 $\pm$ 0.122  & 0 $\pm$ 0.122      \\
                            & NonPar           & -0.017 $\pm$ 0.052 & -0.777 $\pm$ 0.187 & -0.143 $\pm$ 0.148 & -0.192 $\pm$ 0.152 & -0.088 $\pm$ 0.137 & -0.037 $\pm$ 0.046 & -0.389 $\pm$ 0.172 & -0.057 $\pm$ 0.151 & -0.129 $\pm$ 0.154 & -0.214 $\pm$ 0.14  & -0.037 $\pm$ 0.027 & -0.019 $\pm$ 0.144 & 0.108 $\pm$ 0.138  & 0.03 $\pm$ 0.137   & -0.022 $\pm$ 0.138 \\
                            & PV               & -0.018 $\pm$ 0.05  & -0.005 $\pm$ 0.28  & 0.003 $\pm$ 0.13   & -0.001 $\pm$ 0.143 & -0.048 $\pm$ 0.131 & -0.038 $\pm$ 0.045 & -0.002 $\pm$ 0.192 & 0 $\pm$ 0.122      & 0 $\pm$ 0.129      & -0.125 $\pm$ 0.138 & -0.042 $\pm$ 0.027 & 0.001 $\pm$ 0.123  & 0.006 $\pm$ 0.137  & 0.002 $\pm$ 0.122  & 0 $\pm$ 0.122      \\
                            & Semi.PV          & -0.018 $\pm$ 0.052 & -0.008 $\pm$ 0.278 & -0.001 $\pm$ 0.132 & -0.004 $\pm$ 0.14  & -0.049 $\pm$ 0.133 & -0.041 $\pm$ 0.046 & -0.007 $\pm$ 0.19  & -0.004 $\pm$ 0.125 & -0.005 $\pm$ 0.13  & -0.128 $\pm$ 0.142 & -0.045 $\pm$ 0.027 & -0.018 $\pm$ 0.128 & -0.003 $\pm$ 0.14  & -0.014 $\pm$ 0.129 & -0.019 $\pm$ 0.128 \\ \hline
\end{tabular}
\end{adjustbox}
\end{center}
%\end{landscape}
\end{table}

These findings are visually depicted in Figures B.1 - B.6 of the supplementary material for medium sample size, while Tables B.3 - B.4 in supplementary material present biases corresponding to low and high sample sizes. The trends observed in the medium sample size roughly align with those from low and high sample sizes, as demonstrated in Figures B.7 - B.18.

%These findings are visually depicted in Figures~ \ref{Fig: Bias_medSamp_lowAUC} - \ref{Fig: Bias_AUC_medSamp_highAUC} for medium sample size, while Tables~\ref{Tab: bias_nocov_lowSamp} - \ref{Tab: bias_nocov_highSamp} present biases corresponding to low and high sample sizes. The trends observed in the medium sample size roughly align with those from low and high sample sizes, as demonstrated in Figures~\ref{Fig: Bias_lowSamp_lowAUC} - \ref{Fig: Bias_AUC_highSamp_highAUC}.

\subsection{Simulation with covariate} \label{Sec: Sim_Cov}

In simulation with covariates, we consider similar scenarios as without covariates. The simulation settings vary for different data-generating mechanisms and sample sizes. For each of the simulation settings, only the BN, PV, and Semi.PV models are used to estimate AUC and four different cutpoints at different covariate levels because of the inability of the rest of the models (empirical and nonparametric models) to accommodate covariates. The details of the scenarios are tabulated in Table \ref{Tab:Sim_cov_mechanism}.

\begin{table}[!ht]
\begin{center}
\caption{\label{Tab:Sim_cov_mechanism} Details of simulation mechanism with covariate.}
%\begin{adjustbox}{width=\textwidth}
\begin{tabular}{lll}
\hline
\textbf{Data}       & \textbf{Fitting}                              & \textbf{Sample}    \\
\textbf{generating} & \textbf{models}                               & \textbf{size}      \\
\textbf{scenario}   & \textbf{}                                     & \textbf{}          \\ \hline
BN                  & Binormal (BN)                                 & Small ($N = 50$)   \\
Skewed              & Placement value-based parametric (PV)         & Medium ($N = 100$) \\
Mixed               & Placement value-based semiparametric (Semi.PV) & High ($N = 500$)   \\ \hline
\end{tabular}
%\end{adjustbox}
\end{center}
\end{table}

First we generate the healthy and diseased covariates from uniform distributions. Each of $X_0$ and $X_1$ were generated from $U(-0.5, 1.5)$ where $U(a,b)$ is a continuous uniform distribution with support $a \leq x \leq b$. Then we generate 1000 data replicates based on the covariates from different mechanisms and report  median and IQR of biases obtained by different ROC estimation methods to estimate AUC and four different optimal cutpoints obtained at $x=0$ and $x=1$. We also plot the biases incurred by different ROC methods for different simulation mechanisms. The details of the data-generating mechanisms and corresponding true parameter values are tabulated in Table \ref{Tab:Sim_cov_generation}.

\begin{table}[!ht]
\begin{center}
\caption{\label{Tab:Sim_cov_generation} Data generation mechanism details with true parameters, with covariate.}
%\begin{adjustbox}{width=0.5\textwidth}
\begin{tabular}{ccc}
\hline
\textbf{Data}       & \textbf{Data}                                         & \textbf{True}                          \\
\textbf{generating} & \textbf{generation}                                   & \textbf{parameters}                    \\
\textbf{mechanism}  &                                                       &                                        \\ \hline
BN                  & $Y_0 \sim N(b_{00} + b_{01}X_0, \text{ } \sigma_0^2)$ & $b_{00}=1$, $b_{01}=1$, $\sigma_0=1$   \\
                    & $Y_1 \sim N(b_{10} + b_{11}X_1, \text{ } \sigma_1^2)$ & $b_{10}=1.5$, $b_{11}=2$, $\sigma_1=1$ \\ \hline
Skewed              & $Y_0 \sim Gamma(k, \text{ } \theta_0)$                & $b_{00}=3$, $b_{01}=0.1$               \\
                    & $Y_0 \sim Gamma(k, \text{ } \theta_1)$                & $b_{10}=5$, $b_{11}=9$                 \\
                    & $k$: shape, $\theta_j$: scale                         & $k = 2$                                \\
                    & $\theta_0 = b_{00} + b_{01}X_0$                       &                                        \\
                    & $\theta_1 = b_{10} + b_{11}X_1$                       &                                        \\ \hline
Mixed               & $Y_0 \sim N(\mu_0, \text{ } \sigma_0^2)$              & $a_{00}=0$, $a_{01}=1$, $\sigma_0=1$   \\
                    & $Y_1 \sim N(\mu_1, \text{ } \sigma_1^2)$              & $a_{101}=0$, $a_{111}=1$               \\
                    & $\mu_0 = a_{00}+a_{01}X_0$                            & $a_{102}=1$, $a_{112}=5$               \\
                    & $\mu_1 = \pi \mu_{11} + (1-\pi)\mu_{12}$              & $\sigma_1 = 1.5$                       \\
                    & $\mu_{11} = a_{101} + a_{111}X_0$                     & $\pi = 0.5$                            \\
                    & $\mu_{12} = a_{102} + a_{112}X_1$                     &                                        \\ \hline
\end{tabular}
%\end{adjustbox}
\end{center}
\end{table}

Table~\ref{Tab: bias_cov_medSamp} illustrates the median and IQR of biases in estimating AUC and four different optimal cutoff estimates from various ROC fitting models at covariate values $x=0$ and $x=1$. When data are generated from the \enquote{BN} mechanism, it becomes evident that the BN and PV models exhibit the least biases across both covariate levels, owing to their correct model specification. Similar trends are observed for the \enquote{Mixed} data generating mechanism. In these cases, the Semi.PV model's performance closely rivals that of the BN and PV models, particularly when estimating ER at $x=0$. However, when the data generating mechanism deviates from the BN and PV models, as seen in the \enquote{Skewed} data generating mechanism, the Semi.PV model consistently demonstrates the least median biases, except when estimating IU. Interestingly, IU is estimated most effectively by the BN model at both covariate levels. Additionally, it's noted that the estimation of AUC and optimal cutoff points from the Semi.PV model results in higher variability, as evidenced by the elevated IQR estimates across all data generating mechanisms.

%In Figure \ref{Fig: Bias_medSamp_cov}, bias plots illustrate simulation frameworks with medium sample size and with covariate. In the \enquote{BN} framework (panel A1), all models exhibit similar performance when estimating AUC and cutoffs at $x = 0$. However, at $x = 1$ (panel A2), the BN model emerges as the top performer, with the Semi.PV model following closely. Surprisingly, the PV model shows subpar performance. The BN model's efficacy diminishes when biomarkers are generated from the \enquote{Skewed} framework. At $x = 0$, the Semi.PV model excels in cutoff estimation, although AUC estimation is less satisfactory (panel B1). At $x = 1$ (panel B2), both PV and Semi.PV models perform comparably. In the \enquote{Mixed} framework, at $x = 0$ (panel C1), the BN and PV models deliver strong performances, while the Semi.PV model experiences higher bias. At $x = 1$ (panel C2), the BN model maintains its top performance, followed by the other models. Notably, in this scenario (both panels C1 and C2), IU estimation is inadequate, particularly for $x = 1$.

\begin{table}[!htbp]
\begin{center}
\caption{\label{Tab: bias_cov_medSamp} Biases of estimating AUC and optimal thresholds for different fitting models at different covariate levels for medium sample size, with covariate framework.}
\begin{adjustbox}{width=0.95\textwidth}
\begin{tabular}{cccccccccccc}
\hline
\textbf{Data}           & \textbf{Fitting} & \multicolumn{5}{c}{\textbf{x = 0}}                                                                     & \multicolumn{5}{c}{\textbf{x = 1}}                                                                       \\ \cline{3-12} 
\textbf{generating}     & \textbf{model}   & \multicolumn{5}{c}{\textbf{Median $\pm$ IQR}}                                                          & \multicolumn{5}{c}{\textbf{Median $\pm$ IQR}}                                                            \\ \cline{3-12} 
\textbf{mechanism}      & \textbf{}        & AUC                 & J                  & ER                & CZ                 & IU                 & AUC                  & J                  & ER                 & CZ                 & IU                 \\ \hline
\multirow{3}{*}{BN}     & BN               & 0.001   $\pm$ 0.068 & -0.001 $\pm$ 0.283 & 0.001 $\pm$ 0.128 & 0.002 $\pm$ 0.134  & 0.001 $\pm$ 0.158  & -0.004 $\pm$   0.044 & 0 $\pm$ 0.132      & 0.002 $\pm$ 0.127  & 0 $\pm$ 0.125      & 0 $\pm$ 0.132      \\
                        & PV               & 0.002 $\pm$ 0.068   & -0.002 $\pm$ 0.277 & 0 $\pm$ 0.127     & 0.002 $\pm$ 0.133  & -0.003 $\pm$ 0.276 & -0.002 $\pm$ 0.044   & -0.001 $\pm$ 0.131 & 0.002 $\pm$ 0.128  & 0.001 $\pm$ 0.125  & 0.047 $\pm$ 0.23   \\
                        & Semi.PV          & -0.04 $\pm$ 0.139   & -0.012 $\pm$ 0.587 & 0.002 $\pm$ 0.233 & 0.007 $\pm$ 0.251  & -0.016 $\pm$ 0.402 & -0.031 $\pm$ 0.165   & -0.044 $\pm$ 0.319 & -0.052 $\pm$ 0.355 & -0.05 $\pm$ 0.353  & -0.051 $\pm$ 0.539 \\ \hline
\multirow{3}{*}{Skewed} & BN               & -0.076 $\pm$ 0.044  & 4.958 $\pm$ 1.407  & 2.748 $\pm$ 0.915 & 3.812 $\pm$ 0.998  & -0.527 $\pm$ 0.734 & 0.015 $\pm$ 0.032    & 2.057 $\pm$ 1.289  & 1.694 $\pm$ 1.242  & 2.175 $\pm$ 1.277  & -0.869 $\pm$ 1.613 \\
                        & PV               & -0.035 $\pm$ 0.05   & 3.84 $\pm$ 1.288   & 2.159 $\pm$ 0.786 & 2.818 $\pm$ 0.835  & 1.823 $\pm$ 0.907  & 0.035 $\pm$ 0.028    & 1.354 $\pm$ 1.315  & 1.513 $\pm$ 1.266  & 1.56 $\pm$ 1.295   & -3.494 $\pm$ 1.065 \\
                        & Semi.PV          & -0.033 $\pm$ 0.138  & 2.553 $\pm$ 3.672  & 1.582 $\pm$ 2.8   & 1.88 $\pm$ 2.843   & 1.738 $\pm$ 2.795  & -0.004 $\pm$ 0.104   & 0.326 $\pm$ 4.133  & 0.7 $\pm$ 4.007    & 0.476 $\pm$ 4.05   & -3.004 $\pm$ 4.435 \\ \hline
\multirow{3}{*}{Mixed}  & BN               & 0.001 $\pm$ 0.07    & -0.001 $\pm$ 0.271 & 0.002 $\pm$ 0.154 & -0.002 $\pm$ 0.159 & -0.436 $\pm$ 0.168 & -0.002 $\pm$ 0.032   & 0.007 $\pm$ 0.154  & 0.003 $\pm$ 0.165  & 0.004 $\pm$ 0.156  & -0.185 $\pm$ 0.159 \\
                        & PV               & 0.002 $\pm$ 0.07    & -0.01 $\pm$ 0.271  & 0 $\pm$ 0.154     & -0.003 $\pm$ 0.159 & -0.02 $\pm$ 0.164  & -0.002 $\pm$ 0.031   & 0.004 $\pm$ 0.154  & 0.001 $\pm$ 0.165  & 0.002 $\pm$ 0.156  & -0.645 $\pm$ 0.277 \\
                        & Semi.PV          & -0.027 $\pm$ 0.126  & 0.044 $\pm$ 0.588  & 0.002 $\pm$ 0.285 & -0.014 $\pm$ 0.298 & -0.042 $\pm$ 0.362 & -0.033 $\pm$ 0.149   & -0.052 $\pm$ 0.331 & -0.065 $\pm$ 0.419 & -0.076 $\pm$ 0.403 & -0.72 $\pm$ 0.497  \\ \hline
\end{tabular}
\end{adjustbox}
\end{center}
\end{table}

%These findings are further supported by bias plots depicted in Figures~\ref{Fig: Bias_medSamp_cov} - \ref{Fig: Bias_medSamp_cov_AUC} for medium sample size. Similar simulations have been conducted for low ($N=50$) and high ($N=500$) sample sizes, and the overall conclusions drawn from the medium sample size simulations are consistent across different sample sizes. The corresponding bias tables for other sample sizes can be found in Tables~\ref{Tab: bias_cov_lowSamp} - \ref{Tab: bias_cov_highSamp}, and similar bias plots are available in Figures~\ref{Fig: Bias_lowSamp_cov} - \ref{Fig: Bias_highSamp_cov_AUC}.

These findings are further supported by bias plots depicted in Figures B.19 - B.20 of the supplementary material for medium sample size. Similar simulations have been conducted for low ($N=50$) and high ($N=500$) sample sizes, and the overall conclusions drawn from the medium sample size simulations are consistent across different sample sizes. See the supplementary material for more details. The corresponding bias tables for other sample sizes can be found in Tables B.5 - B.6, and similar bias plots are available in Figures B.21 - B.24.

%Figures \ref{Fig: Bias_lowSamp_cov} and \ref{Fig: Bias_highSamp_cov} present analogous bias plots for low and high sample sizes, mirroring the observations made in the medium sample size scenario.

\section{Data Application} \label{Sec: Application}

Data used in this article were obtained from the Alzheimer’s Disease Neuroimaging Initiative (ADNI) database \footnote{As such, the investigators within the ADNI contributed to the design and implementation of ADNI and/or provided data but did not participate in analysis or writing of this article. A complete listing of ADNI investigators can be found at: \url{https://adni.loni.usc.edu/wp-content/uploads/how_to_apply/ADNI_Acknowledgement_List.pdf}} (\url{adni.loni.usc.edu}). The ADNI was launched in 2003 as a public-private partnership, led by Principal Investigator Michael W. Weiner, MD. The primary goal of ADNI has been to test whether serial magnetic resonance imaging, positron emission tomography, other biological markers, and clinical and neuropsychological assessment can be combined to measure the progression of mild cognitive impairment and early Alzheimer’s disease (AD).  

In this context, our aim is to assess the diagnostic accuracy and determine the optimal thresholds for various fluid biomarkers in AD diagnosis. The focus biomarkers include plasma amyloid-$\beta$ ($A\beta 42$) \citep{teunissen2018plasma}, tau (total-tau or t-tau) \citep{holper2022tau}, and phosphorylated tau (p-tau) \citep{gonzalez2023plasma}. To achieve this, we utilized the dataset from the ADSP Phenotype Harmonization Consortium (PHC), which collected fluid biomarker levels from various studies and cohorts, then merged the biomarker data across these cohorts. The fluid biomarker scores were harmonized across datasets such as ADNI, the National Alzheimer’s Coordinating Center (NACC), and the Memory and Aging Project at Knight Alzheimer’s Disease Research Center (MAP at Knight ADRC). Subsequently, the scores were co-calibrated and standardized to create z-score versions of the biomarkers. Table~\ref{Tab:Tab1} illustrates the overall summary of the ADSP data.

\begin{table}[!htbp]
\begin{center}
\caption{\label{Tab:Tab1} ADSP PHC standardized fluid biomarker data. P-value (\textbf{p}) corresponds to the tests to compare covariates between normal cognition group and AD group.}
\begin{adjustbox}{width=0.9\textwidth}
\begin{tabular}{lcccc}
\hline
\multicolumn{1}{c}{\multirow{2}{*}{\textbf{Covariates}}} & \textbf{Overall} & \textbf{Normal cognition} & \textbf{AD}       & \multirow{2}{*}{\textbf{p}} \\
\multicolumn{1}{c}{}                                     & N = 682          & N = 360 (52.79\%)         & N = 322 (47.21\%) &                             \\ \hline
Continuous                                               & \multicolumn{4}{c}{Mean (SD)}                                                                  \\ \hline
$A\beta 42$                                              & -0.03 (1.01)     & 0.52 (0.88)               & -0.65 (0.77)      & \textless{}0.001            \\
Tau                                                      & 0.03 (1.06)      & -0.50 (0.86)              & 0.63 (0.93)       & \textless{}0.001            \\
pTau                                                     & 0.14 (1.03)      & -0.23 (0.96)              & 0.56 (0.95)       & \textless{}0.001            \\
Age                                                      & 74.95 (7.08)     & 74.53 (6.41)              & 75.41 (7.75)      & 0.107                       \\ \hline
Categorical covariates                                   & \multicolumn{4}{c}{N (\%)}                                                                     \\ \hline
Sex                                                      &                  &                           &                   & 0.001                       \\
\multicolumn{1}{r}{Male}                                 & 362 (53.1)       & 169 (46.9)                & 193 (59.9)        &                             \\
\multicolumn{1}{r}{Female}                               & 320 (46.9)       & 191 (53.1)                & 129 (40.1)        &                             \\
Race                                                     &                  &                           &                   & 0.01                        \\
\multicolumn{1}{r}{American Indian/Alaskan Native}       & 1 (0.1)          & 1 (0.3)                   & 0 (0.0)           &                             \\
\multicolumn{1}{r}{Asian}                                & 9 (1.3)          & 3 (0.8)                   & 6 (1.9)           &                             \\
\multicolumn{1}{r}{African American}                     & 27 (4.0)         & 21 (5.8)                  & 6 (1.9)           &                             \\
\multicolumn{1}{r}{White}                                & 637 (93.4)       & 328 (91.1)                & 309 (96.0)        &                             \\
\multicolumn{1}{r}{$> 1$ Race}                           & 8 (1.2)          & 7 (1.9)                   & 1 (0.3)           &                             \\ \hline
\end{tabular}
\end{adjustbox}
\end{center}
\end{table}

Among the 682 patients in the ADSP dataset, 52.79\% exhibit normal cognition, while the remainder are diagnosed with Alzheimer's disease. Table~\ref{Tab:Tab1} illustrates significant variations in standardized biomarker levels (p $<$ 0.001 for all biomarkers) between the disease groups, with the diseased cohort showing notably lower standardized $A\beta 42$ values and higher standardized Tau and pTau values. While age does not differ significantly between disease groups (p = 0.107), there are significant discrepancies in sex and race distributions (p = 0.001 and 0.01, respectively).

In the subsequent two subsections, we will assess the overall diagnostic accuracy and optimal thresholds for all three standardized biomarkers, as well as explore sex-specific diagnostic accuracy and optimal thresholds.

\subsection{Overall diagnostic accuracy} \label{Sec: Application_nocov}

\begin{figure}[!htbp]%
    \begin{center}
    \includegraphics[width=0.9\textwidth]{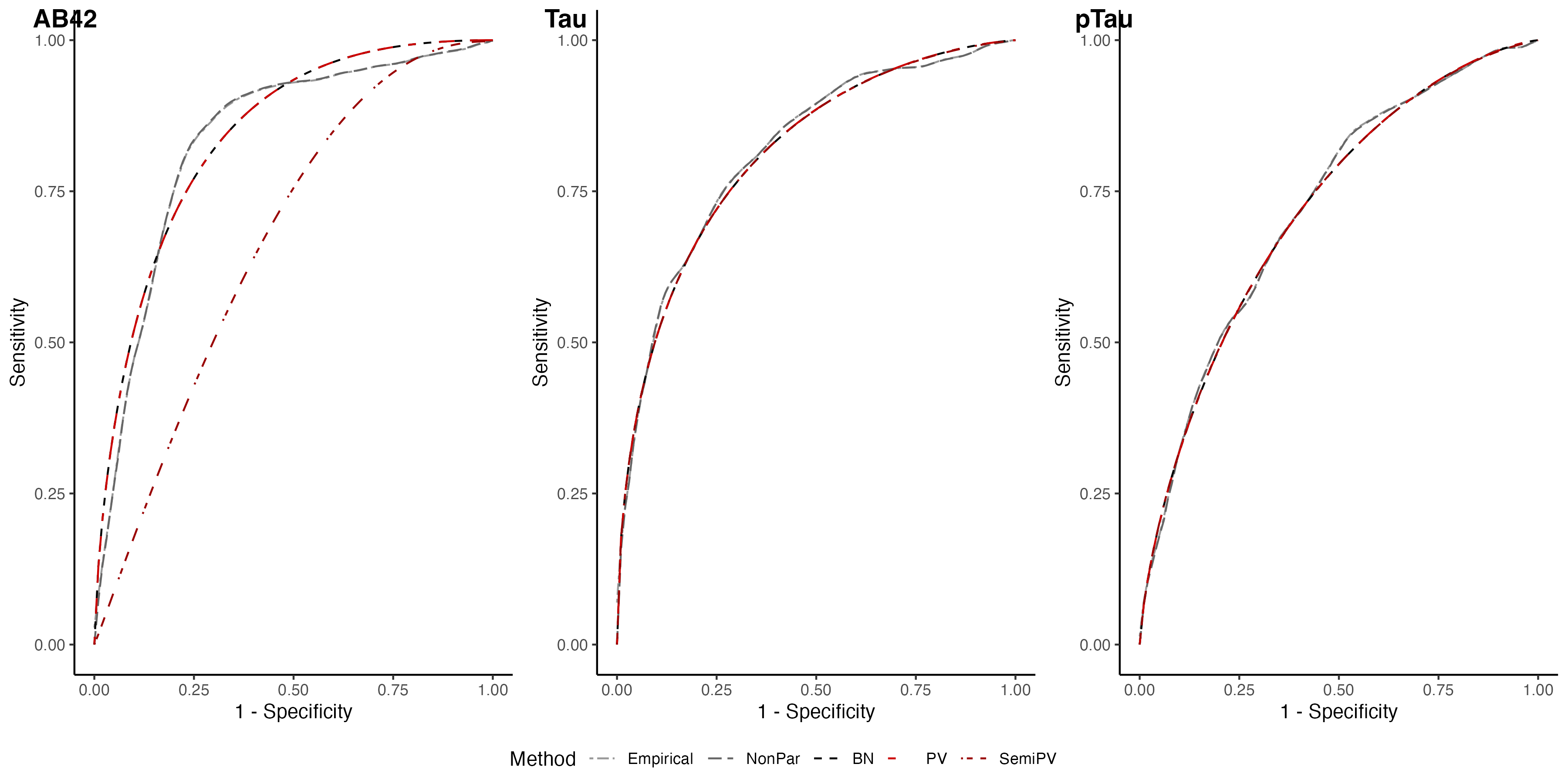}
    \caption{\label{Fig: data_analysis_nocov_ROC} ROC curves for different ROC estimating methods for different biomarkers.}
    \end{center}
\end{figure}

In this section, our focus lies in assessing the comprehensive diagnostic accuracy of the standardized fluid biomarkers $A\beta 42$, Tau, and pTau. We aim to estimate the mean and 95\% credible intervals (CI) of AUC and four distinct optimal cutoffs obtained from the methodologies delineated in Section~\ref{Sec: Sim_noCov}, encompassing Emp, NonPar, BN, PV, and Semi.PV. These results are consolidated in Table~\ref{Tab:nocov_data_analysis}. To derive credible intervals for the Emp and NonPar models, we utilize 1000 bootstrap samples. Additionally, corresponding ROC curves are presented in Figure~\ref{Fig: data_analysis_nocov_ROC}.

\begin{table}[!htbp]
\begin{center}
\caption{\label{Tab:nocov_data_analysis} Estimates of AUC and optimal thresholds for different ROC estimating methods for different biomarkers.}
\begin{adjustbox}{width=\textwidth}
\begin{tabular}{ccccccc}
\hline
\multirow{2}{*}{\textbf{Biomarker}} & \multirow{2}{*}{\textbf{Method}} & \multicolumn{5}{c}{\textbf{Metrics (Mean (95\% CI))}}                                                                      \\ \cline{3-7} 
                                    &                                  & AUC                  & J                      & ER                      & CZ                      & IU                     \\ \hline
\multirow{5}{*}{$A\beta 42$}        & Emp                        & 0.833 (0.801, 0.865) & -0.037 (-0.164, 0.048) & -0.042 (-0.206, 0.037)  & -0.041 (-0.200, 0.041)  & -0.05 (-0.305, 0.030)  \\
                                    & NonPar                           & 0.834 (0.800, 0.865) & -0.038 (-0.171, 0.048) & -0.042 (-0.211, 0.036)  & -0.039 (-0.177, 0.038)  & -0.049 (-0.284, 0.028) \\
                                    & BN                               & 0.841 (0.824, 0.857) & -0.026 (-0.066, 0.015) & -0.074 (-0.110, -0.036) & -0.053 (-0.089, -0.016) & -0.026 (-0.066, 0.015) \\
                                    & PV                               & 0.841 (0.829, 0.851) & -0.026 (-0.091, 0.038) & -0.074 (-0.139, -0.011) & -0.053 (-0.118, 0.010)  & -0.026 (-0.091, 0.038) \\
                                    & Semi.PV                           & 0.667 (0.639, 0.694) & 0.311 (0.006, 0.525)   & 0.093 (-0.136, 0.320)   & 0.127 (-0.116, 0.355)   & 0.098 (-0.135, 0.325)  \\ \hline
\multirow{5}{*}{Tau}                & Emp                        & 0.815 (0.782, 0.847) & 0.034 (-0.038, 0.101)  & 0.033 (-0.033, 0.098)   & 0.034 (-0.035, 0.100)   & 0.033 (-0.038, 0.098)  \\
                                    & NonPar                           & 0.815 (0.783, 0.847) & 0.032 (-0.050, 0.124)  & 0.032 (-0.036, 0.104)   & 0.032 (-0.040, 0.114)   & 0.032 (-0.041, 0.114)  \\
                                    & BN                               & 0.812 (0.793, 0.830) & 0.104 (0.056, 0.153)   & 0.064 (0.024, 0.103)    & 0.079 (0.039, 0.119)    & 0.104 (0.056, 0.153)   \\
                                    & PV                               & 0.812 (0.798, 0.825) & 0.104 (0.038, 0.171)   & 0.064 (0.001, 0.127)    & 0.079 (0.015, 0.143)    & 0.104 (0.038, 0.171)   \\
                                    & Semi.PV                           & 0.812 (0.797, 0.825) & 0.105 (0.038, 0.172)   & 0.064 (0.000, 0.127)    & 0.079 (0.014, 0.143)    & 0.105 (0.038, 0.172)   \\ \hline
\multirow{5}{*}{pTau}               & Emp                        & 0.722 (0.685, 0.760) & 0.143 (-0.114, 0.385)  & 0.144 (0.066, 0.218)    & 0.144 (0.062, 0.218)    & 0.145 (0.063, 0.221)   \\
                                    & NonPar                           & 0.721 (0.682, 0.759) & 0.138 (-0.169, 0.306)  & 0.143 (0.059, 0.223)    & 0.142 (0.052, 0.227)    & 0.142 (0.060, 0.223)   \\
                                    & BN                               & 0.719 (0.697, 0.741) & 0.156 (0.083, 0.232)   & 0.164 (0.123, 0.205)    & 0.162 (0.117, 0.206)    & 0.156 (0.083, 0.232)   \\
                                    & PV                               & 0.719 (0.703, 0.734) & 0.158 (0.078, 0.236)   & 0.164 (0.099, 0.232)    & 0.163 (0.095, 0.230)    & 0.158 (0.078, 0.236)   \\
                                    & Semi.PV                           & 0.719 (0.702, 0.736) & 0.160 (0.077, 0.241)    & 0.166 (0.095, 0.234)    & 0.164 (0.093, 0.235)    & 0.160 (0.077, 0.241)   \\ \hline
\end{tabular}
\end{adjustbox}
\end{center}
\end{table}

From Table~\ref{Tab:nocov_data_analysis}, it is evident that the diagnostic accuracy of $A\beta 42$ is notably high, ranging from 0.833 to 0.841, as estimated by most models, with the exception of Semi.PV, which yields barely moderate AUC estimate. The optimal cutoff metrics for $A\beta 42$ are consistently estimated similarly by the Emp and NonPar models. The BN and PV models also produce comparable yet distinct estimates compared to Emp and NonPar. However, the estimates from Semi.PV are notably different. It is important to note that for the biomarker $A\beta 42$, values lower than the estimated cutoff would be classified as diseased. When comparing all the cutoffs produced by different ROC models, it's observed that the Emp and NonPar models exhibit low variability in their estimates, ranging from -0.059 to -0.037. Conversely, for BN and PV models, the variability in estimating different cutoffs is higher, spanning between -0.073 and -0.026. Notably, the Semi.PV model produces markedly different positive cutoff estimates, ranging from 0.093 to 0.311. For a better visual understanding, please refer to the Figure C.25. %For a better understanding, please refer to the Figure~\ref{Fig: data_analysis_nocov_density}.

When analyzing the biomarker Tau, we observe two distinct groups of estimates. The Emp and NonPar models comprise one group, while the remaining models form another. Despite the consistency in AUC estimates between groups (ranging from 0.812 to 0.815), the cutoff estimates differ notably. Similar to $A\beta 42$, for Tau as well, the cutoffs exhibit minimal variability for Emp and NonPar across the four optimal cutoff metrics (ranging from 0.032 to 0.034). Conversely, for BN, PV, and Semi.PV, the cutoff estimates are higher. While ER and CZ have estimates of 0.064 and 0.079 respectively, the estimates for J and IU are even higher, ranging from around 0.104 to 0.105. For Tau, biomarker values higher than the cutoff would be classified as diseased.

The trends observed in the Tau results closely mirror those of pTau. However, pTau exhibits lower diagnostic accuracy, ranging from 0.719 to 0.722, compared to $A\beta 42$ or Tau. The cutoff estimates obtained by Emp and NonPar range between 0.138 and 0.145, while those from the other models vary between 0.156 and 0.166. In this instance, unlike Tau, the estimates for ER and CZ from BN, PV, and Semi.PV (ranging from 0.162 to 0.166) are higher than the estimates for J and IU (ranging from 0.156 to 0.160). For pTau too, biomarker values higher than the cutoff would be classified as diseased.

\subsection{Sex-specific diagnostic accuracy} \label{Sec: Application_cov}

\begin{figure}[!htbp]%
    \begin{center}
    \includegraphics[width=0.8\textwidth]{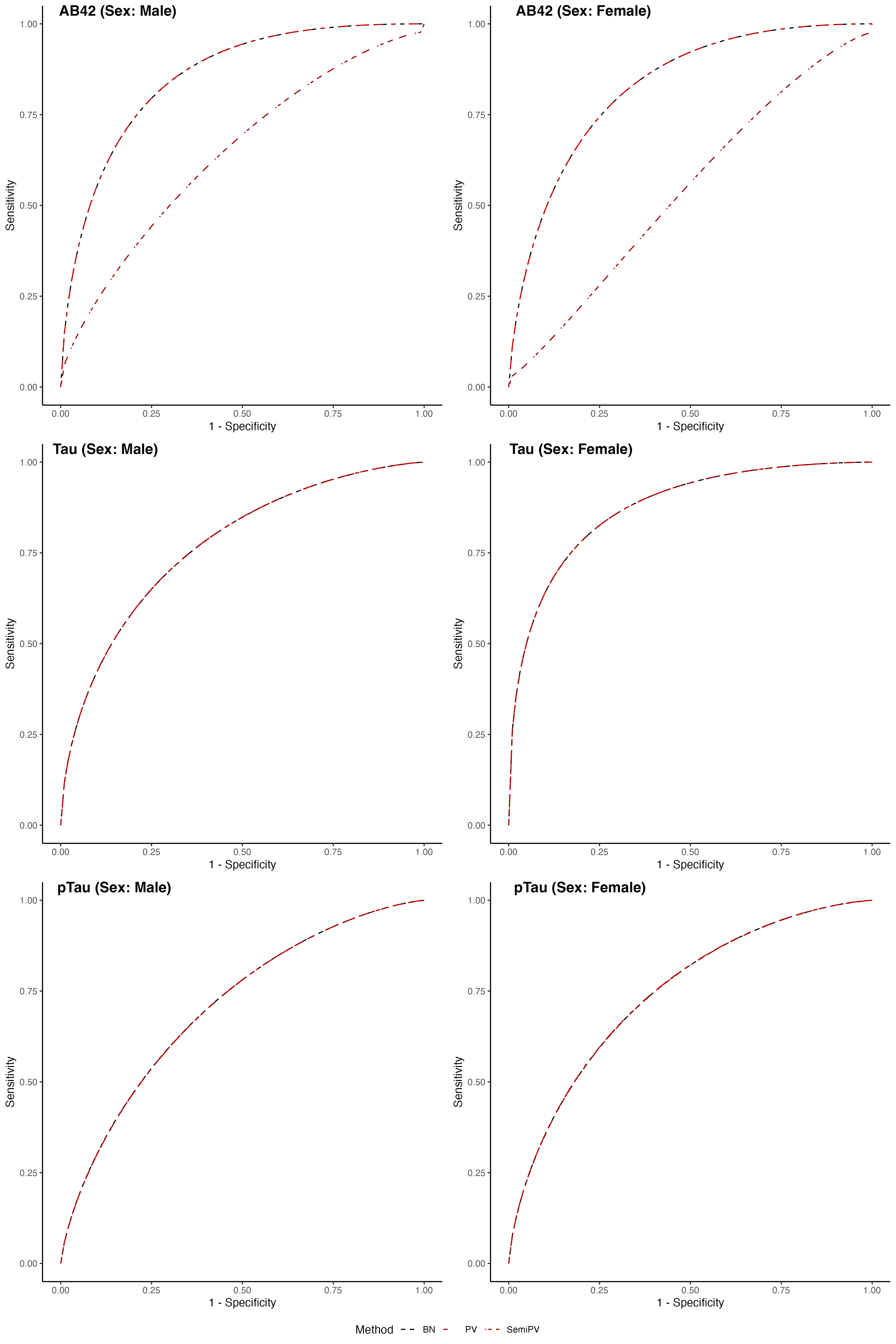}
    \caption{\label{Fig: data_analysis_cov_ROC} Sex-specific ROC curves for different ROC estimating methods for different biomarkers.}
    \end{center}
\end{figure}

In this section, we aim to examine whether potential covariates significantly impact the diagnostic performances of biomarkers. Specifically, we aim to determine if biological sex influences diagnostic accuracy, assessing whether there are significant differences in diagnostic accuracy and optimal thresholds between male and female patients. Similar to the previous section, our objective here is to estimate the mean and 95\% CI of the AUC and four different cutoff metrics obtained from the methodologies employed in Section~\ref{Sec: Sim_Cov}, namely the BN, PV, and Semi.PV models, chosen for their ability to accommodate covariates. These results will be consolidated in Table~\ref{Tab:cov_data_analysis}, and the corresponding ROC curves will be presented in Figure~\ref{Fig: data_analysis_cov_ROC}.

\begin{table}[!htbp]
\begin{center}
\caption{\label{Tab:cov_data_analysis} Sex-specific estimates of AUC and optimal thresholds for different ROC estimating methods for different biomarkers.}
\begin{adjustbox}{width=\textwidth}
\begin{tabular}{cccccccc}
\hline
\textbf{Biomarker}           & \textbf{Covariate}           & \multirow{2}{*}{\textbf{Method}} & \multicolumn{5}{c}{\textbf{Metrics (Mean (95\% CI))}}                                                                        \\ \cline{4-8} 
                             & \textbf{level}               &                                  & AUC                  & J                        & ER                      & CZ                      & IU                     \\ \hline
\multirow{6}{*}{$A\beta 42$} & \multirow{3}{*}{Sex: Male}   & BN                               & 0.853 (0.833, 0.873) & -0.014 (-0.067, 0.037)   & -0.058 (-0.109, -0.01)  & -0.038 (-0.089, 0.011)  & -0.014 (-0.067, 0.037) \\
                             &                              & PV                               & 0.853 (0.840, 0.866) & -0.014 (-0.102, 0.076)   & -0.059 (-0.146, 0.029)  & -0.038 (-0.125, 0.050)  & -0.014 (-0.102, 0.076) \\
                             &                              & Semi.PV                           & 0.642 (0.505, 0.695) & -0.215 (-1.488, 0.617)   & -0.168 (-1.470, 0.624)  & -0.172 (-1.446, 0.622)  & -0.186 (-1.761, 0.618) \\ \cline{2-8} 
                             & \multirow{3}{*}{Sex: Female} & BN                               & 0.826 (0.801, 0.850) & -0.032 (-0.089, 0.028)   & -0.083 (-0.135, -0.029) & -0.062 (-0.115, -0.008) & -0.032 (-0.089, 0.028) \\
                             &                              & PV                               & 0.826 (0.808, 0.843) & -0.031 (-0.121, 0.055)   & -0.083 (-0.171, 0.003)  & -0.062 (-0.150, 0.023)  & -0.031 (-0.121, 0.055) \\
                             &                              & Semi.PV                           & 0.541 (0.436, 0.631) & -0.124 (-10.847, 10.322) & 0.321 (-0.984, 2.104)   & 0.378 (-0.975, 2.452)   & 0.204 (-1.550, 1.873)  \\ \hline
\multirow{6}{*}{Tau}         & \multirow{3}{*}{Sex: Male}   & BN                               & 0.772 (0.744, 0.798) & -0.018 (-0.084, 0.048)   & -0.044 (-0.096, 0.008)  & -0.037 (-0.090, 0.017)  & -0.018 (-0.084, 0.048) \\
                             &                              & PV                               & 0.772 (0.753, 0.790) & -0.020 (-0.109, 0.071)   & -0.046 (-0.130, 0.040)  & -0.038 (-0.122, 0.048)  & -0.02 (-0.109, 0.071)  \\
                             &                              & Semi.PV                           & 0.771 (0.752, 0.790) & -0.018 (-0.107, 0.072)   & -0.044 (-0.129, 0.046)  & -0.036 (-0.122, 0.054)  & -0.018 (-0.107, 0.072) \\ \cline{2-8} 
                             & \multirow{3}{*}{Sex: Female} & BN                               & 0.874 (0.852, 0.894) & 0.239 (0.181, 0.298)     & 0.224 (0.166, 0.282)    & 0.232 (0.176, 0.289)    & 0.239 (0.181, 0.298)   \\
                             &                              & PV                               & 0.873 (0.856, 0.889) & 0.240 (0.156, 0.330)     & 0.225 (0.138, 0.314)    & 0.233 (0.147, 0.322)    & 0.240 (0.156, 0.330)   \\
                             &                              & Semi.PV                           & 0.873 (0.856, 0.888) & 0.240 (0.151, 0.331)     & 0.224 (0.137, 0.318)    & 0.232 (0.145, 0.324)    & 0.240 (0.151, 0.331)   \\ \hline
\multirow{6}{*}{pTau}        & \multirow{3}{*}{Sex: Male}   & BN                               & 0.707 (0.677, 0.736) & 0.096 (0.006, 0.180)     & 0.108 (0.050, 0.166)    & 0.106 (0.044, 0.165)    & 0.096 (0.006, 0.180)   \\
                             &                              & PV                               & 0.707 (0.686, 0.727) & 0.097 (-0.009, 0.200)    & 0.108 (0.014, 0.201)    & 0.106 (0.012, 0.199)    & 0.097 (-0.009, 0.200)  \\
                             &                              & Semi.PV                           & 0.707 (0.687, 0.727) & 0.095 (-0.013, 0.203)    & 0.107 (0.010, 0.204)    & 0.104 (0.008, 0.203)    & 0.095 (-0.013, 0.203)  \\ \cline{2-8} 
                             & \multirow{3}{*}{Sex: Female} & BN                               & 0.742 (0.711, 0.771) & 0.229 (0.147, 0.311)     & 0.239 (0.177, 0.302)    & 0.236 (0.173, 0.302)    & 0.229 (0.147, 0.311)   \\
                             &                              & PV                               & 0.742 (0.718, 0.765) & 0.229 (0.124, 0.336)     & 0.239 (0.144, 0.338)    & 0.236 (0.141, 0.335)    & 0.229 (0.124, 0.336)   \\
                             &                              & Semi.PV                           & 0.741 (0.717, 0.764) & 0.228 (0.129, 0.334)     & 0.238 (0.147, 0.334)    & 0.236 (0.143, 0.333)    & 0.228 (0.129, 0.334)   \\ \hline
\end{tabular}
\end{adjustbox}
\end{center}
\end{table}

%The sex-specific AUC and the optimal cutoff estimates in Table~\ref{Tab:cov_data_analysis} across all scenarios track the results in Table~\ref{Tab:nocov_data_analysis}. For the biomarker $A\beta 42$, the performance of the BN and PV are almost identical in estimating AUC and optimal cutoffs, while the Semi.PV produces very different estimates. Based on the BN and PV models, we see that the diagnostic performance of $A\beta 42$ is significantly higher for males (AUC = 0.853) than that of females (AUC = 0.826). The slope parameter (b = 0.15, 95\% CI = (0.037, 0.264)) corresponding to the sex variable in the PV-model denotes the significance. On the other hand, the Semi.PV model yields smaller AUC (0.642 for male and 0.541 for female). The estimates of the cutpoints are also different between these models. The BN, PV models yield same estimates of J, IU (-0.014), but the ER (around -0.058) and CZ (-0.038) estimates are different. The Semi.PV yields very different estimates of cutpoints for different sex  groups, which are tabulated in Table~\ref{Tab:cov_data_analysis}.

The sex-specific AUC and optimal cutoff estimates in Table~\ref{Tab:cov_data_analysis} echo the trends observed in Table~\ref{Tab:nocov_data_analysis}. For the biomarker $A\beta 42$, the BN and PV models yield nearly identical estimates of AUC and optimal cutoffs, while the Semi.PV model produces markedly different results. According to the BN and PV models, the diagnostic performance of $A\beta 42$ is notably higher for males (AUC = 0.853) compared to females (AUC = 0.826). The slope parameter ($\beta_1$ = 0.15, 95\% CI = (0.037, 0.264)) corresponding to the sex variable in the PV regression model in Section~\ref{Sec: Cov_PV_ROC} confirms the significance of this difference. In contrast, the Semi.PV model yields lower AUC values (0.642 for males and 0.541 for females). Moreover, the estimates of the cutoff points vary between these models. While the BN and PV models yield identical estimates for J and IU (-0.014), the ER (approximately -0.058) and CZ (-0.038) estimates differ. In contrast, the Semi.PV model yields distinct estimates of cutoff points for different sex groups, as detailed in the rows corresponding to the biomarker $A\beta 42$ of Table~\ref{Tab:cov_data_analysis}.

For the biomarker Tau, the performance of all models exhibits notable similarity. However, based on the AUC estimates corresponding to Tau, we observe significantly higher diagnostic capacity for females (AUC: 0.873 - 0.874) compared to males (AUC: 0.771 - 0.772), as evidenced by the slope parameter ($\beta_1$ = -0.582, 95\% CI = (-0.716, -0.449)) corresponding to the sex variable from the PV regression model. Consequently, the cutoffs for males and females also differ significantly. For males, all ROC models yield similar estimates of J, IU (-0.020 to -0.018), ER (-0.046 to -0.044), and CZ (-0.038 to -0.036). Conversely, for females, the pattern remains the same but with different cutoff estimates of J, IU (0.239 - 0.240), ER (0.224 - 0.225), and CZ (0.232 - 0.233).

In the case of the biomarker pTau, once again, the ROC models behave similarly as compared to Tau. Here, we observe marginally higher performance of pTau for females (AUC: 0.741 - 0.742) than males (AUC: 0.707). However, similar to $A\beta 42$, the statistical difference is marginal, as the 95\% CI of the slope parameter ($\beta_1$ = -0.148, 95\% CI = (-0.278, -0.024)) corresponding to the sex variable from the PV model barely includes 0. The cutoff estimates remain consistent across all ROC models, with J, IU (0.095 - 0.097) estimates being similar, as well as ER (0.107 - 0.108) and CZ (0.104 - 0.106) estimates. The pattern remains consistent for females, with similar estimates of J, IU (0.228 - 0.229), ER (0.238 - 0.239), and CZ (0.236).

All of these cutoffs are visually represented on the healthy and diseased density plots of the three biomarkers in Figures C.26 - C.28 in the supplementary material.

%All of these cutoffs are visually represented on the healthy and diseased density plots of the three biomarkers in Figures~\ref{Fig: data_analysis_cov_density_AB42} - \ref{Fig: data_analysis_cov_density_pTau}.

\section{Discussion} \label{Sec: Discussion}

In this article, our aim was to evaluate the performance of various existing optimal threshold techniques. While this task has been previously explored using a single ROC estimation technique, typically the empirical one, we hypothesized that such an approach might not provide a comprehensive understanding. Although the empirical ROC model is simple to fit, widely used, and often the default method for most of the available softwares, extensive simulations revealed few instances where its performance was inadequate, and alternative ROC modeling techniques proved to be more effective. Additionally, the absence of well-established options for estimating covariate-specific optimal thresholds prompted our investigation. This article not only presents some of these options but also compares their performances under various settings.

Through extensive simulation exercises, we found the empirical model (Emp) to be a powerful tool for estimating AUC and optimal cutoffs. Despite concerns regarding the smoothness of the empirical ROC curve, using a moderately dense set of points to estimate empirical ROC curve might mitigate this issue. However, the Kernel-based nonparametric (NonPar) model often outperformed the Emp model in estimating cutoffs. Moreover, the BN and PV models demonstrated robustness, particularly outperforming the Emp and NonPar models unless the biomarkers were generated from highly skewed models. The Semi.PV model, although the most complex, consistently performed well and competed with both parametric and nonparametric models, especially in situations with adequate sample sizes. Similar observations were made in simulations involving covariates: correctly specified parametric models exhibited superior performance, but as biomarkers became more skewed, the Semi.PV model demonstrated its superiority.

Regarding the choice of optimal thresholds, in many scenarios, the estimation of J resulted in high bias and variability, particularly evident at low AUC levels with parametric and semiparametric models. Conversely, IU, CZ, and ER estimates remained more consistent across different scenarios. Although concave models were mentioned, they were consciously avoided in the simulation and data analysis due to their strict shape restrictions and tendency for bias in the event of model misspecification. We also deliberately excluded another framework for modeling ROC curves from our analysis, namely the Lehmann family of ROC curves \citep{gonen2010lehmann}, due to the prerequisite of adhering to the Lehmann assumption.

%The data analysis revealed distinct patterns in the behaviors of different ROC estimating models and cutoff estimators. For the standardized $A\beta 42$ biomarker, three groups of ROC estimating models were identified based on their estimation characteristics, each exhibiting unique performance in terms of AUC and cutoff estimation. Similar patterns were observed for Tau and pTau biomarkers, albeit with slight variations, suggesting potentially different performances between males and females.

The data analysis revealed distinct patterns in the behavior of various ROC estimating models and cutoff estimators. For the standardized $A\beta 42$ biomarker, we identified three distinct groups of ROC estimating models based on their estimation characteristics. The first group, comprising the Emp and NonPar models, exhibited similar estimates of AUC and produced different optimal cutoffs with less variability. In contrast, the second group, consisting of the parametric models (BN and PV), yielded AUC estimates similar to the first group but different cutoff estimates, albeit with similar J and IU estimates. Notably, the third group exclusively comprised the Semi.PV model due to its markedly different AUC and cutoff estimates. The notably smaller AUC observed in this group stemmed from the clear bimodality in both healthy and diseased group densities, as depicted in the first row of Figure C.25 of the supplementary material. Consequently, the estimated cutoffs from the Semi.PV model exhibited significant divergence. As for the biomarkers Tau and pTau, the majority of the aforementioned patterns persisted, albeit with the Semi.PV model exhibiting behavior more akin to the parametric models. This similarity may be attributed to the more bell-shaped densities of the biomarkers. This pattern in the data analysis was also observed in the simulation scenarios as well.

%However, there are some caveats to consider. Firstly, covariate impacts were assumed to be linear, which might not fully capture the complexity of relationships, particularly for continuous covariates like age. Additionally, while significant differences in performance were observed for males and females, caution is warranted, especially considering the limitations of linear covariate impact assumptions and potential model misspecification.

The covariate-specific analysis revealed significant differences in biomarker performance between males and females. However, it is important to note a couple of caveats. Firstly, our model only accounted for linear covariate effects, which may be limiting for continuous covariates like age. Future research could explore non-linear covariate effects for more comprehensive insights. Additionally, for the $A\beta 42$ and pTau biomarkers, the slope parameters associated with sex in the PV regression models had credible intervals that barely included 0, warranting cautious interpretation of findings for these biomarkers. Nevertheless, for Tau, there's strong evidence indicating divergent performance between males and females.

Another limitation concerns the estimation of PV-based models, which may underestimate variability in placement value estimation, potentially affecting cutoff estimation. Strategies such as the Bayesian bootstrap proposed by \citet{de2013bayesian} offer potential solutions, but their translation to the cutoff estimation process remains unclear. These considerations highlight avenues for future research and refinement of methodologies in biomarker diagnostic accuracy assessment.

\section*{Acknowledgements}
%This research was supported by the Intramural Research Program of {\it Eunice Kennedy Shriver} National Institute of Child Health and Human Development. 
The authors acknowledge Research Computing at The University of Virginia for providing computational resources and technical support that have contributed to the results reported within this publication. (URL: \url{https://rc.virginia.edu})

\section*{Conflict of interest statement}
The author has declared no conflict of interest.

\section*{Data availability statement}
The ADNI data is publicly available online. Users can access the datasets and biosamples  following an approval process. To request approval, please visit \url{https://adni.loni.usc.edu/data-samples/access-data/}. 

\section*{Code availability statement}
R code of implementing data analysis will be made available online.

%\bibliographystyle{apalike}
%\bibliography{biblio_cutoff}
 \newcommand{\noop}[1]{}

\begin{appendices}
\renewcommand\thefigure{\thesection.\arabic{figure}}
\renewcommand\thetable{\thesection.\arabic{table}}
\setcounter{figure}{0}
\setcounter{table}{0}

\section{Noted ROC frameworks} \label{AppSec: Noted_ROC_models}
\subsection{General ROC model} \label{AppSec: Noted_general_ROC_models}

The prevailing ROC curve frameworks, including the empirical, parametric (such as the binormal model, bigamma model, etc.), and nonparametric (such as kernel-based ROC models), are all derivatives of the general ROC model framework. Within this framework, the distributions $F_0$ and $F_1$ are characterized either empirically, parametrically, or nonparametrically. Below, we briefly describe some of these widely recognized models:

\begin{itemize}

    \item \textbf{Empirical model}: The empirical model assumes no particular form of the ROC curve, rather is characterized by the empirical CDFs estimated as
    \begin{align}
        F_j(x) &= \frac{1}{n_j} \sum_{i = 1}^{n_j} I(Y_{ji} \leq x), \text{ } j=0,1, \label{Eq: EMPmodel}
    \end{align}
    \noindent
    and the corresponding ROC curve can be obtained from equations (\ref{Eq: ROCdef}) and the AUC can be estimated from \citet{delong1988comparing} as:
    \begin{align*}
        AUC = \frac{1}{n_0n_1} \sum_{i=1}^{n_0} \sum_{j=1}^{n_1} \Psi(Y_{0i}, Y_{1j}),
    \end{align*}
    \noindent
    where
    \begin{eqnarray}\label{Eq: empAUC}
& \Psi(a,b) = 
\begin{dcases}
	1, & a < b \\
	\frac{1}{2}, & a=b \\
        0, & a> b.
\end{dcases} \\ \nonumber
\end{eqnarray}

    \item \textbf{Nonparametric models}: While the empirical estimates are useful because of simplicity in structure and no requirement of model assumptions, the empirical estimates of ROC curve are not smooth. To overcome the lack of smoothness, the nonparametric models especially the Kernel-based smooth models became more popular \citep{zou1997smooth, lloyd1998using}. Based on Kernel estimators, we can specify $F_0$ and $F_1$ as
    \begin{align}\label{Eq: Kernelmodel}
        F_j(x) &= \frac{1}{n_j} \sum_{i = 1}^{n_j} \kappa \left(\frac{x-Y_{ji}}{h_j} \right),  \\ \nonumber
        h_j &= 0.9\cdot min(SD(Y_j),\text{ } IQR(Y_j)/1.34)\cdot n_j^{-1/5}, \text{ } j=0,1,
    \end{align}
    \noindent
    where, $\kappa$ is the kernel, $h_j$'s are the bandwidths, $SD(x)$ and $IQR(x)$ corresponds to respectively the standard deviation and interquartile range of $x$. In this article, Gaussian kernels were employed for estimating nonparametric models.
    
    \item \textbf{Binormal model}: Consider $Y_0 \sim N(\mu_0, \sigma_0^2)$ and $Y_1 \sim N(\mu_1, \sigma_1^2)$, where $N(a,b)$ denotes normal distribution with mean $a$ and variance $b$. Then according to the binormal model, we can estimate 
    \begin{align} \label{Eq: BN_ab}
        a = \frac{\mu_1 - \mu_0}{\sigma_1}, \text{ } b = \frac{\sigma_0}{\sigma_1}
    \end{align}
    \noindent
    and the binormal ROC and AUC have the closed-form given by
    \begin{align}\label{Eq: BNmodel}
        ROC(t) &= \Phi(a + b \cdot\Phi^{-1}(t)), \text{ } t \in (0,1), \\  \nonumber 
        AUC &= \Phi \left( \frac{a}{\sqrt{1+b^2}} \right), \nonumber
    \end{align}
    \noindent
    where $\Phi(y)$ corresponds to a standard normal CDF obtained at $y$. More on this can be found in \citet{metz1980statistical}.
    \item \textbf{Bigamma model}: The bigamma model characterizes healthy and diseased biomarkers by modeling them using gamma distributions. Consider $Y_0 \sim Gam(k,\phi_0)$ and $Y_1 \sim Gam(k,\phi_1)$, where $Gam(k,\phi)$ denotes gamma distribution with mean $k\phi$. Then the bigamma ROC and AUC are given by:
    \begin{align}\label{Eq: BGmodel}
 ROC(t) &= 1-\mathbb{G}_1 \left(\mathbb{G}_0^{-1}(1-t)\right), \text{ }t\in (0,1), \\  \nonumber
 AUC &= 1-H_{(2k, 2k)} \left( \frac{\phi_0}{\phi_1}\right), \nonumber
\end{align}
\noindent 
where $\mathbb{G}_l(\cdot)\equiv Gam(k,\phi_l), \text{ }l=0,1,$ and $H_{\nu_1,\nu_2}$ is CDF of the F-distribution with degrees of freedom $\nu_1$ and $\nu_2$. The shared shape parameter ensures that the bigamma ROC curve maintains a strictly concave shape. \citet{dorfman1996proper} has further details on this.

\item \textbf{Bichi-squared model}: The bichi-squared model represents another instance of a concave ROC framework. In particular, the bichi-squared model reparameterizes the conventional binormal model in (\ref{Eq: BN_ab}) using an equivalence in proper binormal and bichi-square distribution (see \citet{hillis2016equivalence} for more details), and calculate $\lambda$ and $\theta$ as 
\begin{align*}
\lambda  = \frac{1}{b^2} \text{ and } \theta = \frac{a^2 b^2}{(1-b^2)^2},
\end{align*}
so that the true ROC curve (for $t \in (0,1)$) and true AUC have the following form:
\begin{eqnarray}\label{Eq: BiChimodel}
& ROC(t) = 
\begin{dcases}
	1-F_{\lambda\theta}\left(\frac{1}{\lambda}F_{\theta}^{-1}(1-t)\right), & \lambda > 1 \\
	F_{\lambda\theta}\left(\frac{1}{\lambda}F_{\theta}^{-1}(1t)\right), & \lambda < 1
\end{dcases} \\ \nonumber
& AUC = \Phi\left( \frac{\sqrt(\theta)\sqrt(\lambda-1)}{\sqrt(\lambda+1)}\right)+2F_{BVN} \left( -\frac{\sqrt(\theta)\sqrt(\lambda-1)}{\sqrt(\lambda+1)},0; -\frac{2\sqrt(\lambda)}{\lambda+1}\right), \nonumber
\end{eqnarray}
\noindent
where $F_{\nu}$ denotes CDF of a chi-squared distribution with noncentrality parameter $\nu$ and $F_{BVN}(\cdot,\cdot; \rho)$ denotes CDF of a standardized bivariate normal distribution with correlation $\rho$.
\end{itemize}

\subsection{PV-based ROC model} \label{AppSec: Noted_PV_ROC_models}

Some special PV-based ROC models are:

\begin{itemize}
    \item \textbf{Parametric PV model}: Following the transformed normal model in \citet{ghosal2019discriminatory}, we can come up with a parametric version of a PV-based model, where $Y_0 \sim F_0(\cdot) \equiv N(\mu_0, \sigma_0^2)$ similar to the BN model, and PV can be estimated as $Z = 1-\Phi \left( Y_1; \mu_0, \sigma_0 \right)$. We can model $Z$ as $\Phi^{-1}(Z) \sim N(\mu, \sigma)$, and the corresponding ROC curve and AUC can be estimated as
        \begin{align}\label{Eq: parametricPV}
        ROC(t) &= \Phi \left( \Phi^{-1}(t); \mu, \sigma \right), \text{ } t \in (0,1), \\  \nonumber 
        AUC &= \int_0^1 ROC(t) dt. \nonumber
    \end{align}

    \item \textbf{Semiparametric PV model}: The PV-based semiparametric model is based on the Dirichlet process mixture (DPM) model which models the data as a mixture of normals where the DP prior is assumed on mixing distribution. For example, we can write $F_0$ in terms of DPM as:
    \begin{align*}
        F_0(x) &= \int \Phi\left( x; \mu_{0}, \sigma_0 \right) G_0(\mu_0), \\
        G_0 &\sim DP(\alpha_0, G^*_0),
    \end{align*}
    \noindent
    where $\alpha_0$ is the concentration parameter, $G^*_0(\mu_0)$ is the baseline distribution with appropriate normal priors. Note that, $G_0$ can include both $\mu_0$ and $\sigma_0$. We opt to solely incorporate the location parameter in the DP prior, guided by literature \citep{ghosal1999posterior,lijoi2005hierarchical}, which demonstrates that any density on the real line can be approximated utilizing a Dirichlet process with location mixtures of normals. Considering the stick-breaking representation of the DP \citep{sethuraman1994constructive} and truncated DP process \citep{ishwaran2002approximate}, we can write the infinite mixtures of normal representation of DPM as a finite mixture of normals given by:
    \begin{align*}
        F_0(x) &= \sum_{k=1}^{K} \pi_{0k} \Phi \left( x; \mu_{0k}, \sigma_0 \right),
    \end{align*}
    \noindent
    where $K$ is a finite number of clusters, say $K=30$, $\pi_0 = (\pi_{01},\pi_{02},\ldots,\pi_{0K})$ corresponds to the mixing weights correspond to $F_0$. Now, based on the above specification, we can estimate the PV as
    \begin{align*}
        z_j &= 1 - \sum_{k=1}^{K} \pi_{0k} \Phi \left( y_{1j}; \mu_{0k}, \sigma_0 \right), \text{ }j=1,\ldots,n_j,
    \end{align*}
    \noindent
    and fit the $\eta^{-1}(z)$ as a separate yet similar DPM structure as above given as before, which results in the estimation of the CDF of PV as $F$, i.e. the ROC curve and the corresponding AUC as:
    \begin{align} \label{Eq: semiparametricPV}
        F(t) = ROC(t) &= \sum_{k=1}^{K} \pi_{k} \Phi \left( \eta^{-1}(t); \mu_{k}, \sigma \right), \text{ } t \in (0,1), \\ \nonumber
        AUC &= \int_0^1 ROC(t)dt, \nonumber
    \end{align}
    \noindent
    where the interpretations of $\mu$, $\sigma$, $\pi$ are similar for the estimation of $F$ or the ROC curve.

%\begin{align} \label{Eq: semiparametricPV}
%    \eta(z_i) \sim N(\mu_i, \sigma^2)
%\end{align}
    
\end{itemize}

\end{appendices}

\end{document}

\title{Supplement for \enquote{Impact of methodological assumptions and covariates on the cutoff estimation in ROC analysis}}

% \author{Author name(s)}
% \address{Author affiliation and full address}
% \email{e-mail address}
%%Uncomment the following line to override copyright year from the default current year.
%\copyrightyear{2022}

%\author{Soutik Ghosal,\authormark{1} Author Two,\authormark{2,*} and Author Three\authormark{2,3}}
\author{Soutik Ghosal}
\address{Division of Biostatistics, Department of Public Health Sciences, School of Medicine, University of Virginia, Charlottesville, VA 22903}
%\address{\authormark{1} Publications Department, Optica Publishing Group, 2010 Massachusetts Avenue NW, Washington, DC 20036\\
%\authormark{2}Publications Department, Optica Publishing Group, 2010 Massachusetts Avenue NW, Washington, DC 20036\\
%\authormark{3}Currently with the Department of Electronic Journals, Optica Publishing Group, 2010 Massachusetts Avenue NW, Washington, DC 20036}

\email{\authormark{*}soutik.ghosal@virginia.edu} %% email address is required

%\begin{appendices}

\renewcommand{\thesection}{\Alph{section}}
\renewcommand\thefigure{\thesection.\arabic{figure}}
\renewcommand\thetable{\thesection.\arabic{table}}
\setcounter{figure}{0}
\setcounter{table}{0}

In this supplementary material accompanying the main paper, we provide technical information along with supporting tables and figures. Section~\ref{AppSec: True cutoff} details the true values of AUC and optimal cutoffs for simulated data with and without covariates. Section~\ref{AppSec: Other_sim} presents bias plots for simulations across all sample sizes, along with tables for low and high sample sizes. Finally, Section~\ref{AppSec: Other_data_analysis} showcases density plots of the biomarkers used in the data analyses, with different cutoff estimates obtained from various ROC methodologies overlaid on the plots.

\section{True cutoff} \label{AppSec: True cutoff}

Table~\ref{Tab:Sim_nocov_true_data} and Table~\ref{Tab:Sim_cov_true_data} shows the true values of AUC, cutoff estimates for different data generating mechanism under no covariate and with covariate framework respectively.

\begin{table}[hbt!]
\begin{center}
\caption{\label{Tab:Sim_nocov_true_data} True AUC and cutoff estimates under no covariate framework.}
\begin{adjustbox}{width=\textwidth}
\begin{tabular}{ccccc|ccccc}
\hline
\textbf{Data generating}     & \textbf{AUC}            & \textbf{Method} & \textbf{True}          & \textbf{True}   & \textbf{Data generating}     & \textbf{AUC}            & \textbf{Method} & \textbf{True}          & \textbf{True}   \\
\textbf{mechanism}           & \textbf{level}          & \textbf{}       & \textbf{AUC}           & \textbf{Cutoff} & \textbf{mechanism}           & \textbf{level}          & \textbf{}       & \textbf{AUC}           & \textbf{Cutoff} \\ \hline
\multirow{12}{*}{BN.equal}   & \multirow{4}{*}{Low}    & J               & \multirow{4}{*}{0.556} & 0.100           & \multirow{12}{*}{Skewed.III} & \multirow{4}{*}{Low}    & J               & \multirow{4}{*}{0.564} & 0.061           \\
                             &                         & ER              &                        & 0.100           &                              &                         & ER              &                        & 0.031           \\
                             &                         & CZ              &                        & 0.100           &                              &                         & CZ              &                        & 0.031           \\
                             &                         & IU              &                        & 0.100           &                              &                         & IU              &                        & 0.030           \\ \cline{2-5} \cline{7-10} 
                             & \multirow{4}{*}{Medium} & J               & \multirow{4}{*}{0.760} & 0.500           &                              & \multirow{4}{*}{Medium} & J               & \multirow{4}{*}{0.753} & 0.108           \\
                             &                         & ER              &                        & 0.500           &                              &                         & ER              &                        & 0.065           \\
                             &                         & CZ              &                        & 0.500           &                              &                         & CZ              &                        & 0.077           \\
                             &                         & IU              &                        & 0.500           &                              &                         & IU              &                        & 0.067           \\ \cline{2-5} \cline{7-10} 
                             & \multirow{4}{*}{High}   & J               & \multirow{4}{*}{0.961} & 1.250           &                              & \multirow{4}{*}{High}   & J               & \multirow{4}{*}{0.924} & 0.216           \\
                             &                         & ER              &                        & 1.250           &                              &                         & ER              &                        & 0.149           \\
                             &                         & CZ              &                        & 1.250           &                              &                         & CZ              &                        & 0.199           \\
                             &                         & IU              &                        & 1.250           &                              &                         & IU              &                        & 0.157           \\ \hline
\multirow{12}{*}{BN.unequal} & \multirow{4}{*}{Low}    & J               & \multirow{4}{*}{0.555} & -0.636          & \multirow{12}{*}{Mixed.I}    & \multirow{4}{*}{Low}    & J               & \multirow{4}{*}{0.572} & 1.415           \\
                             &                         & ER              &                        & -0.069          &                              &                         & ER              &                        & 0.658           \\
                             &                         & CZ              &                        & -0.107          &                              &                         & CZ              &                        & 0.847           \\
                             &                         & IU              &                        & 0.089           &                              &                         & IU              &                        & 0.182           \\ \cline{2-5} \cline{7-10} 
                             & \multirow{4}{*}{Medium} & J               & \multirow{4}{*}{0.779} & 0.325           &                              & \multirow{4}{*}{Medium} & J               & \multirow{4}{*}{0.767} & 1.389           \\
                             &                         & ER              &                        & 0.548           &                              &                         & ER              &                        & 0.911           \\
                             &                         & CZ              &                        & 0.436           &                              &                         & CZ              &                        & 1.151           \\
                             &                         & IU              &                        & 0.616           &                              &                         & IU              &                        & 0.730           \\ \cline{2-5} \cline{7-10} 
                             & \multirow{4}{*}{High}   & J               & \multirow{4}{*}{0.928} & 1.804           &                              & \multirow{4}{*}{High}   & J               & \multirow{4}{*}{0.928} & 1.650           \\
                             &                         & ER              &                        & 1.667           &                              &                         & ER              &                        & 1.360           \\
                             &                         & CZ              &                        & 1.772           &                              &                         & CZ              &                        & 1.584           \\
                             &                         & IU              &                        & 1.730           &                              &                         & IU              &                        & 1.461           \\ \hline
\multirow{12}{*}{Skewed.I}   & \multirow{4}{*}{Low}    & J               & \multirow{4}{*}{0.556} & 0.010           & \multirow{12}{*}{Mixed.II}   & \multirow{4}{*}{Low}    & J               & \multirow{4}{*}{0.600} & 1.505           \\
                             &                         & ER              &                        & 0.010           &                              &                         & ER              &                        & 0.864           \\
                             &                         & CZ              &                        & 0.010           &                              &                         & CZ              &                        & 0.913           \\
                             &                         & IU              &                        & 0.010           &                              &                         & IU              &                        & 0.783           \\ \cline{2-5} \cline{7-10} 
                             & \multirow{4}{*}{Medium} & J               & \multirow{4}{*}{0.794} & 0.128           &                              & \multirow{4}{*}{Medium} & J               & \multirow{4}{*}{0.737} & 1.408           \\
                             &                         & ER              &                        & 0.258           &                              &                         & ER              &                        & 1.065           \\
                             &                         & CZ              &                        & 0.203           &                              &                         & CZ              &                        & 1.140           \\
                             &                         & IU              &                        & 0.183           &                              &                         & IU              &                        & 1.208           \\ \cline{2-5} \cline{7-10} 
                             & \multirow{4}{*}{High}   & J               & \multirow{4}{*}{0.910} & 9.851           &                              & \multirow{4}{*}{High}   & J               & \multirow{4}{*}{0.944} & 1.768           \\
                             &                         & ER              &                        & 8.946           &                              &                         & ER              &                        & 1.634           \\
                             &                         & CZ              &                        & 9.070           &                              &                         & CZ              &                        & 1.716           \\
                             &                         & IU              &                        & 7.620           &                              &                         & IU              &                        & 1.768           \\ \hline
\multirow{12}{*}{Skewed.II}  & \multirow{4}{*}{Low}    & J               & \multirow{4}{*}{0.556} & 1.105           & \multicolumn{5}{c}{\multirow{12}{*}{}}                                                                              \\
                             &                         & ER              &                        & 1.105           & \multicolumn{5}{c}{}                                                                                                \\
                             &                         & CZ              &                        & 1.105           & \multicolumn{5}{c}{}                                                                                                \\
                             &                         & IU              &                        & 1.105           & \multicolumn{5}{c}{}                                                                                                \\ \cline{2-5}
                             & \multirow{4}{*}{Medium} & J               & \multirow{4}{*}{0.794} & 1.430           & \multicolumn{5}{c}{}                                                                                                \\
                             &                         & ER              &                        & 1.662           & \multicolumn{5}{c}{}                                                                                                \\
                             &                         & CZ              &                        & 1.570           & \multicolumn{5}{c}{}                                                                                                \\
                             &                         & IU              &                        & 1.532           & \multicolumn{5}{c}{}                                                                                                \\ \cline{2-5}
                             & \multirow{4}{*}{High}   & J               & \multirow{4}{*}{0.910} & 5.994           & \multicolumn{5}{c}{}                                                                                                \\
                             &                         & ER              &                        & 6.762           & \multicolumn{5}{c}{}                                                                                                \\
                             &                         & CZ              &                        & 6.222           & \multicolumn{5}{c}{}                                                                                                \\
                             &                         & IU              &                        & 6.229           & \multicolumn{5}{c}{}                                                                                                \\ \cline{1-5}
\end{tabular}
\end{adjustbox}
\end{center}
\end{table}

\begin{table}[hbt!]
\begin{center}
\caption{\label{Tab:Sim_cov_true_data} True AUC and cutoff estimates under covariate framework.}
\begin{adjustbox}{width=0.65\textwidth}
\begin{tabular}{ccccc}
\hline
\textbf{Data generating} & \textbf{Covariate}     & \textbf{Method} & \textbf{True}          & \textbf{True}   \\
\textbf{mechanism}       & \textbf{level}         & \textbf{}       & \textbf{AUC}           & \textbf{cutoff} \\ \hline
\multirow{8}{*}{BN}      & \multirow{4}{*}{$X=0$} & J               & \multirow{4}{*}{0.638} & 1.250           \\
                         &                        & ER              &                        & 1.250           \\
                         &                        & CZ              &                        & 1.250           \\
                         &                        & IU              &                        & 1.250           \\ \cline{2-5} 
                         & \multirow{4}{*}{$X=1$} & J               & \multirow{4}{*}{0.856} & 2.750           \\
                         &                        & ER              &                        & 2.750           \\
                         &                        & CZ              &                        & 2.750           \\
                         &                        & IU              &                        & 2.750           \\ \hline
\multirow{8}{*}{Skewed}  & \multirow{4}{*}{$X=0$} & J               & \multirow{4}{*}{0.683} & 7.663           \\
                         &                        & ER              &                        & 6.731           \\
                         &                        & CZ              &                        & 6.874           \\
                         &                        & IU              &                        & 7.663           \\ \cline{2-5} 
                         & \multirow{4}{*}{$X=1$} & J               & \multirow{4}{*}{0.911} & 12.006          \\
                         &                        & ER              &                        & 10.844          \\
                         &                        & CZ              &                        & 11.588          \\
                         &                        & IU              &                        & 13.355          \\ \hline
\multirow{8}{*}{Mixed}   & \multirow{4}{*}{$X=0$} & J               & \multirow{4}{*}{0.609} & 0.949           \\
                         &                        & ER              &                        & 0.405           \\
                         &                        & CZ              &                        & 0.472           \\
                         &                        & IU              &                        & 0.716           \\ \cline{2-5} 
                         & \multirow{4}{*}{$X=1$} & J               & \multirow{4}{*}{0.917} & 2.234           \\
                         &                        & ER              &                        & 2.095           \\
                         &                        & CZ              &                        & 2.186           \\
                         &                        & IU              &                        & 2.424           \\ \hline
\end{tabular}
\end{adjustbox}
\end{center}
\end{table}

\section{Other simulations} \label{AppSec: Other_sim}
This section contains figures and tables corresponding to the simulation that could not be accommodated in the main paper due to space constraints.

\subsection{Without covariates} \label{AppSec: Other_sim_nocov}

\begin{figure}[hbt!]%
    \includegraphics[width=0.95\textwidth]{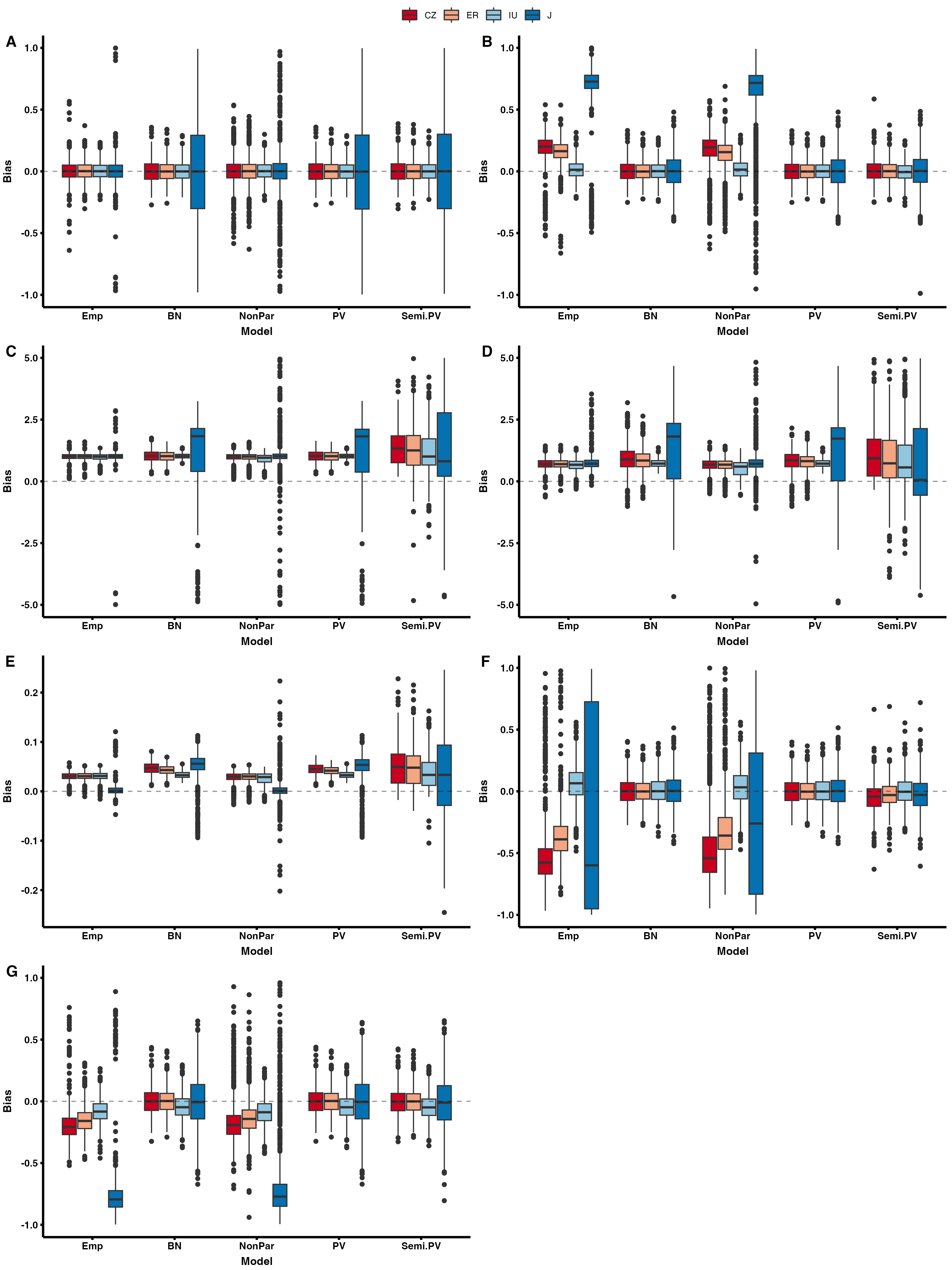}
   \caption{\label{Fig: Bias_medSamp_lowAUC}Bias in estimation AUC and optimal cut-points for medium sample size, when AUC level is low. Panels A - G respectively correspond to the simulation scenarios: BN equal, BN unequal, Skewed I, Skewed II, Skewed III, Mixed I, and Mixed II.}
\end{figure}

\begin{figure}[hbt!]%
    \includegraphics[width=\textwidth]{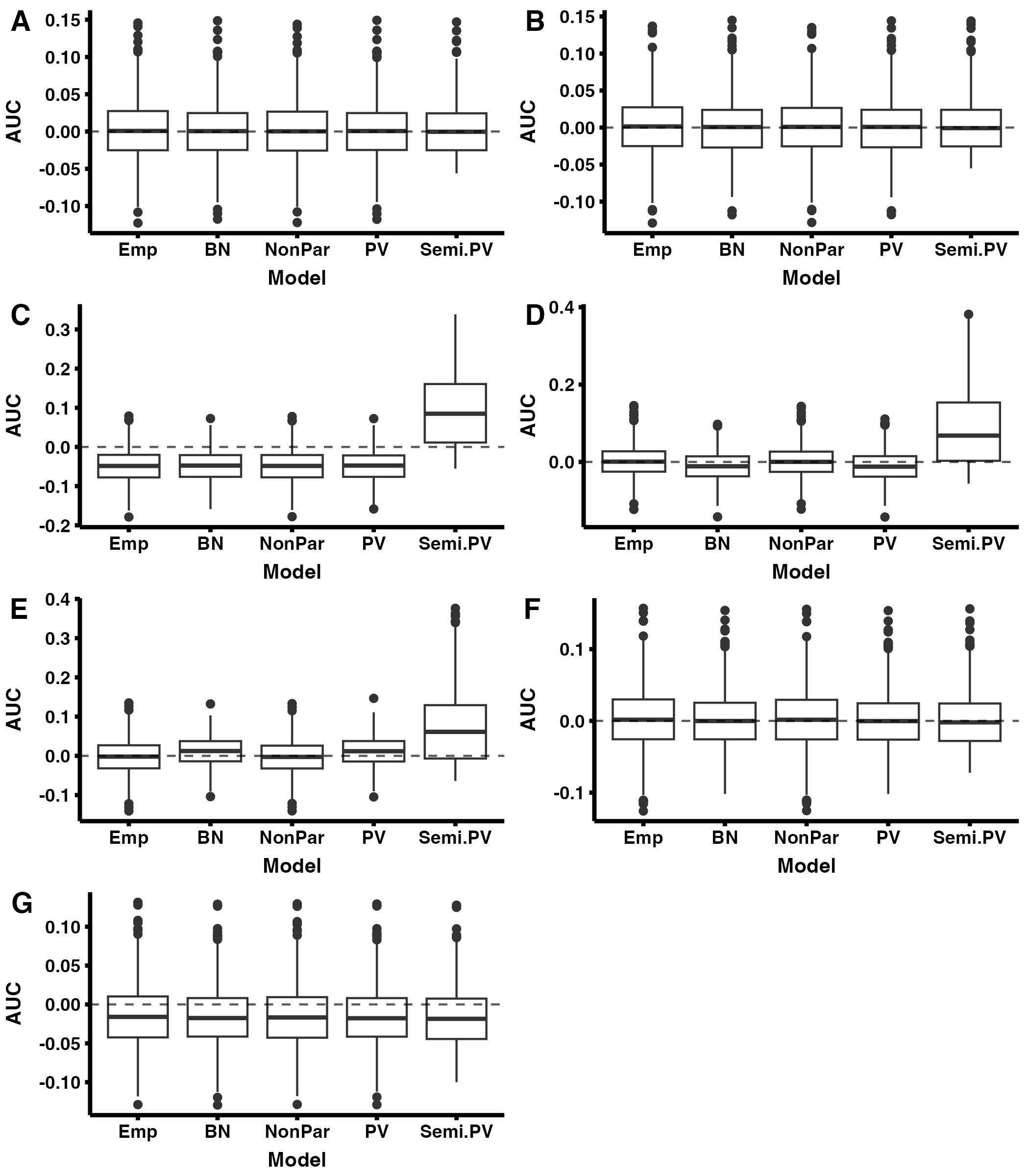}
    \caption{\label{Fig: Bias_AUC_medSamp_lowAUC}Bias in estimation of AUC for medium sample size, when AUC level is low. Panels A - G respectively correspond to the simulation scenarios: BN equal, BN unequal, Skewed I, Skewed II, Skewed III, Mixed I, and Mixed II.}
\end{figure}

\begin{figure}[hbt!]%
    \includegraphics[width=0.95\textwidth]{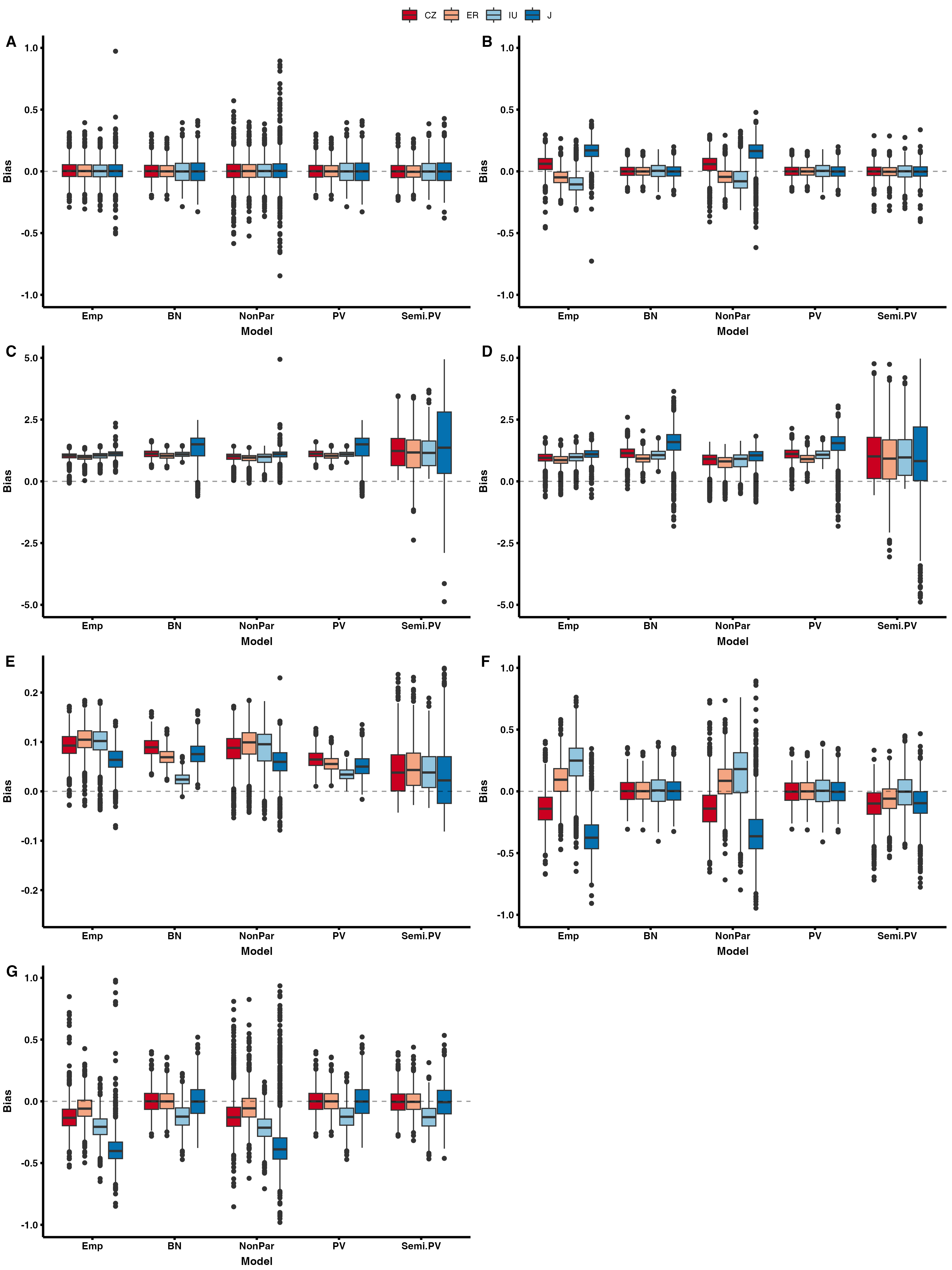}
   \caption{\label{Fig: Bias_medSamp_medAUC}Bias in estimation AUC and optimal cut-points for medium sample size, when AUC level is medium. Panels A - G respectively correspond to the simulation scenarios: BN equal, BN unequal, Skewed I, Skewed II, Skewed III, Mixed I, and Mixed II.}
\end{figure}

\begin{figure}[hbt!]%
    \includegraphics[width=\textwidth]{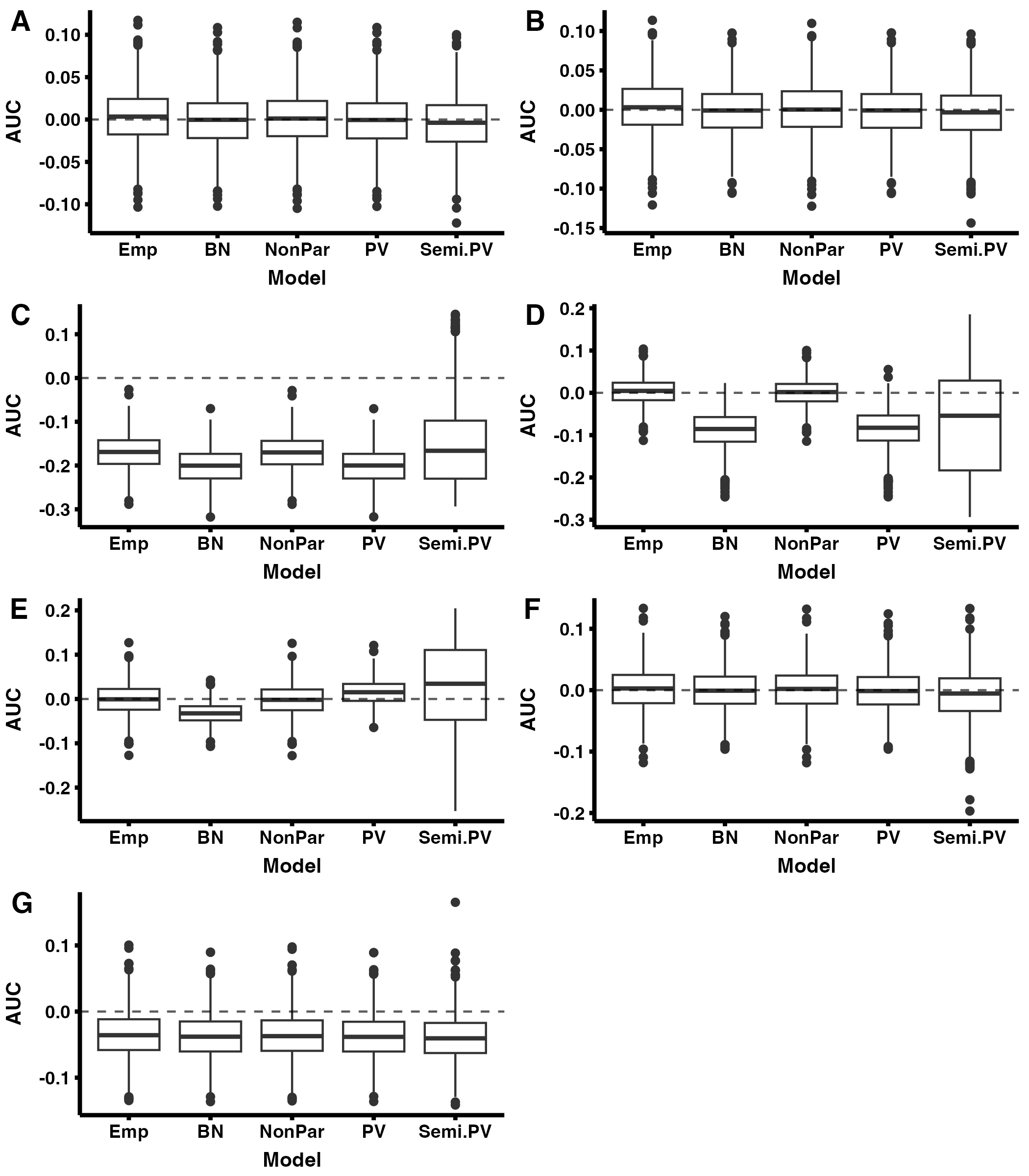}
    \caption{\label{Fig: Bias_AUC_medSamp_medAUC}Bias in estimation of AUC for medium sample size, when AUC level is medium. Panels A - G respectively correspond to the simulation scenarios: BN equal, BN unequal, Skewed I, Skewed II, Skewed III, Mixed I, and Mixed II.}
\end{figure}

\begin{figure}[hbt!]%
    \includegraphics[width=0.95\textwidth]{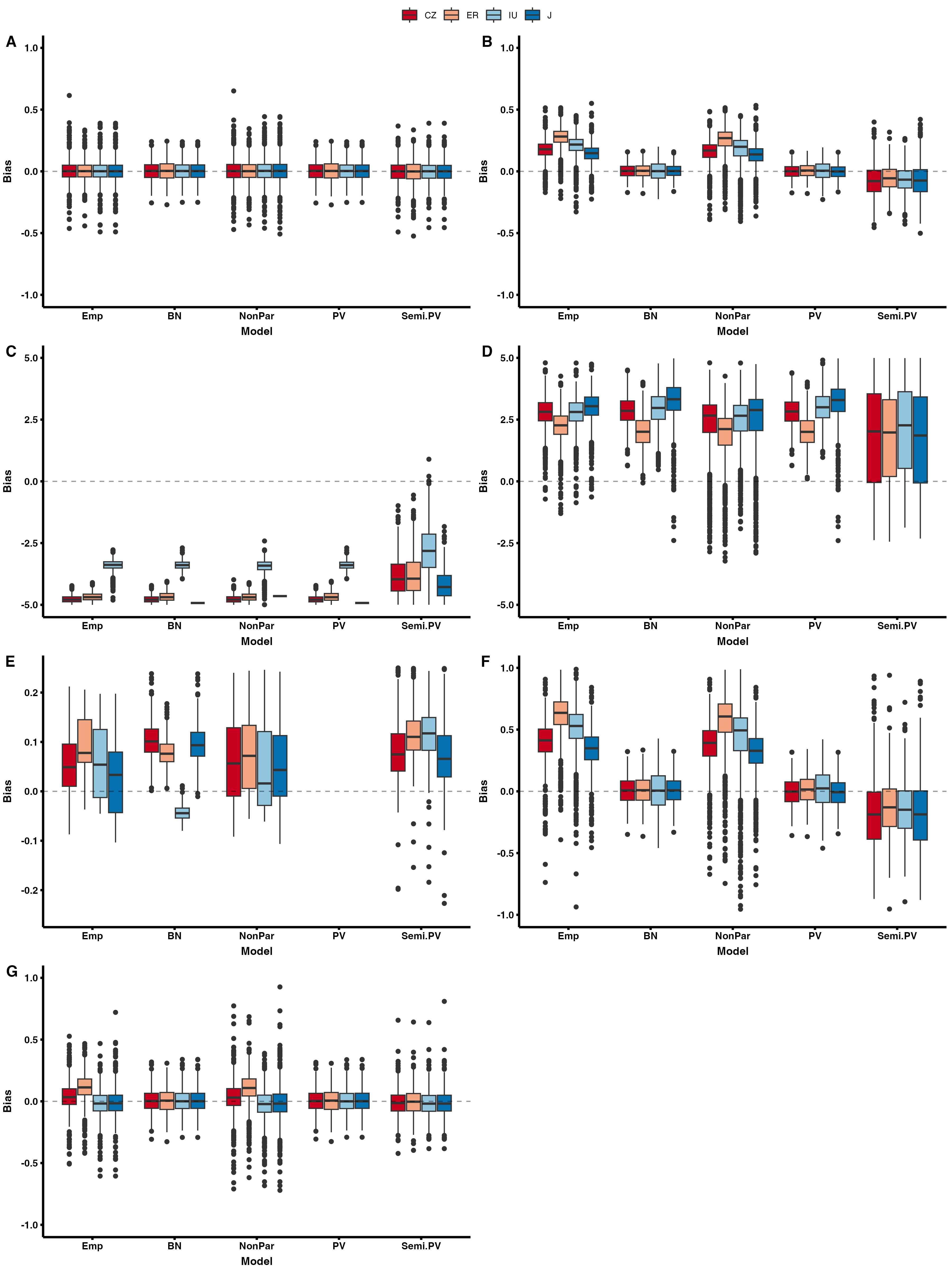}
   \caption{\label{Fig: Bias_medSamp_highAUC}Bias in estimation AUC and optimal cut-points for medium sample size, when AUC level is high. Panels A - G respectively correspond to the simulation scenarios: BN equal, BN unequal, Skewed I, Skewed II, Skewed III, Mixed I, and Mixed II.}
\end{figure}

\begin{figure}[hbt!]%
    \includegraphics[width=\textwidth]{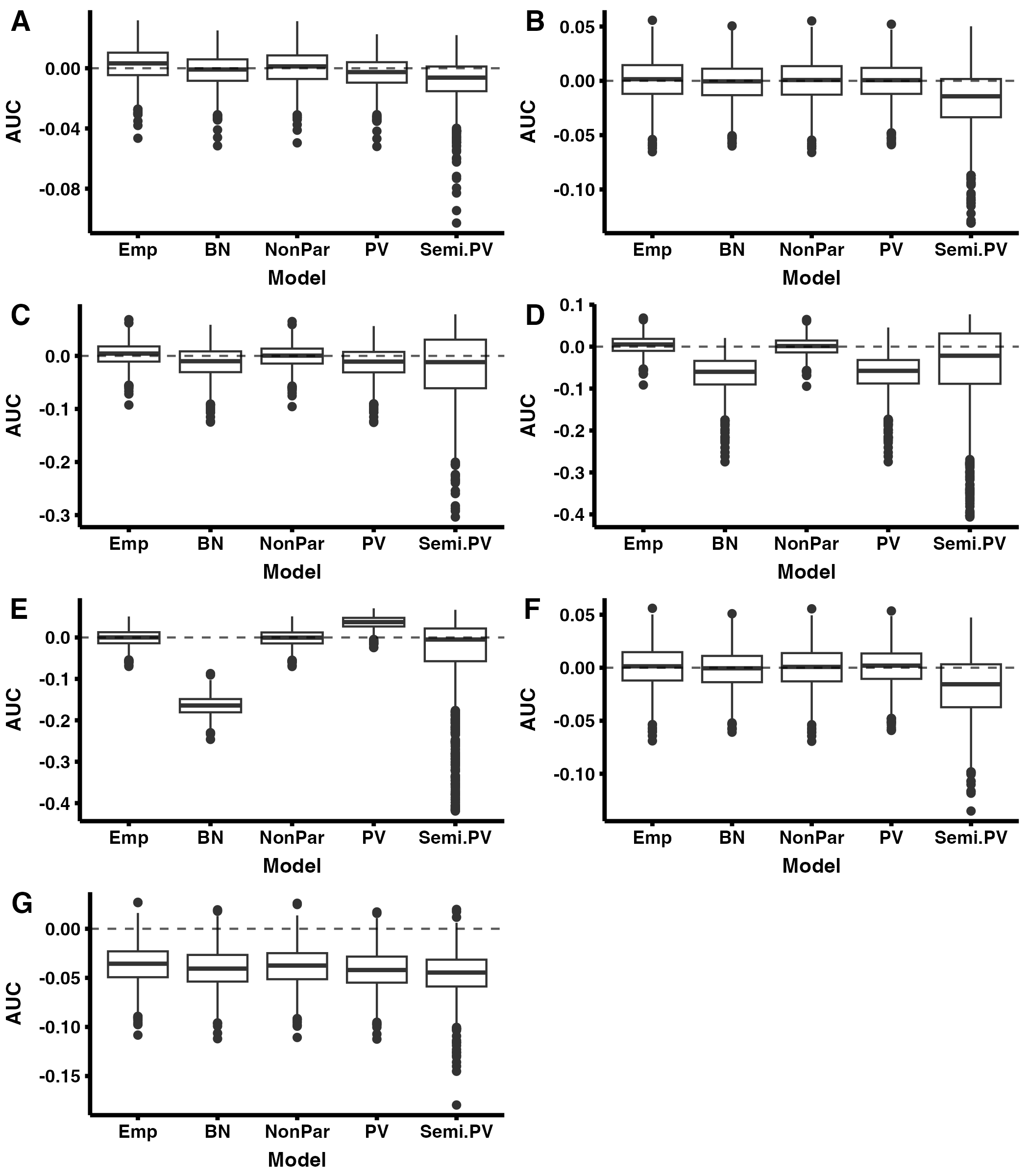}
    \caption{\label{Fig: Bias_AUC_medSamp_highAUC}Bias in estimation of AUC for medium sample size, when AUC level is high. Panels A - G respectively correspond to the simulation scenarios: BN equal, BN unequal, Skewed I, Skewed II, Skewed III, Mixed I, and Mixed II.}
\end{figure}

Figures~\ref{Fig: Bias_medSamp_lowAUC}, \ref{Fig: Bias_medSamp_medAUC}, and \ref{Fig: Bias_medSamp_highAUC} respectively shows the biases for different AUC levels, obtained from different ROC methodologies in estimating four different cutpoints under different data generating mechanism, when covariates don't impact the data generating process and the sample size is medium, i.e. $N=100$ for both healthy and diseased biomarker. Figures~\ref{Fig: Bias_AUC_medSamp_lowAUC}, \ref{Fig: Bias_AUC_medSamp_medAUC}, and \ref{Fig: Bias_AUC_medSamp_highAUC} respectively shows the corresponding biases for estimating AUCs. The bias plots for cutoffs and AUCs are separated because the bias for estimating AUCs is bounded and on a different scale compared to that for estimating the cutoffs.

Table~\ref{Tab: bias_nocov_lowSamp} presents the biases in estimating AUCs and different cutoffs from different ROC methodologies under different data generating mechanism and different AUC levels for no covariate framework with low sample size.

%\newgeometry{margin=1cm} % modify this if you need even more space
\begin{table}[hbt!]
%\begin{landscape}
\begin{center}
\caption{\label{Tab: bias_nocov_lowSamp} Biases of estimating AUC and optimal thresholds for different fitting models and different AUC levels for low sample size.}
\begin{adjustbox}{angle=90, width=0.5\textwidth}
\begin{tabular}{ccccccccccccccccc}
\hline
\textbf{Data}               & \textbf{Fitting} & \multicolumn{5}{c}{\textbf{Low AUC}}                                                                                 & \multicolumn{5}{c}{\textbf{Medium AUC}}                                                                                  & \multicolumn{5}{c}{\textbf{High AUC}}                                                                                     \\ \cline{3-17} 
\textbf{generating}         & \textbf{model}   & \multicolumn{5}{c}{\textbf{Median  $\pm$    IQR}}                                                                    & \multicolumn{5}{c}{\textbf{Median  $\pm$  IQR}}                                                                          & \multicolumn{5}{c}{\textbf{Median  $\pm$  IQR}}                                                                           \\ \cline{3-17} 
\textbf{mechanism}          &                  & AUC                    & J                    & ER                   & CZ                   & IU                     & AUC                    & J                      & ER                     & CZ                   & IU                     & AUC                   & J                      & ER                     & CZ                     & IU                     \\ \hline
\multirow{5}{*}{BN equal}   & Emp              & 0.004  $\pm$    0.08   & 0  $\pm$    0.139    & 0  $\pm$    0.138    & 0  $\pm$    0.138    & -0.001  $\pm$    0.134 & 0.007  $\pm$    0.065  & -0.001  $\pm$    0.136 & -0.001  $\pm$    0.136 & 0  $\pm$    0.137    & -0.001  $\pm$    0.133 & 0.006  $\pm$    0.021 & -0.003  $\pm$    0.136 & -0.002  $\pm$    0.135 & -0.003  $\pm$    0.136 & -0.003  $\pm$    0.136 \\
                            & BN               & 0.001  $\pm$    0.077  & 0.016  $\pm$  0.813  & 0.001  $\pm$  0.165  & 0.002  $\pm$  0.178  & 0.004  $\pm$  0.144    & -0.002  $\pm$    0.065 & 0.001  $\pm$  0.204    & -0.002  $\pm$  0.128   & 0  $\pm$  0.143      & 0.001  $\pm$  0.183    & -0.001  $\pm$  0.022  & -0.003  $\pm$  0.141   & -0.002  $\pm$  0.159   & -0.003  $\pm$  0.143   & -0.003  $\pm$  0.141   \\
                            & NonPar           & 0.003  $\pm$    0.079  & 0.001  $\pm$  0.214  & -0.001  $\pm$  0.175 & 0.001  $\pm$  0.178  & -0.003  $\pm$  0.147   & 0.003    $\pm$  0.065  & -0.001  $\pm$  0.18    & 0  $\pm$  0.161        & 0  $\pm$  0.165      & 0.001  $\pm$  0.156    & 0.002  $\pm$  0.023   & 0  $\pm$  0.153        & -0.001  $\pm$  0.144   & 0  $\pm$  0.147        & -0.001  $\pm$  0.151   \\
                            & PV               & 0.001  $\pm$    0.077  & 0.021  $\pm$  0.813  & 0.001  $\pm$  0.166  & 0.002  $\pm$  0.177  & 0.004  $\pm$  0.143    & -0.002  $\pm$    0.065 & 0.001  $\pm$  0.204    & -0.001  $\pm$  0.128   & 0  $\pm$  0.143      & 0.001  $\pm$  0.184    & -0.003  $\pm$  0.021  & -0.004  $\pm$  0.141   & -0.002  $\pm$  0.16    & -0.004  $\pm$  0.145   & -0.004  $\pm$  0.141   \\
                            & Semi.PV          & 0.002  $\pm$    0.065  & 0.017  $\pm$  0.818  & 0  $\pm$  0.167      & 0  $\pm$  0.178      & 0.001  $\pm$  0.158    & -0.009  $\pm$    0.065 & -0.001  $\pm$  0.207   & -0.006  $\pm$  0.133   & -0.002  $\pm$  0.146 & -0.001  $\pm$  0.182   & -0.011  $\pm$  0.025  & -0.008  $\pm$  0.148   & -0.008  $\pm$  0.169   & -0.01  $\pm$  0.15     & -0.008  $\pm$  0.148   \\ \hline
\multirow{5}{*}{BN unequal} & Emp              & 0.002  $\pm$    0.086  & 0.73  $\pm$  0.141   & 0.164  $\pm$  0.139  & 0.203  $\pm$  0.138  & 0.01  $\pm$  0.133     & 0.007    $\pm$  0.07   & 0.166  $\pm$  0.124    & -0.052  $\pm$  0.121   & 0.058  $\pm$  0.125  & -0.112  $\pm$  0.128   & 0.004  $\pm$  0.037   & 0.147  $\pm$  0.119    & 0.281  $\pm$  0.121    & 0.178  $\pm$  0.121    & 0.216  $\pm$  0.123    \\
                            & BN               & 0  $\pm$    0.078      & -0.002  $\pm$  0.297 & 0.001  $\pm$  0.147  & 0.003  $\pm$  0.161  & -0.003  $\pm$  0.141   & -0.001  $\pm$    0.066 & -0.004  $\pm$  0.111   & -0.006  $\pm$  0.09    & -0.004  $\pm$  0.092 & -0.002  $\pm$  0.128   & -0.002  $\pm$  0.035  & 0.003  $\pm$  0.103    & 0  $\pm$  0.106        & 0.001  $\pm$  0.103    & -0.004  $\pm$  0.142   \\
                            & NonPar           & 0.001  $\pm$    0.084  & 0.716  $\pm$  0.216  & 0.152  $\pm$  0.173  & 0.193  $\pm$  0.17   & 0.021  $\pm$  0.143    & 0.002    $\pm$  0.068  & 0.154  $\pm$  0.155    & -0.043  $\pm$  0.136   & 0.057  $\pm$  0.141  & -0.067  $\pm$  0.174   & 0.003  $\pm$  0.037   & 0.134  $\pm$  0.137    & 0.263  $\pm$  0.151    & 0.165  $\pm$  0.141    & 0.187  $\pm$  0.174    \\
                            & PV               & 0  $\pm$    0.079      & -0.001  $\pm$  0.298 & 0.001  $\pm$  0.146  & 0.004  $\pm$  0.161  & -0.004  $\pm$  0.14    & -0.001  $\pm$    0.066 & -0.005  $\pm$  0.111   & -0.006  $\pm$  0.091   & -0.004  $\pm$  0.092 & -0.003  $\pm$  0.127   & -0.001  $\pm$  0.036  & -0.002  $\pm$  0.106   & 0.002  $\pm$  0.108    & -0.003  $\pm$  0.103   & -0.001  $\pm$  0.136   \\
                            & Semi.PV          & 0.001  $\pm$    0.067  & -0.001  $\pm$  0.308 & 0.004  $\pm$  0.149  & 0.005  $\pm$  0.161  & -0.025  $\pm$  0.141   & -0.005  $\pm$    0.066 & -0.006  $\pm$  0.111   & -0.01  $\pm$  0.091    & -0.004  $\pm$  0.093 & -0.01  $\pm$  0.13     & -0.013  $\pm$  0.046  & -0.063  $\pm$  0.138   & -0.033  $\pm$  0.144   & -0.062  $\pm$  0.143   & -0.032  $\pm$  0.161   \\ \hline
\multirow{5}{*}{Skewed I}   & Emp              & -0.053  $\pm$    0.079 & 1.005  $\pm$  0.212  & 1  $\pm$  0.211      & 1  $\pm$  0.209      & 0.991  $\pm$  0.217    & -0.171  $\pm$    0.078 & 1.105  $\pm$  0.201    & 0.972  $\pm$  0.201    & 1.029  $\pm$  0.201  & 1.044  $\pm$  0.21     & 0.009  $\pm$  0.042   & -5.616  $\pm$  0.329   & -4.711  $\pm$  0.33    & -4.835  $\pm$  0.33    & -3.384  $\pm$  0.328   \\
                            & BN               & -0.046  $\pm$    0.072 & 1.545  $\pm$  2.981  & 1.017  $\pm$  0.373  & 1.028  $\pm$  0.389  & 1.016  $\pm$  0.21     & -0.199  $\pm$    0.077 & 1.51  $\pm$  1.021     & 1.025  $\pm$  0.262    & 1.12  $\pm$  0.306   & 1.078  $\pm$  0.214    & -0.01  $\pm$  0.055   & -5.663  $\pm$  0.365   & -4.767  $\pm$  0.443   & -4.876  $\pm$  0.364   & -3.423  $\pm$  0.342   \\
                            & NonPar           & -0.053  $\pm$    0.078 & 1.004  $\pm$  0.265  & 0.986  $\pm$  0.235  & 0.984  $\pm$  0.232  & 0.905  $\pm$  0.366    & -0.173  $\pm$    0.077 & 1.081  $\pm$  0.259    & 0.933  $\pm$  0.269    & 0.991  $\pm$  0.266  & 0.945  $\pm$  0.372    & 0.002  $\pm$  0.042   & -5.677  $\pm$  0.423   & -4.769  $\pm$  0.414   & -4.892  $\pm$  0.419   & -3.441  $\pm$  0.405   \\
                            & PV               & -0.046  $\pm$    0.072 & 1.534  $\pm$  2.954  & 1.014  $\pm$  0.374  & 1.023  $\pm$  0.382  & 1.015  $\pm$  0.208    & -0.199  $\pm$    0.076 & 1.51  $\pm$  1.021     & 1.024  $\pm$  0.263    & 1.12  $\pm$  0.306   & 1.076  $\pm$  0.214    & -0.011  $\pm$  0.054  & -5.663  $\pm$  0.364   & -4.768  $\pm$  0.445   & -4.875  $\pm$  0.36    & -3.422  $\pm$  0.346   \\
                            & Semi.PV          & 0.064  $\pm$    0.138  & 0.939  $\pm$  2.628  & 1.152  $\pm$  1.061  & 1.267  $\pm$  0.982  & 0.995  $\pm$  0.904    & -0.18    $\pm$  0.132  & 1.458  $\pm$  2.34     & 1.104  $\pm$  0.985    & 1.209  $\pm$  0.964  & 1.04  $\pm$  0.892     & -0.023  $\pm$  0.092  & -5.386  $\pm$  1.584   & -4.52  $\pm$  1.246    & -4.619  $\pm$  1.419   & -3.129  $\pm$  1.323   \\ \hline
\multirow{5}{*}{Skewed II}  & Emp              & 0.003  $\pm$    0.08   & 0.698  $\pm$  0.343  & 0.689  $\pm$  0.334  & 0.691  $\pm$  0.333  & 0.677  $\pm$  0.342    & 0.007    $\pm$  0.064  & 1.107  $\pm$  0.351    & 0.872  $\pm$  0.351    & 0.965  $\pm$  0.351  & 0.992  $\pm$  0.357    & 0.01  $\pm$  0.042    & 3.08  $\pm$  0.902     & 2.305  $\pm$  0.903    & 2.846  $\pm$  0.902    & 2.845  $\pm$  0.901    \\
                            & BN               & -0.008  $\pm$    0.078 & 1.599  $\pm$  3.009  & 0.807  $\pm$  0.617  & 0.874  $\pm$  0.74   & 0.691  $\pm$  0.309    & -0.081  $\pm$    0.077 & 1.517  $\pm$  0.742    & 0.873  $\pm$  0.37     & 1.088  $\pm$  0.434  & 0.989  $\pm$  0.412    & -0.056  $\pm$  0.067  & 3.137  $\pm$  1.293    & 1.886  $\pm$  1.237    & 2.713  $\pm$  1.066    & 2.724  $\pm$  1.225    \\
                            & NonPar           & 0.003  $\pm$    0.079  & 0.686  $\pm$  0.408  & 0.654  $\pm$  0.38   & 0.664  $\pm$  0.364  & 0.552  $\pm$  0.518    & 0.002    $\pm$  0.063  & 1.013  $\pm$  0.575    & 0.772  $\pm$  0.56     & 0.868  $\pm$  0.591  & 0.849  $\pm$  0.595    & 0.003  $\pm$  0.042   & 2.713  $\pm$  1.622    & 1.937  $\pm$  1.462    & 2.493  $\pm$  1.57     & 2.462  $\pm$  1.424    \\
                            & PV               & -0.01  $\pm$    0.079  & 1.56  $\pm$  2.907   & 0.781  $\pm$  0.568  & 0.845  $\pm$  0.669  & 0.689  $\pm$  0.296    & -0.076  $\pm$    0.083 & 1.468  $\pm$  0.718    & 0.859  $\pm$  0.364    & 1.057  $\pm$  0.418  & 1.002  $\pm$  0.397    & -0.051  $\pm$  0.071  & 3.073  $\pm$  1.293    & 1.874  $\pm$  1.231    & 2.686  $\pm$  1.057    & 2.773  $\pm$  1.155    \\
                            & Semi.PV          & 0.064  $\pm$    0.158  & 0.432  $\pm$  3.454  & 0.581  $\pm$  1.246  & 0.755  $\pm$  1.36   & 0.517  $\pm$  1.074    & -0.041  $\pm$    0.179 & 0.771  $\pm$  1.924    & 0.685  $\pm$  1.222    & 0.776  $\pm$  1.356  & 0.754  $\pm$  1.152    & -0.02  $\pm$  0.123   & 3.022  $\pm$  4.869    & 2.479  $\pm$  4.274    & 3.088  $\pm$  4.483    & 3.035  $\pm$  4.2      \\ \hline
\multirow{5}{*}{Skewed III} & Emp              & 0.004  $\pm$    0.076  & 0.002  $\pm$  0.012  & 0.032  $\pm$  0.013  & 0.031  $\pm$  0.012  & 0.032  $\pm$  0.013    & 0.004    $\pm$  0.064  & 0.067  $\pm$  0.042    & 0.109  $\pm$  0.043    & 0.097  $\pm$  0.043  & 0.106  $\pm$  0.044    & 0.004  $\pm$  0.041   & 1.549  $\pm$  0.491    & 1.616  $\pm$  0.491    & 1.566  $\pm$  0.491    & 1.607  $\pm$  0.491    \\
                            & BN               & 0.016  $\pm$    0.068  & 0.05  $\pm$  0.032   & 0.042  $\pm$  0.017  & 0.046  $\pm$  0.021  & 0.032  $\pm$  0.013    & -0.031  $\pm$    0.043 & 0.071  $\pm$  0.042    & 0.067  $\pm$  0.03     & 0.086  $\pm$  0.036  & 0.022  $\pm$  0.024    & -0.16  $\pm$  0.042   & 0.132  $\pm$  0.074    & 0.103  $\pm$  0.051    & 0.138  $\pm$  0.071    & -0.032  $\pm$  0.027   \\
                            & NonPar           & 0.003  $\pm$    0.075  & 0.002  $\pm$  0.015  & 0.03  $\pm$  0.015   & 0.03  $\pm$  0.015   & 0.027  $\pm$  0.021    & 0.002    $\pm$  0.062  & 0.058  $\pm$  0.046    & 0.097  $\pm$  0.054    & 0.086  $\pm$  0.05   & 0.092  $\pm$  0.064    & 0.003  $\pm$  0.041   & 1.368  $\pm$  0.693    & 1.435  $\pm$  0.691    & 1.384  $\pm$  0.693    & 1.423  $\pm$  0.697    \\
                            & PV               & 0.016  $\pm$    0.069  & 0.049  $\pm$  0.03   & 0.041  $\pm$  0.017  & 0.045  $\pm$  0.02   & 0.032  $\pm$  0.013    & 0.016    $\pm$  0.055  & 0.047  $\pm$  0.041    & 0.053  $\pm$  0.029    & 0.062  $\pm$  0.033  & 0.032  $\pm$  0.023    & 0.038  $\pm$  0.03    & 1.567  $\pm$  0.464    & 1.633  $\pm$  0.464    & 1.583  $\pm$  0.464    & 1.625  $\pm$  0.464    \\
                            & Semi.PV          & 0.064  $\pm$    0.149  & 0.014  $\pm$  0.106  & 0.042  $\pm$  0.052  & 0.044  $\pm$  0.055  & 0.035  $\pm$  0.043    & 0.085    $\pm$  0.134  & 0.007  $\pm$  0.074    & 0.035  $\pm$  0.054    & 0.029  $\pm$  0.063  & 0.034  $\pm$  0.05     & -0.019  $\pm$  0.054  & 0.199  $\pm$  0.362    & 0.215  $\pm$  0.242    & 0.197  $\pm$  0.317    & 0.231  $\pm$  0.196    \\ \hline
\multirow{5}{*}{Mixed I}    & Emp              & 0.002  $\pm$    0.082  & -1.143  $\pm$  0.273 & -0.388  $\pm$  0.267 & -0.576  $\pm$  0.27  & 0.067  $\pm$  0.257    & 0.006    $\pm$  0.065  & -0.379  $\pm$  0.258   & 0.092  $\pm$  0.254    & -0.145  $\pm$  0.259 & 0.259  $\pm$  0.256    & 0.004  $\pm$  0.038   & 0.352  $\pm$  0.257    & 0.64  $\pm$  0.257     & 0.418  $\pm$  0.256    & 0.538  $\pm$  0.262    \\
                            & BN               & 0  $\pm$    0.075      & 0.013  $\pm$  0.267  & -0.004  $\pm$  0.182 & -0.005  $\pm$  0.206 & -0.004  $\pm$  0.206   & -0.002  $\pm$    0.063 & 0.008  $\pm$  0.226    & -0.003  $\pm$  0.18    & -0.005  $\pm$  0.189 & -0.003  $\pm$  0.223   & -0.002  $\pm$  0.035  & 0.007  $\pm$  0.212    & 0.002  $\pm$  0.214    & 0.001  $\pm$  0.213    & -0.005  $\pm$  0.301   \\
                            & NonPar           & 0.002  $\pm$    0.081  & -1.068  $\pm$  0.616 & -0.35  $\pm$  0.349  & -0.534  $\pm$  0.364 & 0.023  $\pm$  0.256    & 0.004    $\pm$  0.064  & -0.366  $\pm$  0.308   & 0.076  $\pm$  0.292    & -0.137  $\pm$  0.297 & 0.157  $\pm$  0.358    & 0.003  $\pm$  0.038   & 0.313  $\pm$  0.282    & 0.589  $\pm$  0.31     & 0.376  $\pm$  0.287    & 0.47  $\pm$  0.347     \\
                            & PV               & 0  $\pm$    0.075      & 0.013  $\pm$  0.271  & -0.005  $\pm$  0.182 & -0.007  $\pm$  0.205 & -0.004  $\pm$  0.204   & -0.002  $\pm$    0.064 & 0.002  $\pm$  0.229    & -0.006  $\pm$  0.177   & -0.01  $\pm$  0.191  & -0.007  $\pm$  0.224   & 0.001  $\pm$  0.036   & -0.008  $\pm$  0.221   & 0.006  $\pm$  0.219    & -0.008  $\pm$  0.215   & 0.01  $\pm$  0.286     \\
                            & Semi.PV          & -0.003  $\pm$    0.07  & -0.016  $\pm$  0.279 & -0.032  $\pm$  0.187 & -0.056  $\pm$  0.199 & -0.001  $\pm$  0.202   & -0.011  $\pm$    0.07  & -0.077  $\pm$  0.243   & -0.062  $\pm$  0.197   & -0.102  $\pm$  0.204 & -0.019  $\pm$  0.251   & -0.014  $\pm$  0.048  & -0.144  $\pm$  0.309   & -0.081  $\pm$  0.297   & -0.145  $\pm$  0.305   & -0.069  $\pm$  0.341   \\ \hline
\multirow{5}{*}{Mixed II}   & Emp              & -0.013  $\pm$    0.079 & -0.801  $\pm$  0.186 & -0.159  $\pm$  0.181 & -0.21  $\pm$  0.181  & -0.084  $\pm$  0.176   & -0.033  $\pm$    0.071 & -0.405  $\pm$  0.179   & -0.064  $\pm$  0.175   & -0.137  $\pm$  0.177 & -0.208  $\pm$  0.176   & -0.033  $\pm$  0.041  & -0.015  $\pm$  0.177   & 0.116  $\pm$  0.178    & 0.036  $\pm$  0.178    & -0.018  $\pm$  0.178   \\
                            & BN               & -0.018  $\pm$    0.075 & 0.002  $\pm$  0.397  & -0.002  $\pm$  0.195 & 0.002  $\pm$  0.211  & -0.051  $\pm$  0.179   & -0.04    $\pm$  0.071  & 0.003  $\pm$  0.261    & -0.006  $\pm$  0.174   & -0.002  $\pm$  0.186 & -0.137  $\pm$  0.185   & -0.041  $\pm$  0.04   & -0.004  $\pm$  0.175   & -0.005  $\pm$  0.184   & -0.003  $\pm$  0.17    & -0.011  $\pm$  0.173   \\
                            & NonPar           & -0.014  $\pm$    0.078 & -0.781  $\pm$  0.287 & -0.151  $\pm$  0.225 & -0.2  $\pm$  0.234   & -0.093  $\pm$  0.192   & -0.035  $\pm$    0.071 & -0.39  $\pm$  0.263    & -0.056  $\pm$  0.218   & -0.131  $\pm$  0.225 & -0.225  $\pm$  0.195   & -0.037  $\pm$  0.041  & -0.022  $\pm$  0.217   & 0.105  $\pm$  0.208    & 0.028  $\pm$  0.217    & -0.03  $\pm$  0.214    \\
                            & PV               & -0.018  $\pm$    0.074 & 0.002  $\pm$  0.398  & -0.003  $\pm$  0.194 & 0  $\pm$  0.212      & -0.053  $\pm$  0.18    & -0.04    $\pm$  0.071  & 0.004  $\pm$  0.26     & -0.006  $\pm$  0.174   & -0.002  $\pm$  0.186 & -0.137  $\pm$  0.186   & -0.043  $\pm$  0.039  & -0.004  $\pm$  0.176   & -0.006  $\pm$  0.183   & -0.005  $\pm$  0.171   & -0.012  $\pm$  0.172   \\
                            & Semi.PV          & -0.018  $\pm$    0.071 & -0.012  $\pm$  0.405 & -0.008  $\pm$  0.2   & -0.01  $\pm$  0.212  & -0.051  $\pm$  0.178   & -0.045  $\pm$    0.068 & -0.006  $\pm$  0.26    & -0.014  $\pm$  0.172   & -0.012  $\pm$  0.191 & -0.147  $\pm$  0.195   & -0.05  $\pm$  0.043   & -0.024  $\pm$  0.184   & -0.018  $\pm$  0.196   & -0.026  $\pm$  0.186   & -0.03  $\pm$  0.189    \\ \hline
\end{tabular}
\end{adjustbox}
\end{center}
%\end{landscape}
\end{table}
%\restoregeometry

\begin{figure}[hbt!]%
    \includegraphics[width=0.95\textwidth]{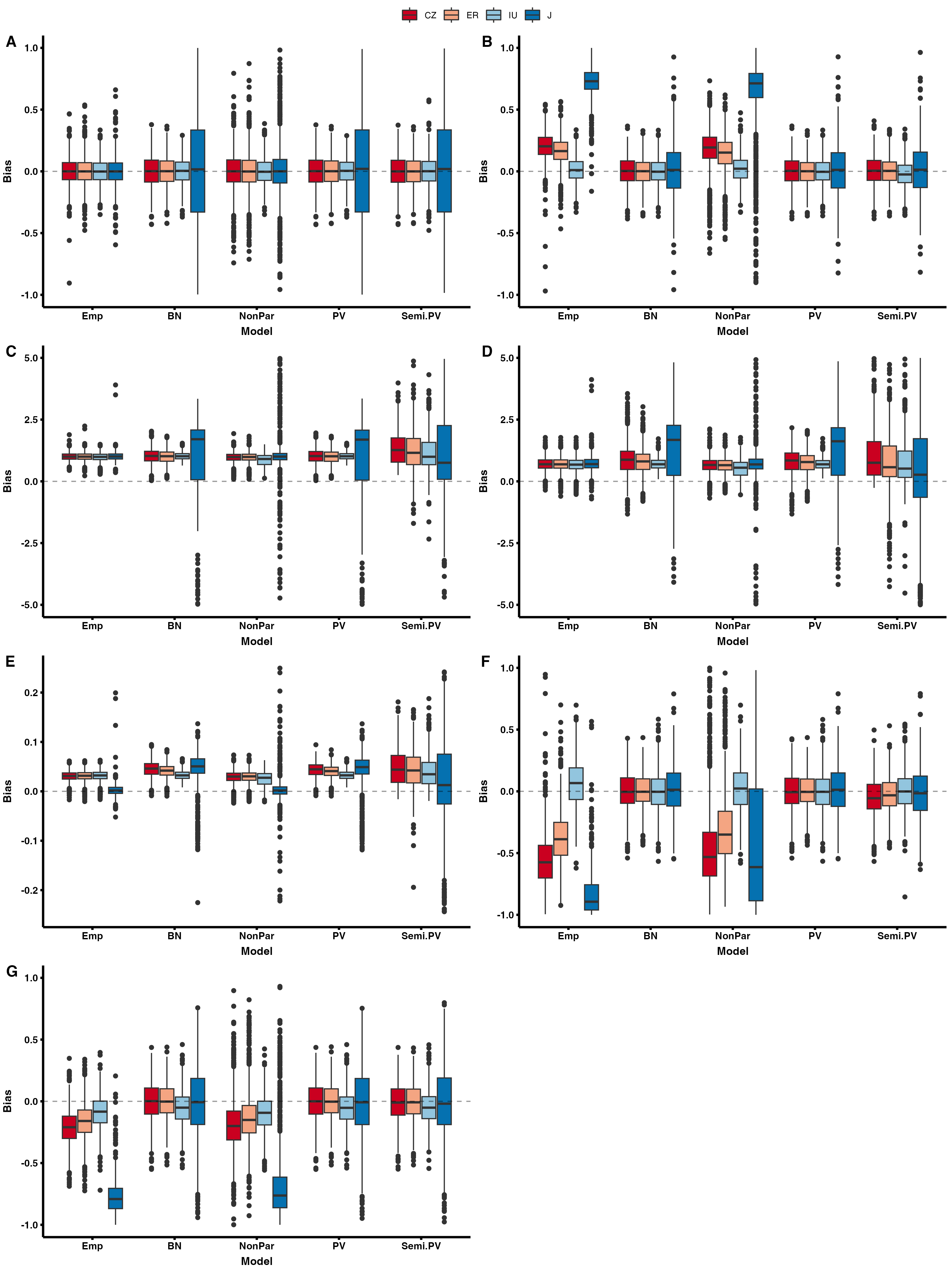}
    \caption{\label{Fig: Bias_lowSamp_lowAUC}Bias in estimation optimal cut-points for low sample size, when AUC level is low. Panels A - G respectively correspond to the simulation scenarios: BN equal, BN unequal, Skewed I, Skewed II, Skewed III, Mixed I, and Mixed II.}
\end{figure}

\begin{figure}[hbt!]%
    \includegraphics[width=\textwidth]{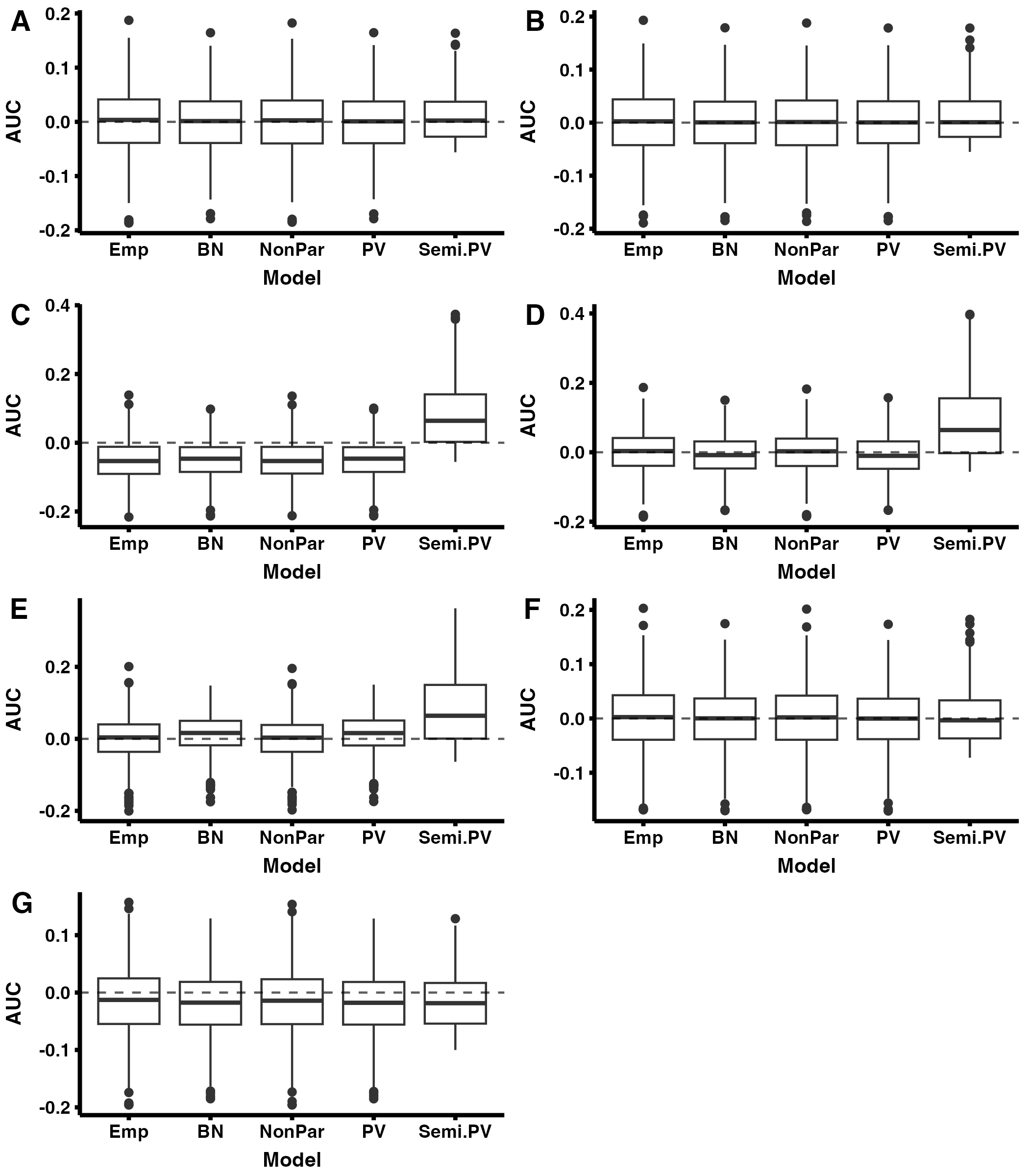}
    \caption{\label{Fig: Bias_AUC_lowSamp_lowAUC}Bias in estimation of AUC for low sample size, when AUC level is low. Panels A - G respectively correspond to the simulation scenarios: BN equal, BN unequal, Skewed I, Skewed II, Skewed III, Mixed I, and Mixed II.}
\end{figure}

\begin{figure}[hbt!]%
    \includegraphics[width=0.95\textwidth]{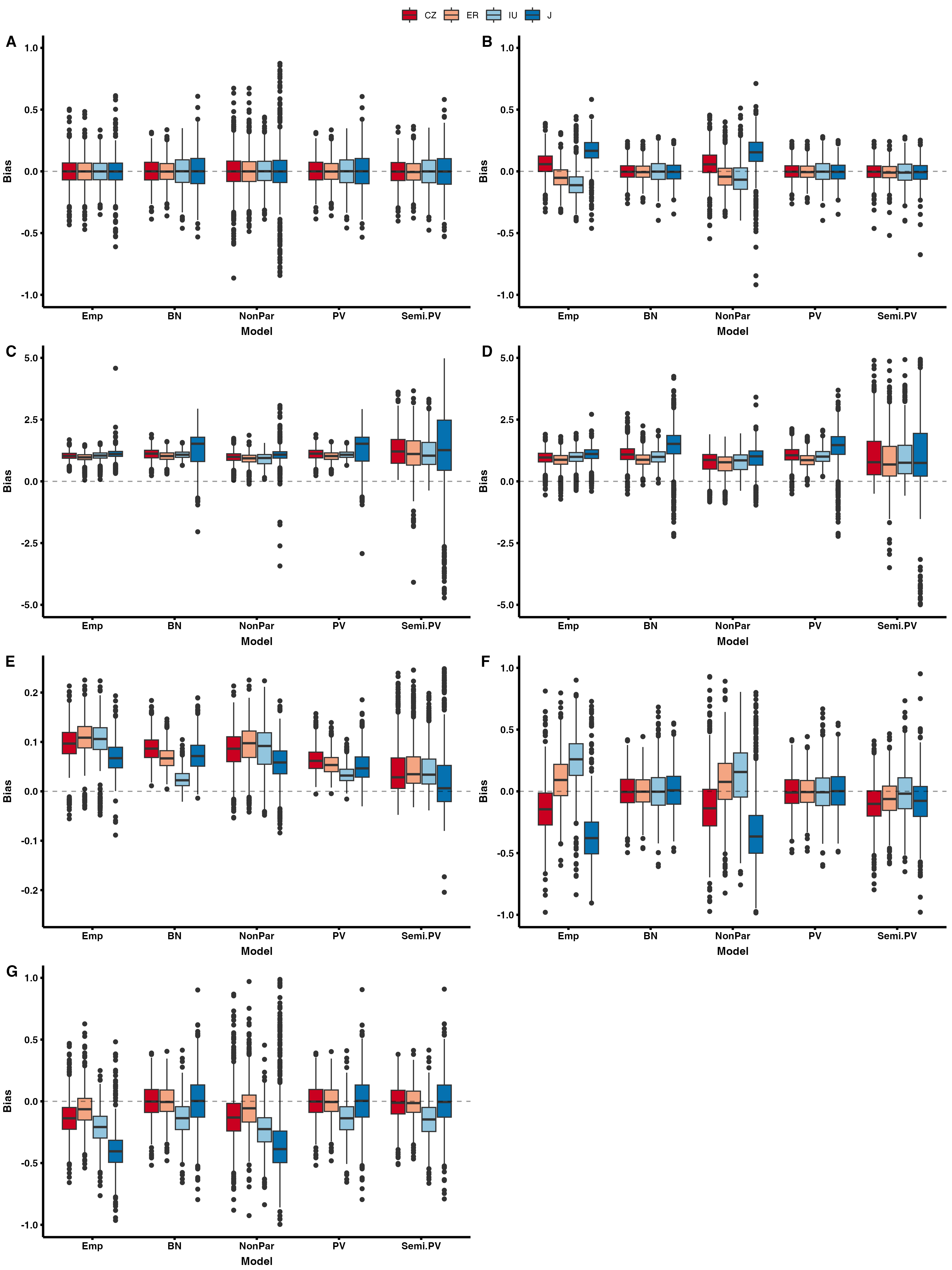}
    \caption{\label{Fig: Bias_lowSamp_medAUC}Bias in estimation optimal cut-points for low sample size, when AUC level is medium. Panels A - G respectively correspond to the simulation scenarios: BN equal, BN unequal, Skewed I, Skewed II, Skewed III, Mixed I, and Mixed II.}
\end{figure}

\begin{figure}[hbt!]%
    \includegraphics[width=\textwidth]{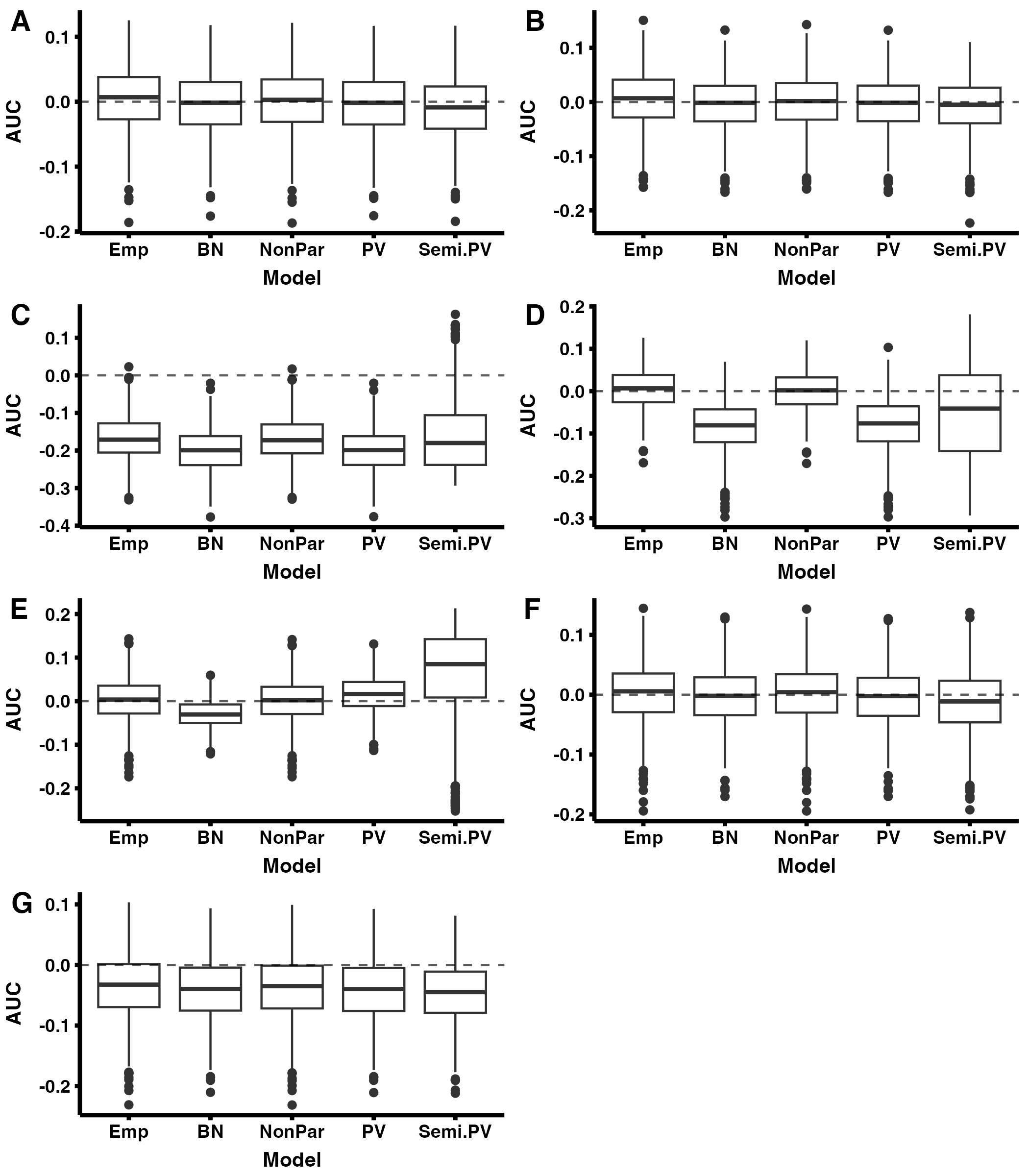}
    \caption{\label{Fig: Bias_AUC_lowSamp_medAUC}Bias in estimation of AUC for low sample size, when AUC level is medium. Panels A - G respectively correspond to the simulation scenarios: BN equal, BN unequal, Skewed I, Skewed II, Skewed III, Mixed I, and Mixed II.}
\end{figure}

\begin{figure}[hbt!]%
    \includegraphics[width=0.95\textwidth]{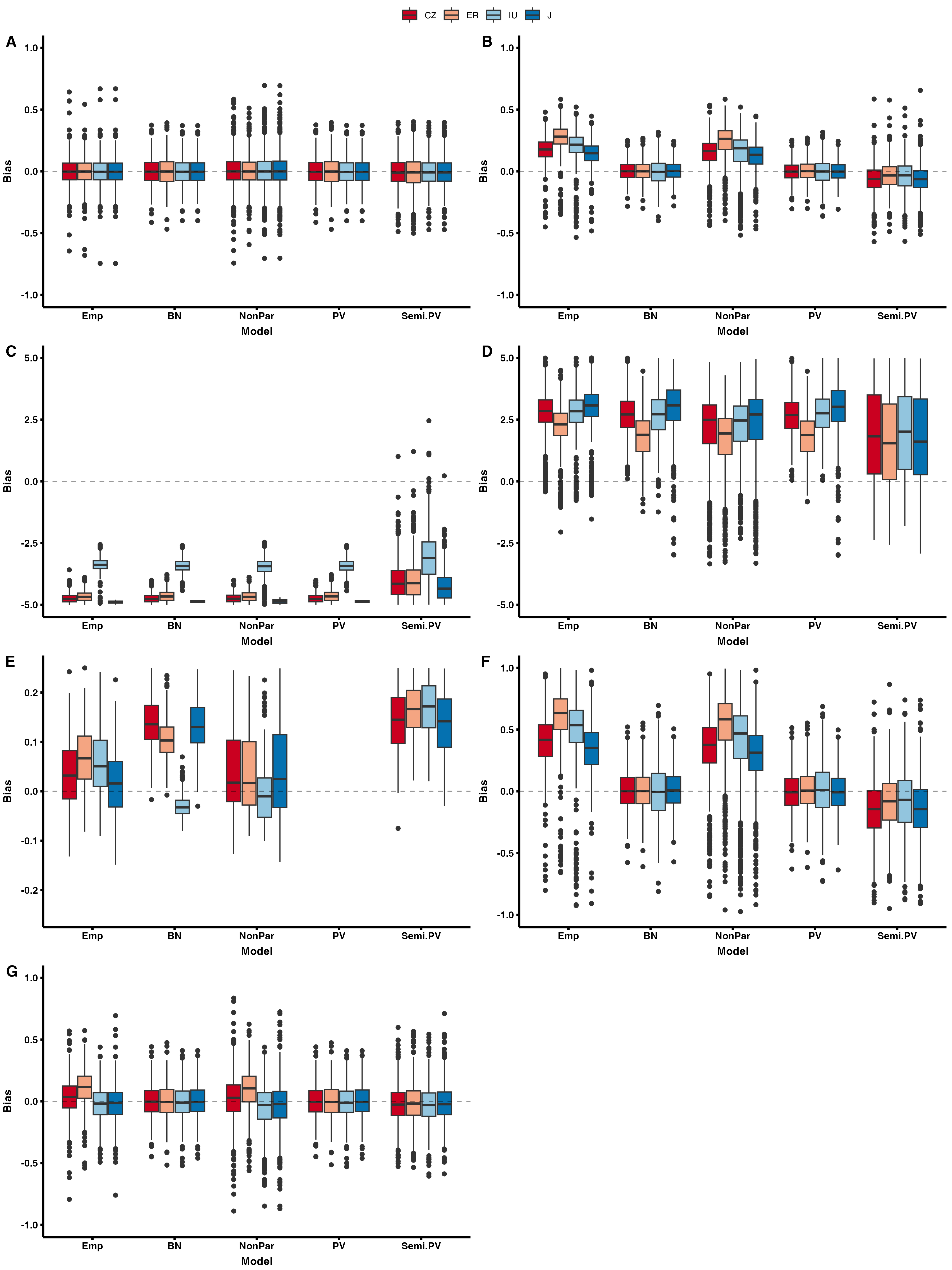}
    \caption{\label{Fig: Bias_lowSamp_highAUC}Bias in estimation optimal cut-points for low sample size, when AUC level is high. Panels A - G respectively correspond to the simulation scenarios: BN equal, BN unequal, Skewed I, Skewed II, Skewed III, Mixed I, and Mixed II.}
\end{figure}

\begin{figure}[hbt!]%
    \includegraphics[width=\textwidth]{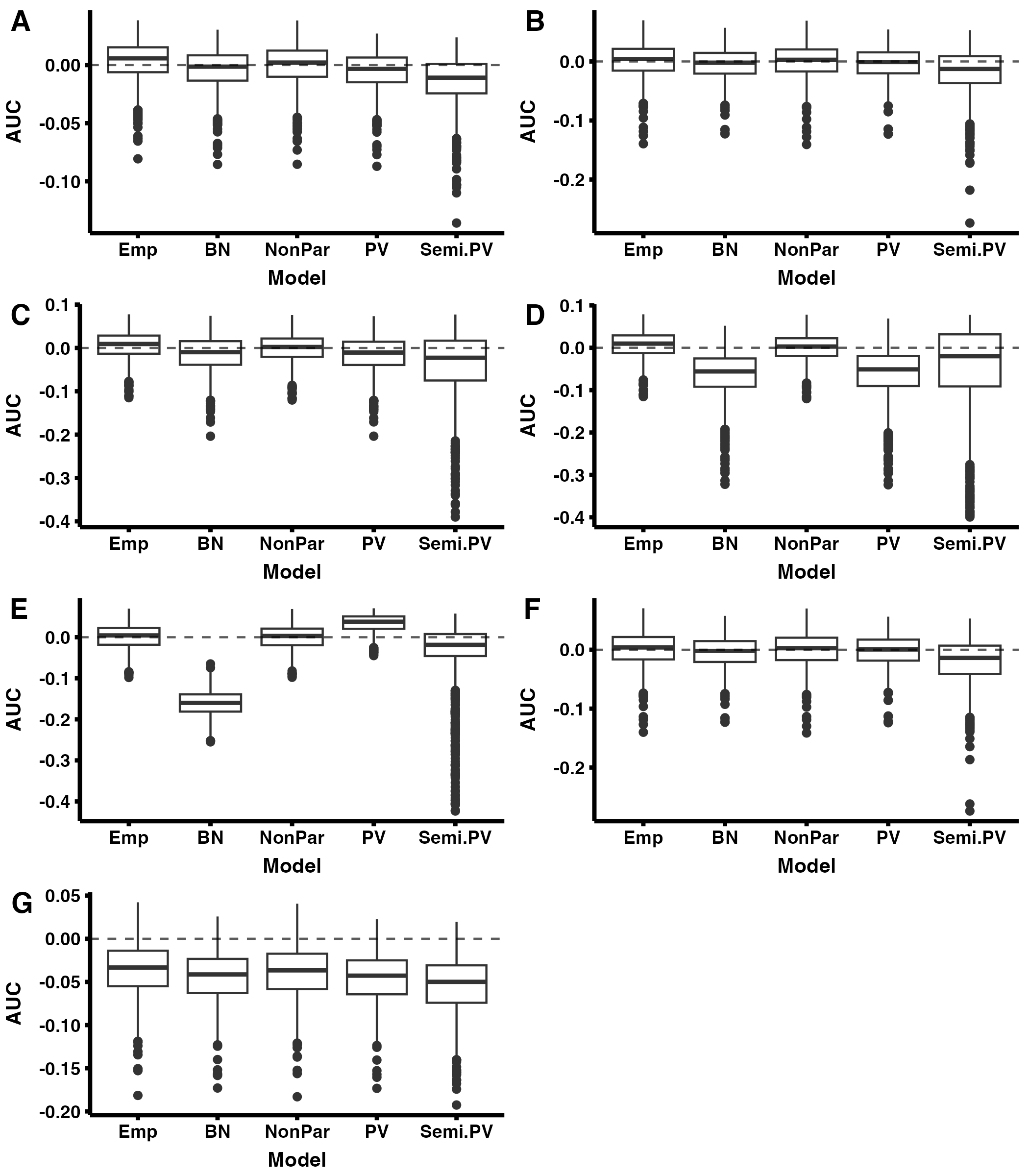}
    \caption{\label{Fig: Bias_AUC_lowSamp_highAUC}Bias in estimation of AUC for low sample size, when AUC level is high. Panels A - G respectively correspond to the simulation scenarios: BN equal, BN unequal, Skewed I, Skewed II, Skewed III, Mixed I, and Mixed II.}
\end{figure}

Figures~\ref{Fig: Bias_lowSamp_lowAUC}, \ref{Fig: Bias_lowSamp_medAUC}, and \ref{Fig: Bias_lowSamp_highAUC} respectively shows the biases for different AUC levels, obtained from different ROC methodologies in estimating four different cutpoints under different data generating mechanism, when covariates don't impact the data generating process and the sample size is low, i.e. $N=50$ for both healthy and diseased biomarker. Figures~\ref{Fig: Bias_AUC_lowSamp_lowAUC}, \ref{Fig: Bias_AUC_lowSamp_medAUC}, and \ref{Fig: Bias_AUC_lowSamp_highAUC} respectively shows the corresponding biases for estimating AUCs.

Table~\ref{Tab: bias_nocov_highSamp} presents the biases in estimating AUCs and different cutoffs from different ROC methodologies under different data generating mechanism and different AUC levels for no covariate framework with high sample size.

\begin{table}[hbt!]
%\begin{landscape}
\begin{center}
\caption{\label{Tab: bias_nocov_highSamp} Biases of estimating AUC and optimal thresholds for different fitting models and different AUC levels for high sample size.}
\begin{adjustbox}{angle=90, width=0.5\textwidth}
\begin{tabular}{ccccccccccccccccc}
\hline
\textbf{Data}               & \textbf{Fitting} & \multicolumn{5}{c}{\textbf{Low AUC}}                                                                   & \multicolumn{5}{c}{\textbf{Medium AUC}}                                                                & \multicolumn{5}{c}{\textbf{High AUC}}                                                                  \\ \cline{3-17} 
\textbf{generating}         & \textbf{model}   & \multicolumn{5}{c}{\textbf{Median  $\pm$    IQR}}                                                      & \multicolumn{5}{c}{\textbf{Median  $\pm$  IQR}}                                                        & \multicolumn{5}{c}{\textbf{Median  $\pm$  IQR}}                                                        \\ \cline{3-17} 
\textbf{mechanism}          &                  & AUC                & J                  & ER                 & CZ                 & IU                 & AUC                & J                  & ER                 & CZ                 & IU                 & AUC                & J                  & ER                 & CZ                 & IU                 \\ \hline
\multirow{5}{*}{BN equal}   & Emp              & 0 $\pm$   0.025    & 0.002 $\pm$ 0.07   & 0.001 $\pm$ 0.06   & 0.002 $\pm$ 0.063  & 0.001 $\pm$ 0.047  & 0.001 $\pm$   0.02 & 0.004 $\pm$ 0.057  & 0 $\pm$ 0.055      & 0.002 $\pm$ 0.057  & 0.004 $\pm$ 0.053  & 0 $\pm$ 0.007      & 0.002 $\pm$ 0.054  & 0.001 $\pm$ 0.053  & 0.002 $\pm$ 0.054  & 0.002 $\pm$ 0.054  \\
                            & BN               & 0 $\pm$ 0.024      & 0.004 $\pm$ 0.298  & -0.001 $\pm$ 0.052 & -0.001 $\pm$ 0.055 & 0.001 $\pm$ 0.073  & 0 $\pm$ 0.02       & 0.001 $\pm$ 0.063  & 0 $\pm$ 0.044      & 0 $\pm$ 0.047      & 0.001 $\pm$ 0.063  & 0 $\pm$ 0.007      & 0.003 $\pm$ 0.046  & 0.002 $\pm$ 0.052  & 0.002 $\pm$ 0.047  & 0.003 $\pm$ 0.046  \\
                            & NonPar           & 0 $\pm$ 0.025      & 0 $\pm$ 0.081      & 0.003 $\pm$ 0.063  & 0.002 $\pm$ 0.066  & 0.001 $\pm$ 0.049  & 0 $\pm$ 0.02       & 0.003 $\pm$ 0.062  & 0.003 $\pm$ 0.057  & 0.003 $\pm$ 0.06   & 0.003 $\pm$ 0.056  & 0 $\pm$ 0.007      & 0.002 $\pm$ 0.055  & 0 $\pm$ 0.053      & 0.001 $\pm$ 0.056  & 0.002 $\pm$ 0.055  \\
                            & PV               & 0 $\pm$ 0.024      & 0.006 $\pm$ 0.299  & -0.001 $\pm$ 0.053 & -0.001 $\pm$ 0.055 & 0.001 $\pm$ 0.073  & 0 $\pm$ 0.02       & 0.001 $\pm$ 0.063  & 0.001 $\pm$ 0.043  & 0 $\pm$ 0.047      & 0.001 $\pm$ 0.063  & -0.002 $\pm$ 0.007 & 0.003 $\pm$ 0.046  & 0.002 $\pm$ 0.052  & 0.002 $\pm$ 0.048  & 0.003 $\pm$ 0.046  \\
                            & Semi.PV          & -0.001 $\pm$ 0.025 & 0.005 $\pm$ 0.297  & -0.002 $\pm$ 0.055 & -0.003 $\pm$ 0.058 & 0 $\pm$ 0.075      & -0.002 $\pm$ 0.021 & 0 $\pm$ 0.065      & -0.002 $\pm$ 0.045 & -0.001 $\pm$ 0.048 & 0 $\pm$ 0.065      & -0.004 $\pm$ 0.008 & -0.001 $\pm$ 0.05  & -0.003 $\pm$ 0.058 & -0.001 $\pm$ 0.052 & -0.001 $\pm$ 0.05  \\ \hline
\multirow{5}{*}{BN unequal} & Emp              & 0 $\pm$ 0.024      & 0.716 $\pm$ 0.879  & 0.154 $\pm$ 0.113  & 0.193 $\pm$ 0.108  & 0.015 $\pm$ 0.048  & 0 $\pm$ 0.021      & 0.16 $\pm$ 0.136   & -0.042 $\pm$ 0.051 & 0.057 $\pm$ 0.061  & -0.029 $\pm$ 0.17  & 0 $\pm$ 0.012      & 0.136 $\pm$ 0.058  & 0.267 $\pm$ 0.225  & 0.166 $\pm$ 0.079  & 0.191 $\pm$ 0.263  \\
                            & BN               & 0 $\pm$ 0.023      & -0.002 $\pm$ 0.092 & -0.002 $\pm$ 0.049 & -0.001 $\pm$ 0.053 & 0.002 $\pm$ 0.045  & 0 $\pm$ 0.02       & 0 $\pm$ 0.036      & 0 $\pm$ 0.03       & 0 $\pm$ 0.031      & 0.002 $\pm$ 0.04   & 0 $\pm$ 0.011      & 0.001 $\pm$ 0.034  & 0.001 $\pm$ 0.036  & 0.001 $\pm$ 0.034  & 0.002 $\pm$ 0.05   \\
                            & NonPar           & 0 $\pm$ 0.024      & 0.704 $\pm$ 0.801  & 0.15 $\pm$ 0.126   & 0.187 $\pm$ 0.129  & 0.016 $\pm$ 0.049  & -0.001 $\pm$ 0.021 & 0.158 $\pm$ 0.154  & -0.042 $\pm$ 0.052 & 0.057 $\pm$ 0.061  & -0.015 $\pm$ 0.163 & 0 $\pm$ 0.012      & 0.134 $\pm$ 0.065  & 0.263 $\pm$ 0.237  & 0.162 $\pm$ 0.099  & 0.183 $\pm$ 0.266  \\
                            & PV               & 0 $\pm$ 0.023      & -0.001 $\pm$ 0.092 & -0.002 $\pm$ 0.049 & -0.001 $\pm$ 0.052 & 0.002 $\pm$ 0.045  & 0 $\pm$ 0.02       & 0 $\pm$ 0.036      & 0 $\pm$ 0.03       & 0 $\pm$ 0.031      & 0.002 $\pm$ 0.04   & 0.001 $\pm$ 0.011  & -0.004 $\pm$ 0.036 & 0.004 $\pm$ 0.036  & -0.002 $\pm$ 0.035 & 0.005 $\pm$ 0.046  \\
                            & Semi.PV          & -0.001 $\pm$ 0.024 & -0.001 $\pm$ 0.096 & -0.001 $\pm$ 0.05  & 0 $\pm$ 0.054      & 0.002 $\pm$ 0.047  & -0.002 $\pm$ 0.022 & -0.001 $\pm$ 0.037 & -0.002 $\pm$ 0.032 & -0.001 $\pm$ 0.033 & -0.002 $\pm$ 0.042 & -0.01 $\pm$ 0.023  & -0.102 $\pm$ 0.221 & -0.071 $\pm$ 0.135 & -0.103 $\pm$ 0.207 & -0.089 $\pm$ 0.125 \\ \hline
\multirow{5}{*}{Skewed I}   & Emp              & -0.05 $\pm$ 0.024  & 1.007 $\pm$ 0.107  & 0.975 $\pm$ 0.283  & 0.975 $\pm$ 0.227  & 0.764 $\pm$ 0.532  & -0.17 $\pm$ 0.025  & 1.103 $\pm$ 0.126  & 0.957 $\pm$ 0.297  & 1.012 $\pm$ 0.245  & 0.954 $\pm$ 0.534  & 0 $\pm$ 0.014      & -5.652 $\pm$ 0.638 & -4.757 $\pm$ 0.508 & -4.878 $\pm$ 0.538 & -3.42 $\pm$ 0.527  \\
                            & BN               & -0.048 $\pm$ 0.023 & 1.819 $\pm$ 2.505  & 1.038 $\pm$ 0.146  & 1.04 $\pm$ 0.151   & 1.019 $\pm$ 0.065  & -0.201 $\pm$ 0.026 & 1.521 $\pm$ 0.374  & 1.041 $\pm$ 0.082  & 1.114 $\pm$ 0.103  & 1.125 $\pm$ 0.085  & -0.011 $\pm$ 0.018 & -5.604 $\pm$ 0.107 & -4.714 $\pm$ 0.149 & -4.823 $\pm$ 0.117 & -3.372 $\pm$ 0.107 \\
                            & NonPar           & -0.05 $\pm$ 0.024  & 1.007 $\pm$ 0.124  & 0.967 $\pm$ 0.303  & 0.969 $\pm$ 0.258  & 0.737 $\pm$ 0.526  & -0.17 $\pm$ 0.025  & 1.096 $\pm$ 0.25   & 0.947 $\pm$ 0.314  & 1.004 $\pm$ 0.27   & 0.8 $\pm$ 0.511    & -0.001 $\pm$ 0.013 & -5.658 $\pm$ 0.647 & -4.763 $\pm$ 0.513 & -4.885 $\pm$ 0.554 & -3.426 $\pm$ 0.552 \\
                            & PV               & -0.048 $\pm$ 0.023 & 1.785 $\pm$ 2.502  & 1.034 $\pm$ 0.141  & 1.036 $\pm$ 0.145  & 1.018 $\pm$ 0.066  & -0.201 $\pm$ 0.026 & 1.519 $\pm$ 0.372  & 1.041 $\pm$ 0.083  & 1.114 $\pm$ 0.103  & 1.125 $\pm$ 0.085  & -0.012 $\pm$ 0.018 & -5.603 $\pm$ 0.107 & -4.713 $\pm$ 0.149 & -4.822 $\pm$ 0.117 & -3.371 $\pm$ 0.107 \\
                            & Semi.PV          & 0.058 $\pm$ 0.183  & 1.308 $\pm$ 4.037  & 1.046 $\pm$ 1.083  & 1.002 $\pm$ 0.817  & 1.069 $\pm$ 1.197  & -0.182 $\pm$ 0.122 & 0.99 $\pm$ 1.944   & 0.896 $\pm$ 0.82   & 0.894 $\pm$ 0.763  & 1.063 $\pm$ 0.724  & -0.012 $\pm$ 0.049 & -5.423 $\pm$ 1.108 & -4.145 $\pm$ 1.032 & -4.527 $\pm$ 1.058 & -2.873 $\pm$ 1.102 \\ \hline
\multirow{5}{*}{Skewed II}  & Emp              & 0 $\pm$ 0.025      & 0.715 $\pm$ 0.165  & 0.644 $\pm$ 0.488  & 0.651 $\pm$ 0.531  & 0.227 $\pm$ 0.689  & 0.001 $\pm$ 0.019  & 1.059 $\pm$ 0.585  & 0.776 $\pm$ 0.728  & 0.885 $\pm$ 0.74   & 0.55 $\pm$ 0.739   & 0.001 $\pm$ 0.013  & 2.919 $\pm$ 1.806  & 2.071 $\pm$ 1.857  & 2.622 $\pm$ 1.743  & 2.684 $\pm$ 1.76   \\
                            & BN               & -0.012 $\pm$ 0.024 & 2.117 $\pm$ 0.886  & 0.924 $\pm$ 0.301  & 0.961 $\pm$ 0.357  & 0.748 $\pm$ 0.113  & -0.092 $\pm$ 0.029 & 1.72 $\pm$ 0.339   & 0.971 $\pm$ 0.136  & 1.178 $\pm$ 0.183  & 1.164 $\pm$ 0.178  & -0.067 $\pm$ 0.03  & 3.54 $\pm$ 0.435   & 2.16 $\pm$ 0.465   & 2.995 $\pm$ 0.343  & 3.272 $\pm$ 0.436  \\
                            & NonPar           & 0 $\pm$ 0.025      & 0.71 $\pm$ 0.196   & 0.621 $\pm$ 0.514  & 0.634 $\pm$ 0.568  & 0.154 $\pm$ 0.675  & 0 $\pm$ 0.019      & 1.041 $\pm$ 0.702  & 0.617 $\pm$ 0.795  & 0.806 $\pm$ 0.799  & 0.529 $\pm$ 0.78   & 0 $\pm$ 0.014      & 2.809 $\pm$ 2.322  & 1.323 $\pm$ 2.144  & 1.877 $\pm$ 2.319  & 2.574 $\pm$ 1.865  \\
                            & PV               & -0.013 $\pm$ 0.024 & 1.864 $\pm$ 1.109  & 0.838 $\pm$ 0.22   & 0.857 $\pm$ 0.252  & 0.742 $\pm$ 0.111  & -0.088 $\pm$ 0.03  & 1.637 $\pm$ 0.28   & 0.951 $\pm$ 0.123  & 1.136 $\pm$ 0.153  & 1.179 $\pm$ 0.173  & -0.066 $\pm$ 0.031 & 3.486 $\pm$ 0.397  & 2.16 $\pm$ 0.465   & 2.968 $\pm$ 0.335  & 3.244 $\pm$ 0.394  \\
                            & Semi.PV          & 0.056 $\pm$ 0.135  & 1.028 $\pm$ 5.926  & 0.529 $\pm$ 1.599  & 0.526 $\pm$ 1.371  & 0.356 $\pm$ 1.228  & -0.08 $\pm$ 0.149  & 0.454 $\pm$ 1.367  & 0.626 $\pm$ 1.038  & 0.569 $\pm$ 1.102  & 0.891 $\pm$ 0.885  & -0.044 $\pm$ 0.075 & 2.227 $\pm$ 2.735  & 2.263 $\pm$ 2.546  & 2.255 $\pm$ 2.558  & 3.05 $\pm$ 2.626   \\ \hline
\multirow{5}{*}{Skewed III} & Emp              & 0 $\pm$ 0.024      & 0.002 $\pm$ 0.006  & 0.03 $\pm$ 0.017   & 0.029 $\pm$ 0.015  & 0.018 $\pm$ 0.033  & 0 $\pm$ 0.02       & 0.063 $\pm$ 0.039  & 0.101 $\pm$ 0.073  & 0.089 $\pm$ 0.066  & 0.064 $\pm$ 0.114  & 0.001 $\pm$ 0.012  & 1.461 $\pm$ 0.9    & 1.528 $\pm$ 0.9    & 1.478 $\pm$ 0.9    & 1.519 $\pm$ 0.905  \\
                            & BN               & 0.013 $\pm$ 0.022  & 0.059 $\pm$ 0.01   & 0.045 $\pm$ 0.006  & 0.048 $\pm$ 0.008  & 0.034 $\pm$ 0.004  & -0.033 $\pm$ 0.015 & 0.078 $\pm$ 0.015  & 0.07 $\pm$ 0.01    & 0.092 $\pm$ 0.013  & 0.024 $\pm$ 0.008  & -0.166 $\pm$ 0.016 & 0.057 $\pm$ 0.024  & 0.051 $\pm$ 0.017  & 0.066 $\pm$ 0.023  & -0.055 $\pm$ 0.01  \\
                            & NonPar           & 0 $\pm$ 0.024      & 0.002 $\pm$ 0.006  & 0.03 $\pm$ 0.018   & 0.029 $\pm$ 0.015  & 0.017 $\pm$ 0.033  & 0 $\pm$ 0.02       & 0.062 $\pm$ 0.051  & 0.099 $\pm$ 0.081  & 0.088 $\pm$ 0.069  & 0.047 $\pm$ 0.113  & 0.001 $\pm$ 0.012  & 1.424 $\pm$ 1.087  & 1.491 $\pm$ 1.102  & 1.441 $\pm$ 1.102  & 1.483 $\pm$ 1.109  \\
                            & PV               & 0.013 $\pm$ 0.022  & 0.056 $\pm$ 0.009  & 0.043 $\pm$ 0.005  & 0.046 $\pm$ 0.006  & 0.034 $\pm$ 0.004  & 0.018 $\pm$ 0.017  & 0.051 $\pm$ 0.014  & 0.056 $\pm$ 0.009  & 0.066 $\pm$ 0.011  & 0.035 $\pm$ 0.008  & 0.04 $\pm$ 0.01    & 1.559 $\pm$ 0.154  & 1.626 $\pm$ 0.154  & 1.576 $\pm$ 0.154  & 1.618 $\pm$ 0.154  \\
                            & Semi.PV          & 0.039 $\pm$ 0.112  & 0.028 $\pm$ 0.14   & 0.035 $\pm$ 0.048  & 0.032 $\pm$ 0.044  & 0.031 $\pm$ 0.042  & 0.019 $\pm$ 0.192  & 0.05 $\pm$ 0.092   & 0.049 $\pm$ 0.058  & 0.045 $\pm$ 0.075  & 0.04 $\pm$ 0.05    & 0.036 $\pm$ 0.08   & 1.526 $\pm$ 0.874  & 1.531 $\pm$ 0.837  & 1.511 $\pm$ 0.829  & 1.156 $\pm$ 1.285  \\ \hline
\multirow{5}{*}{Mixed I}    & Emp              & 0.001 $\pm$ 0.026  & -1.096 $\pm$ 1.605 & -0.328 $\pm$ 0.326 & -0.519 $\pm$ 0.641 & 0.013 $\pm$ 0.1    & 0.001 $\pm$ 0.022  & -0.35 $\pm$ 0.309  & 0.081 $\pm$ 0.11   & -0.128 $\pm$ 0.134 & 0.07 $\pm$ 0.359   & 0 $\pm$ 0.012      & 0.33 $\pm$ 0.121   & 0.61 $\pm$ 0.467   & 0.389 $\pm$ 0.178  & 0.492 $\pm$ 0.551  \\
                            & BN               & 0 $\pm$ 0.024      & 0.001 $\pm$ 0.082  & 0.001 $\pm$ 0.057  & -0.001 $\pm$ 0.064 & 0.001 $\pm$ 0.071  & 0 $\pm$ 0.02       & 0.002 $\pm$ 0.07   & 0.003 $\pm$ 0.062  & 0.001 $\pm$ 0.061  & 0.004 $\pm$ 0.081  & 0 $\pm$ 0.011      & 0.003 $\pm$ 0.072  & 0.003 $\pm$ 0.073  & 0.003 $\pm$ 0.07   & 0.003 $\pm$ 0.101  \\
                            & NonPar           & 0.001 $\pm$ 0.026  & -0.651 $\pm$ 1.727 & -0.272 $\pm$ 0.369 & -0.408 $\pm$ 0.683 & 0.007 $\pm$ 0.099  & 0.001 $\pm$ 0.022  & -0.34 $\pm$ 0.347  & 0.079 $\pm$ 0.11   & -0.127 $\pm$ 0.143 & 0.041 $\pm$ 0.349  & 0 $\pm$ 0.012      & 0.325 $\pm$ 0.142  & 0.602 $\pm$ 0.488  & 0.384 $\pm$ 0.246  & 0.477 $\pm$ 0.561  \\
                            & PV               & 0 $\pm$ 0.024      & -0.001 $\pm$ 0.081 & 0 $\pm$ 0.057      & -0.003 $\pm$ 0.063 & 0 $\pm$ 0.07       & 0 $\pm$ 0.02       & -0.003 $\pm$ 0.07  & 0.001 $\pm$ 0.062  & -0.004 $\pm$ 0.062 & 0.001 $\pm$ 0.08   & 0.002 $\pm$ 0.011  & -0.013 $\pm$ 0.074 & 0.009 $\pm$ 0.073  & -0.009 $\pm$ 0.072 & 0.021 $\pm$ 0.095  \\
                            & Semi.PV          & -0.004 $\pm$ 0.034 & -0.047 $\pm$ 0.113 & -0.042 $\pm$ 0.088 & -0.064 $\pm$ 0.115 & -0.007 $\pm$ 0.097 & -0.014 $\pm$ 0.059 & -0.196 $\pm$ 0.375 & -0.129 $\pm$ 0.239 & -0.213 $\pm$ 0.339 & -0.043 $\pm$ 0.211 & -0.01 $\pm$ 0.023  & -0.232 $\pm$ 0.455 & -0.18 $\pm$ 0.258  & -0.244 $\pm$ 0.401 & -0.2 $\pm$ 0.269   \\ \hline
\multirow{5}{*}{Mixed II}   & Emp              & -0.017 $\pm$ 0.024 & -0.779 $\pm$ 0.552 & -0.147 $\pm$ 0.112 & -0.196 $\pm$ 0.114 & -0.083 $\pm$ 0.066 & -0.037 $\pm$ 0.022 & -0.391 $\pm$ 0.13  & -0.057 $\pm$ 0.081 & -0.128 $\pm$ 0.092 & -0.2 $\pm$ 0.072   & -0.039 $\pm$ 0.013 & -0.016 $\pm$ 0.076 & 0.103 $\pm$ 0.083  & 0.03 $\pm$ 0.084   & -0.017 $\pm$ 0.076 \\
                            & BN               & -0.017 $\pm$ 0.024 & -0.003 $\pm$ 0.127 & 0 $\pm$ 0.06       & 0 $\pm$ 0.067      & -0.049 $\pm$ 0.066 & -0.037 $\pm$ 0.022 & 0.001 $\pm$ 0.09   & 0 $\pm$ 0.057      & 0 $\pm$ 0.061      & -0.122 $\pm$ 0.069 & -0.039 $\pm$ 0.012 & 0.001 $\pm$ 0.056  & 0.003 $\pm$ 0.061  & 0.003 $\pm$ 0.057  & 0.001 $\pm$ 0.056  \\
                            & NonPar           & -0.017 $\pm$ 0.024 & -0.771 $\pm$ 0.698 & -0.143 $\pm$ 0.12  & -0.192 $\pm$ 0.133 & -0.083 $\pm$ 0.067 & -0.037 $\pm$ 0.022 & -0.381 $\pm$ 0.254 & -0.055 $\pm$ 0.084 & -0.122 $\pm$ 0.111 & -0.2 $\pm$ 0.075   & -0.039 $\pm$ 0.013 & -0.015 $\pm$ 0.081 & 0.102 $\pm$ 0.092  & 0.028 $\pm$ 0.086  & -0.017 $\pm$ 0.078 \\
                            & PV               & -0.018 $\pm$ 0.024 & -0.002 $\pm$ 0.127 & 0 $\pm$ 0.061      & 0 $\pm$ 0.067      & -0.049 $\pm$ 0.066 & -0.038 $\pm$ 0.022 & 0.001 $\pm$ 0.09   & 0.001 $\pm$ 0.057  & 0 $\pm$ 0.061      & -0.123 $\pm$ 0.068 & -0.041 $\pm$ 0.012 & 0.001 $\pm$ 0.056  & 0.003 $\pm$ 0.061  & 0.003 $\pm$ 0.057  & 0.001 $\pm$ 0.056  \\
                            & Semi.PV          & -0.018 $\pm$ 0.024 & -0.004 $\pm$ 0.132 & -0.003 $\pm$ 0.063 & -0.004 $\pm$ 0.069 & -0.05 $\pm$ 0.066  & -0.04 $\pm$ 0.023  & -0.002 $\pm$ 0.092 & -0.002 $\pm$ 0.059 & -0.003 $\pm$ 0.065 & -0.126 $\pm$ 0.07  & -0.043 $\pm$ 0.016 & -0.01 $\pm$ 0.061  & -0.003 $\pm$ 0.07  & -0.007 $\pm$ 0.065 & -0.011 $\pm$ 0.062 \\ \hline
\end{tabular}
\end{adjustbox}
\end{center}
%\end{landscape}
\end{table}

\begin{figure}[hbt!]%
    \includegraphics[width=0.95\textwidth]{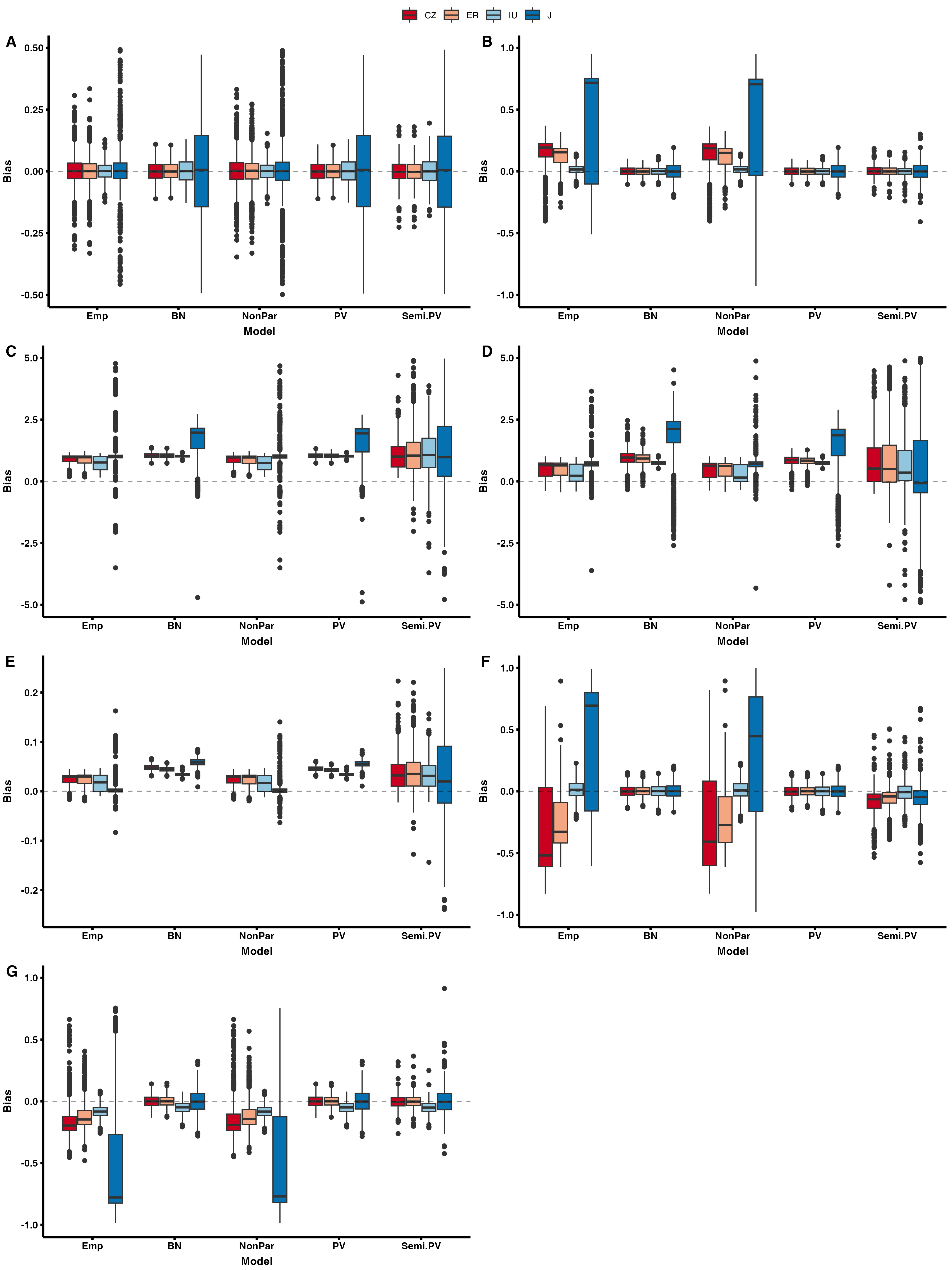}
    \caption{\label{Fig: Bias_highSamp_lowAUC}Bias in estimation optimal cut-points for high sample size, when AUC level is low. Panels A - G respectively correspond to the simulation scenarios: BN equal, BN unequal, Skewed I, Skewed II, Skewed III, Mixed I, and Mixed II.}
\end{figure}

\begin{figure}[hbt!]%
    \includegraphics[width=\textwidth]{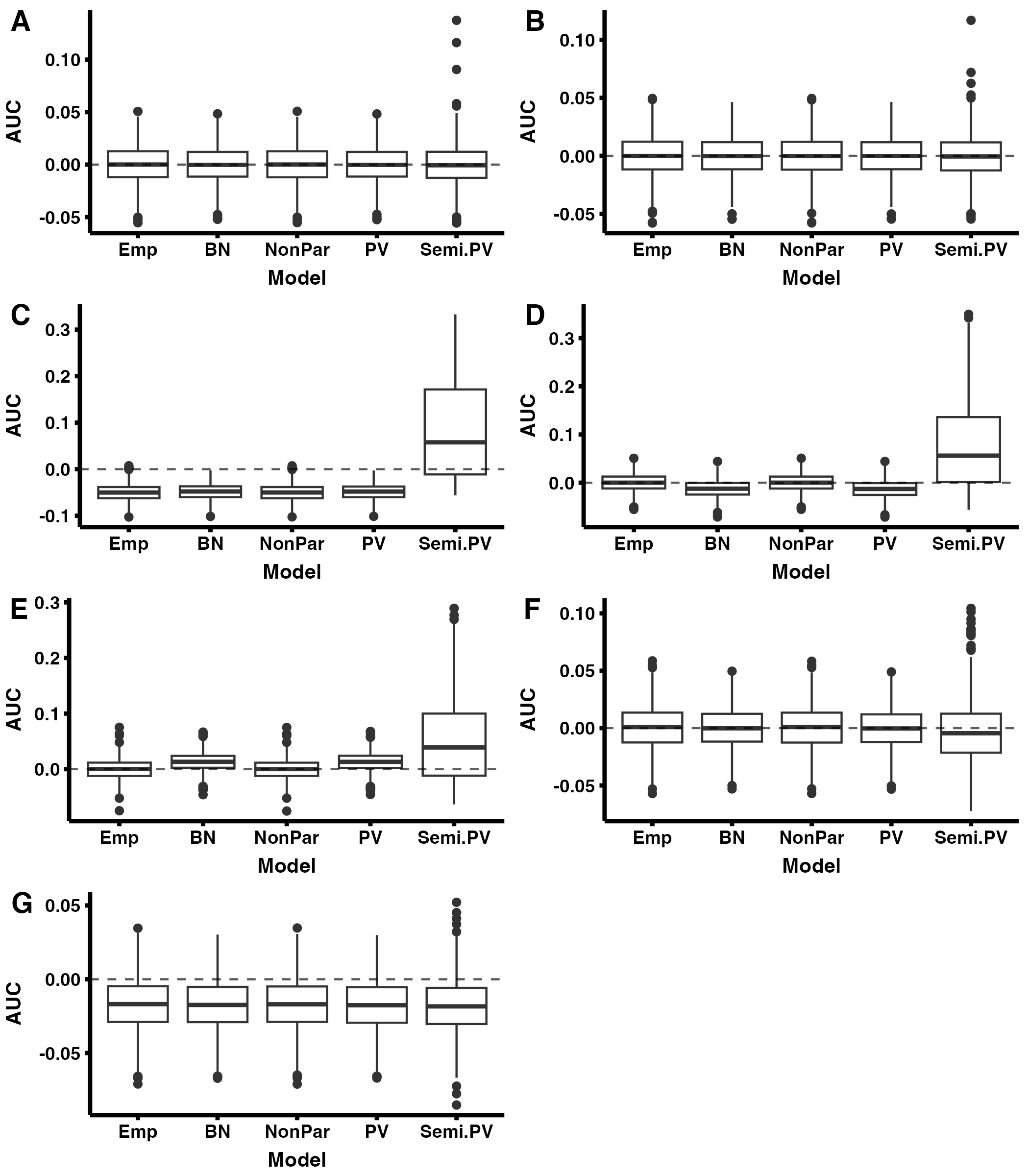}
    \caption{\label{Fig: Bias_AUC_highSamp_lowAUC}Bias in estimation of AUC for high sample size, when AUC level is low. Panels A - G respectively correspond to the simulation scenarios: BN equal, BN unequal, Skewed I, Skewed II, Skewed III, Mixed I, and Mixed II.}
\end{figure}

\begin{figure}[hbt!]%
    \includegraphics[width=0.95\textwidth]{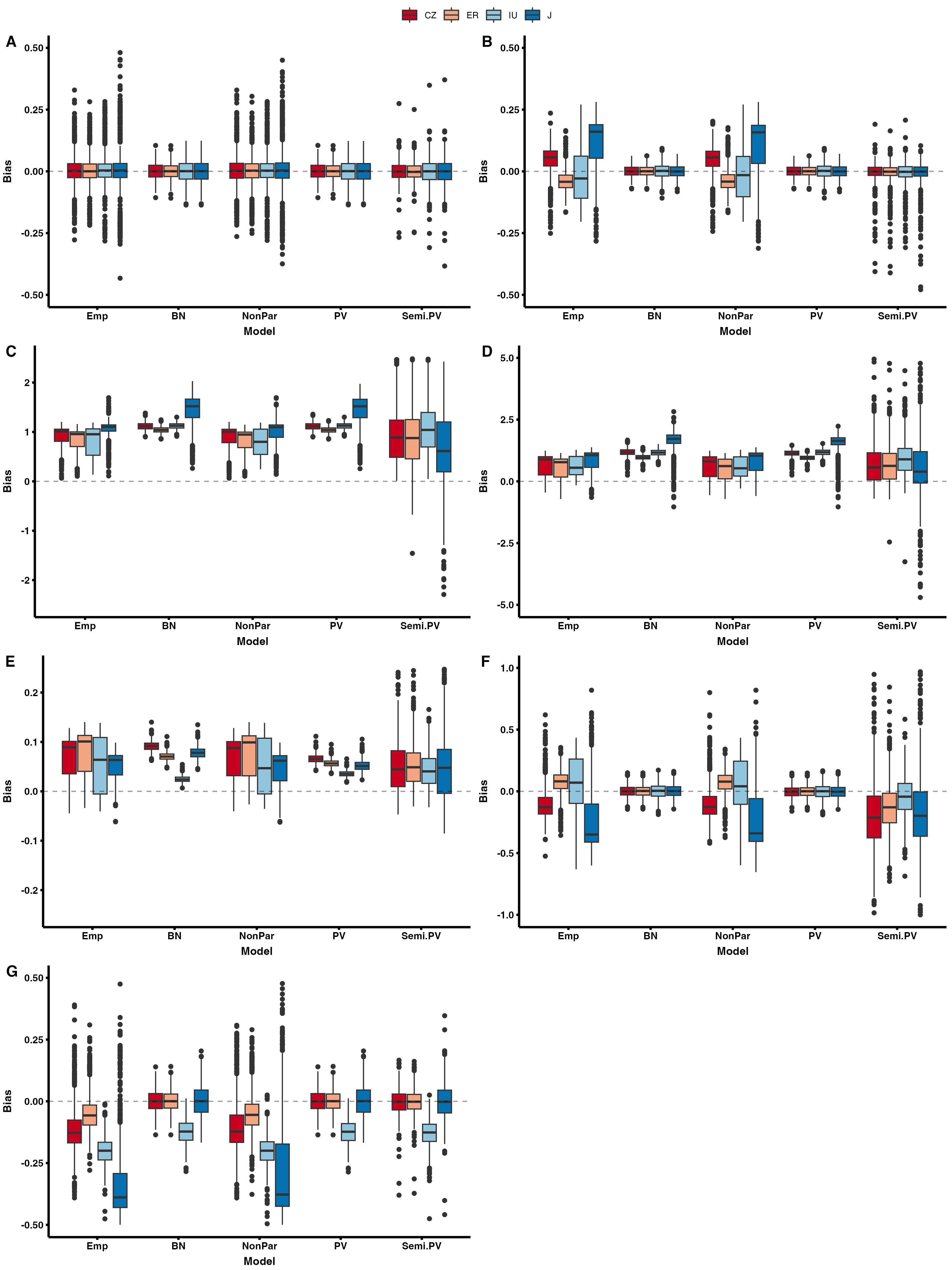}
    \caption{\label{Fig: Bias_highSamp_medAUC}Bias in estimation optimal cut-points for high sample size, when AUC level is medium. Panels A - G respectively correspond to the simulation scenarios: BN equal, BN unequal, Skewed I, Skewed II, Skewed III, Mixed I, and Mixed II.}
\end{figure}

\begin{figure}[hbt!]%
    \includegraphics[width=\textwidth]{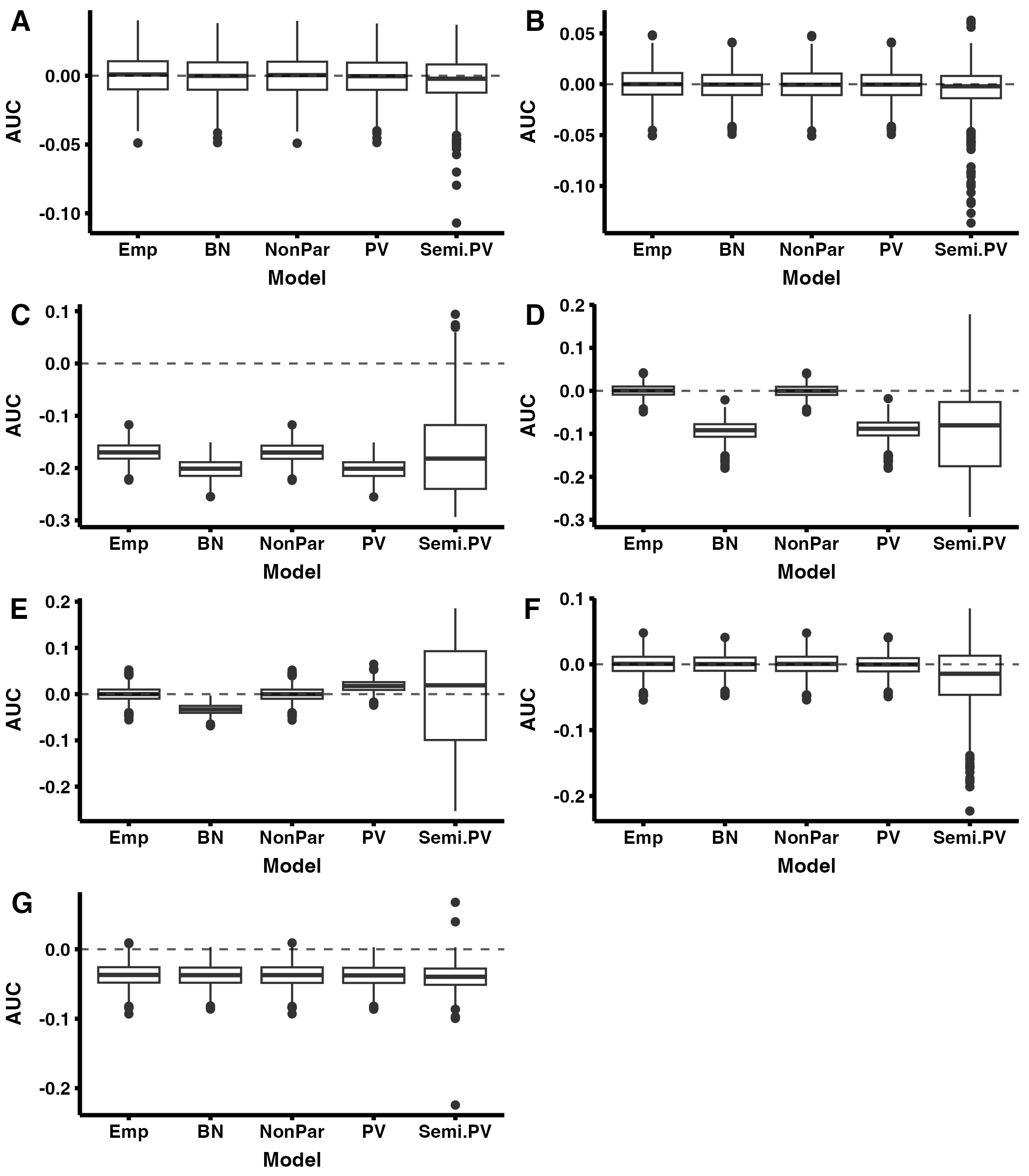}
    \caption{\label{Fig: Bias_AUC_highSamp_medAUC}Bias in estimation of AUC for high sample size, when AUC level is medium. Panels A - G respectively correspond to the simulation scenarios: BN equal, BN unequal, Skewed I, Skewed II, Skewed III, Mixed I, and Mixed II.}
\end{figure}

\begin{figure}[hbt!]%
    \includegraphics[width=0.95\textwidth]{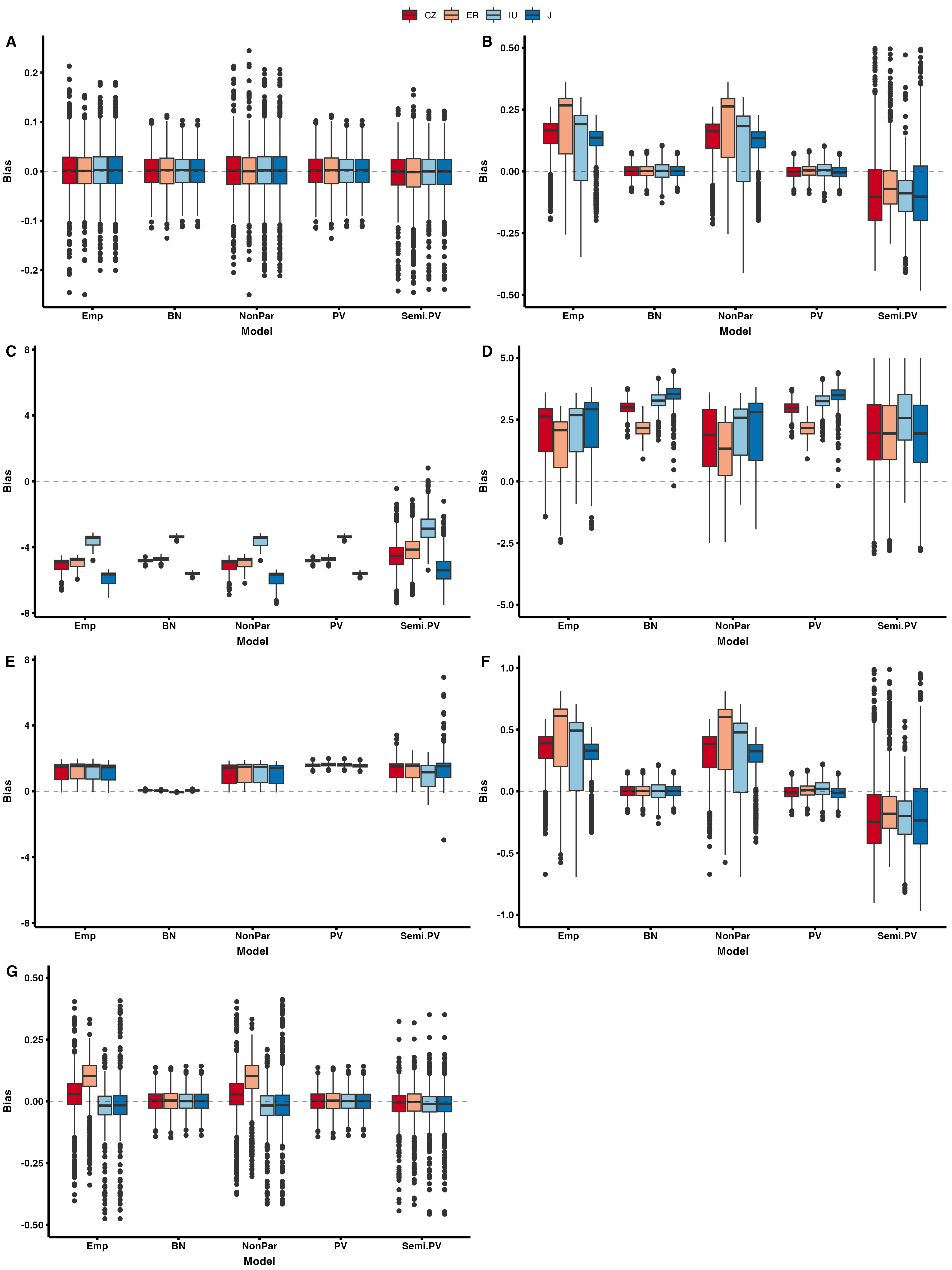}
    \caption{\label{Fig: Bias_highSamp_highAUC}Bias in estimation optimal cut-points for high sample size, when AUC level is high. Panels A - G respectively correspond to the simulation scenarios: BN equal, BN unequal, Skewed I, Skewed II, Skewed III, Mixed I, and Mixed II.}
\end{figure}

\begin{figure}[hbt!]%
    \includegraphics[width=\textwidth]{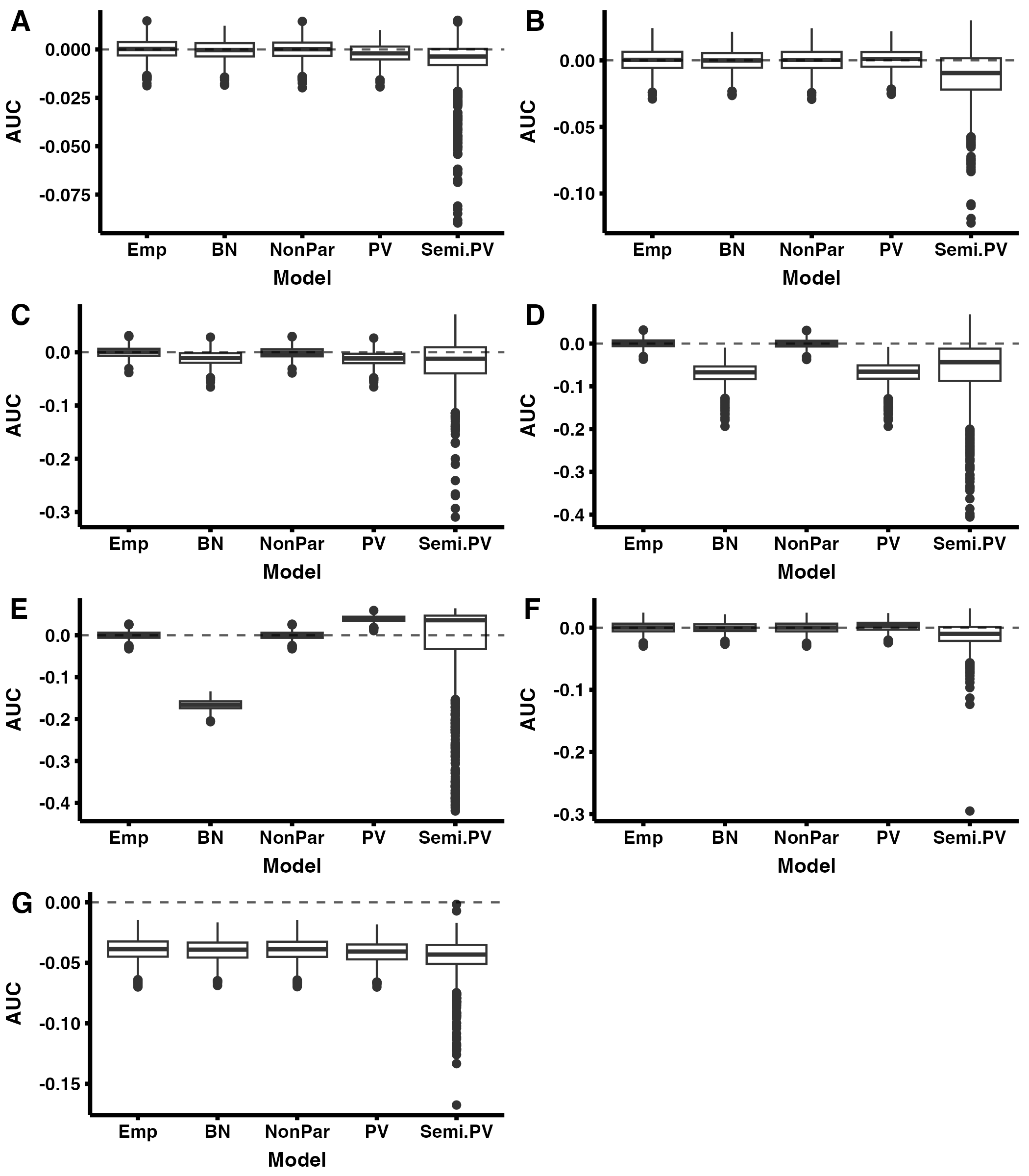}
    \caption{\label{Fig: Bias_AUC_highSamp_highAUC}Bias in estimation of AUC for high sample size, when AUC level is high. Panels A - G respectively correspond to the simulation scenarios: BN equal, BN unequal, Skewed I, Skewed II, Skewed III, Mixed I, and Mixed II.}
\end{figure}

Figures~\ref{Fig: Bias_highSamp_lowAUC}, \ref{Fig: Bias_highSamp_medAUC}, and \ref{Fig: Bias_highSamp_highAUC} respectively shows the biases for different AUC levels, obtained from different ROC methodologies in estimating four different cutpoints under different data generating mechanism, when covariates don't impact the data generating process and the sample size is high, i.e. $N=500$ for both healthy and diseased biomarker. Figures~\ref{Fig: Bias_AUC_highSamp_lowAUC}, \ref{Fig: Bias_AUC_highSamp_medAUC}, and \ref{Fig: Bias_AUC_highSamp_highAUC} respectively shows the corresponding biases for estimating AUCs.

\subsection{With covariates} \label{AppSec: Other_sim_cov}

\begin{figure}[hbt!]%
    \includegraphics[width=0.95\textwidth]{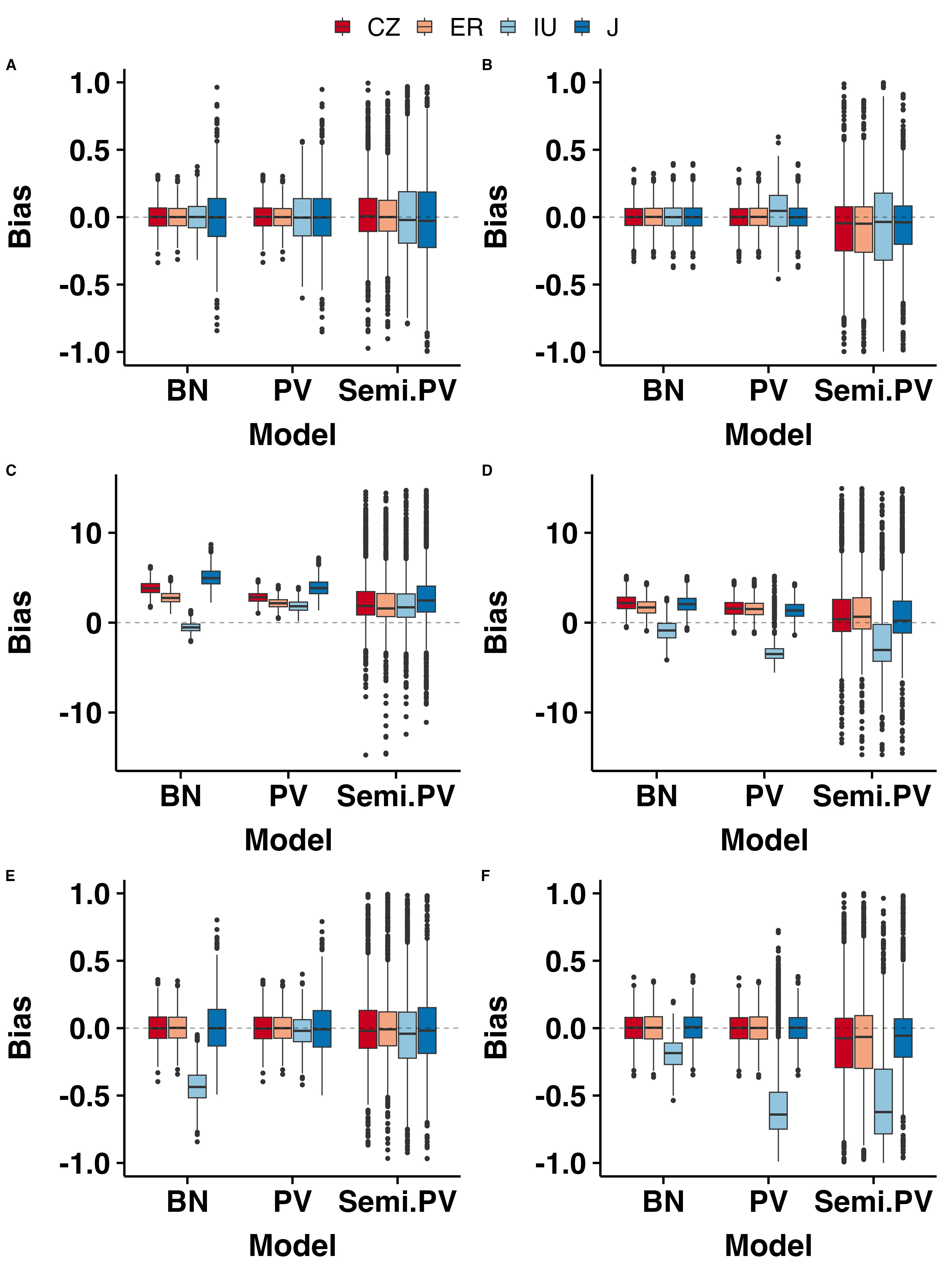}
    \caption{\label{Fig: Bias_medSamp_cov}Bias in estimation of optimal cut-points for medium sample size in the presence of covariates. The left panel (A, C, E) respectively corresponds to simulation scenarios: BN, Skewed, and Mixed for covariate value at 0. The right panel (B, D, F) corresponds to the covariate value at 1.}
\end{figure}

\begin{figure}[hbt!]%
    \includegraphics[width=\textwidth]{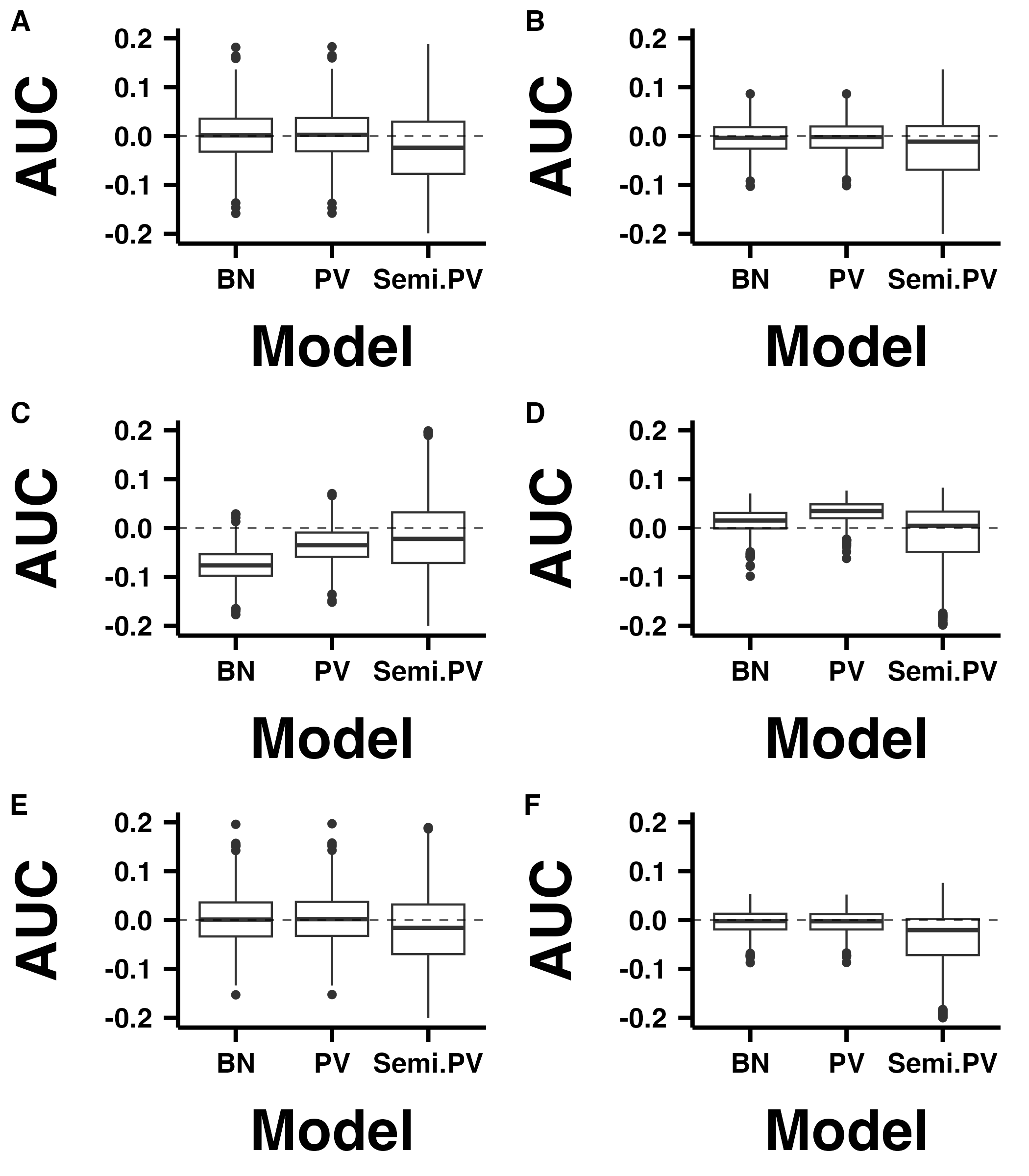}
    \caption{\label{Fig: Bias_medSamp_cov_AUC}Bias in estimation of AUC for medium sample size in the presence of covariates. The left panel (A, C, E) respectively corresponds to simulation scenarios: BN, Skewed, and Mixed for covariate value at 0. The right panel (B, D, F) corresponds to the covariate value at 1.}
\end{figure}

Figure~\ref{Fig: Bias_medSamp_cov} shows the biases for different covariate levels, obtained from different ROC methodologies in estimating four different cutpoints under different data generating mechanism, when covariates impacts the data generating process and the sample size is medium, i.e. $N=100$ for both healthy and diseased biomarker. Figure~\ref{Fig: Bias_medSamp_cov_AUC} shows the corresponding biases for estimating AUCs.

Table~\ref{Tab: bias_cov_lowSamp} presents the biases in estimating AUCs and different cutoffs from different ROC methodologies under different data generating mechanism and different covariate levels with low sample size.

\begin{table}[hbt!]
\begin{center}
\caption{\label{Tab: bias_cov_lowSamp} Biases of estimating AUC and optimal thresholds for different fitting models at different covariate levels for low sample size.}
\begin{adjustbox}{width=0.95\textwidth}
\begin{tabular}{cccccccccccc}
\hline
\textbf{Data}           & \textbf{Fitting} & \multicolumn{5}{c}{\textbf{x = 0}}                                                                      & \multicolumn{5}{c}{\textbf{x = 1}}                                                                       \\ \cline{3-12} 
\textbf{generating}     & \textbf{model}   & \multicolumn{5}{c}{\textbf{Median $\pm$ IQR}}                                                           & \multicolumn{5}{c}{\textbf{Median $\pm$ IQR}}                                                            \\ \cline{3-12} 
\textbf{mechanism}      & \textbf{}        & AUC                  & J                 & ER                 & CZ                 & IU                 & AUC                  & J                  & ER                 & CZ                 & IU                 \\ \hline
\multirow{3}{*}{BN}     & BN               & -0.002   $\pm$ 0.099 & 0.001 $\pm$ 0.402 & -0.007 $\pm$ 0.186 & -0.002 $\pm$ 0.197 & -0.001 $\pm$ 0.2   & -0.002 $\pm$   0.061 & 0.008 $\pm$ 0.19   & 0.007 $\pm$ 0.175  & 0.007 $\pm$ 0.177  & 0.009 $\pm$ 0.189  \\
                        & PV               & 0 $\pm$ 0.101        & 0 $\pm$ 0.399     & -0.006 $\pm$ 0.186 & -0.002 $\pm$ 0.195 & -0.006 $\pm$ 0.37  & 0.002 $\pm$ 0.06     & 0.006 $\pm$ 0.187  & 0.007 $\pm$ 0.175  & 0.007 $\pm$ 0.179  & 0.028 $\pm$ 0.31   \\
                        & Semi.PV          & -0.051 $\pm$ 0.152   & 0.024 $\pm$ 0.798 & -0.011 $\pm$ 0.305 & -0.001 $\pm$ 0.318 & -0.022 $\pm$ 0.437 & -0.046 $\pm$ 0.179   & -0.062 $\pm$ 0.364 & -0.075 $\pm$ 0.387 & -0.076 $\pm$ 0.375 & -0.077 $\pm$ 0.6   \\ \hline
\multirow{3}{*}{Skewed} & BN               & -0.078 $\pm$ 0.059   & 4.987 $\pm$ 2.044 & 2.73 $\pm$ 1.343   & 3.794 $\pm$ 1.463  & -0.586 $\pm$ 1.073 & 0.013 $\pm$ 0.046    & 2.082 $\pm$ 1.865  & 1.642 $\pm$ 1.742  & 2.174 $\pm$ 1.815  & -0.994 $\pm$ 2.068 \\
                        & PV               & -0.036 $\pm$ 0.074   & 3.758 $\pm$ 1.991 & 2.062 $\pm$ 1.274  & 2.68 $\pm$ 1.342   & 1.699 $\pm$ 1.357  & 0.036 $\pm$ 0.042    & 1.298 $\pm$ 1.827  & 1.459 $\pm$ 1.741  & 1.482 $\pm$ 1.794  & -3.324 $\pm$ 1.703 \\
                        & Semi.PV          & -0.053 $\pm$ 0.149   & 2.808 $\pm$ 3.815 & 1.428 $\pm$ 2.378  & 1.697 $\pm$ 2.583  & 1.635 $\pm$ 2.76   & -0.021 $\pm$ 0.126   & 0.255 $\pm$ 4.376  & 0.794 $\pm$ 4.579  & 0.538 $\pm$ 4.629  & -2.97 $\pm$ 5.266  \\ \hline
\multirow{3}{*}{Mixed}  & BN               & -0.001 $\pm$ 0.1     & 0.024 $\pm$ 0.418 & 0.001 $\pm$ 0.223  & -0.005 $\pm$ 0.235 & -0.448 $\pm$ 0.225 & -0.002 $\pm$ 0.045   & 0.012 $\pm$ 0.21   & 0.003 $\pm$ 0.216  & 0.006 $\pm$ 0.21   & -0.194 $\pm$ 0.213 \\
                        & PV               & 0 $\pm$ 0.102        & 0.006 $\pm$ 0.415 & -0.004 $\pm$ 0.221 & -0.01 $\pm$ 0.234  & -0.042 $\pm$ 0.231 & 0 $\pm$ 0.044        & 0.003 $\pm$ 0.211  & 0 $\pm$ 0.217      & 0.002 $\pm$ 0.209  & -0.628 $\pm$ 0.426 \\
                        & Semi.PV          & -0.04 $\pm$ 0.148    & 0.061 $\pm$ 0.868 & -0.011 $\pm$ 0.371 & -0.04 $\pm$ 0.385  & -0.078 $\pm$ 0.476 & -0.045 $\pm$ 0.153   & -0.084 $\pm$ 0.419 & -0.078 $\pm$ 0.491 & -0.094 $\pm$ 0.476 & -0.729 $\pm$ 0.607 \\ \hline
\end{tabular}
\end{adjustbox}
\end{center}
\end{table}

\begin{figure}[hbt!]%
    \includegraphics[width=0.95\textwidth]{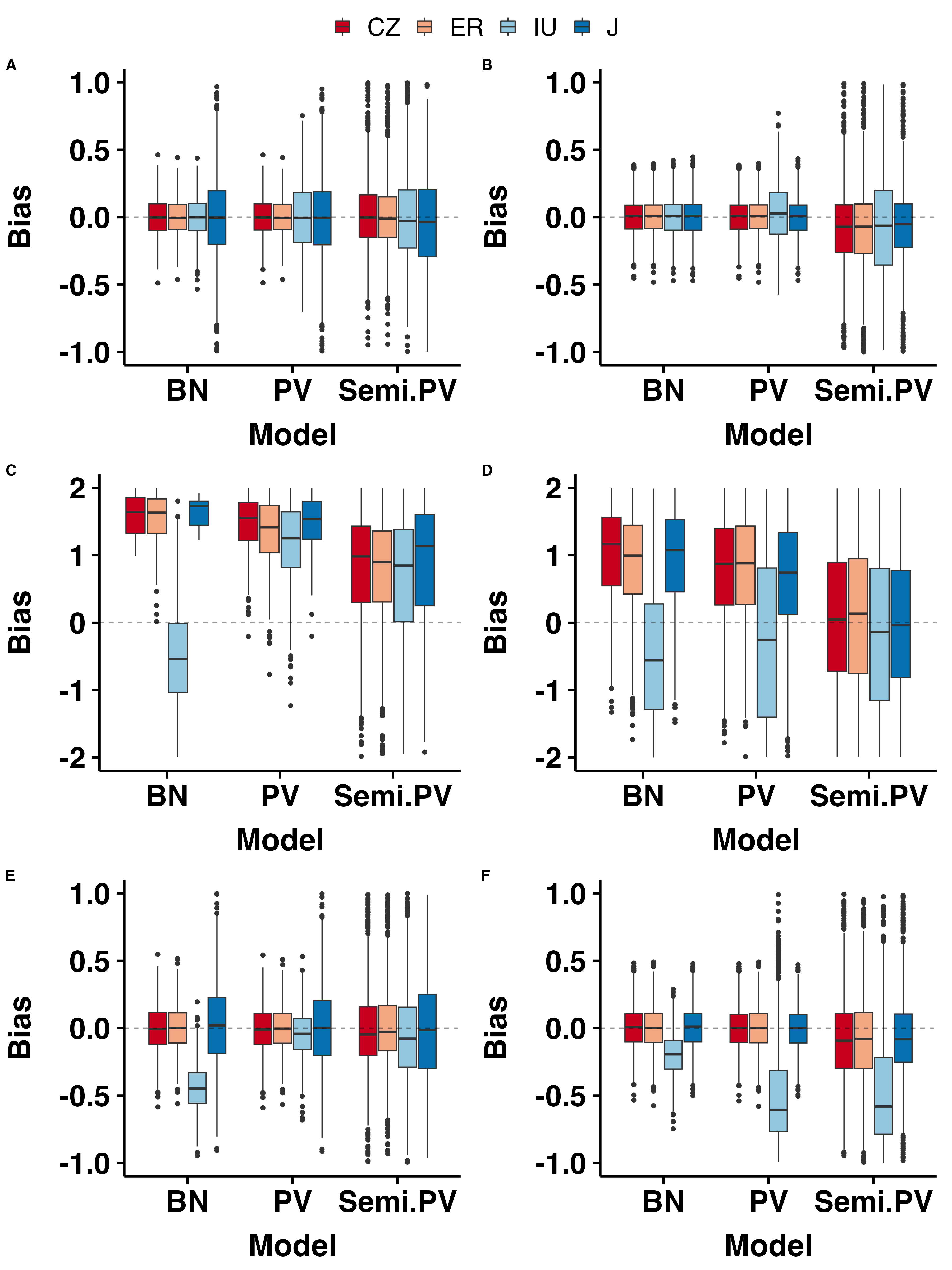}
    \caption{\label{Fig: Bias_lowSamp_cov}Bias in estimation AUC and optimal cut-points for low sample size in the presence of covariates. The left panel (A, C, E) respectively corresponds to simulation scenarios: BN, Skewed, and Mixed for covariate value at 0. The right panel (B, D, F) corresponds to the covariate value at 1.}
\end{figure}

\begin{figure}[hbt!]%
    \includegraphics[width=\textwidth]{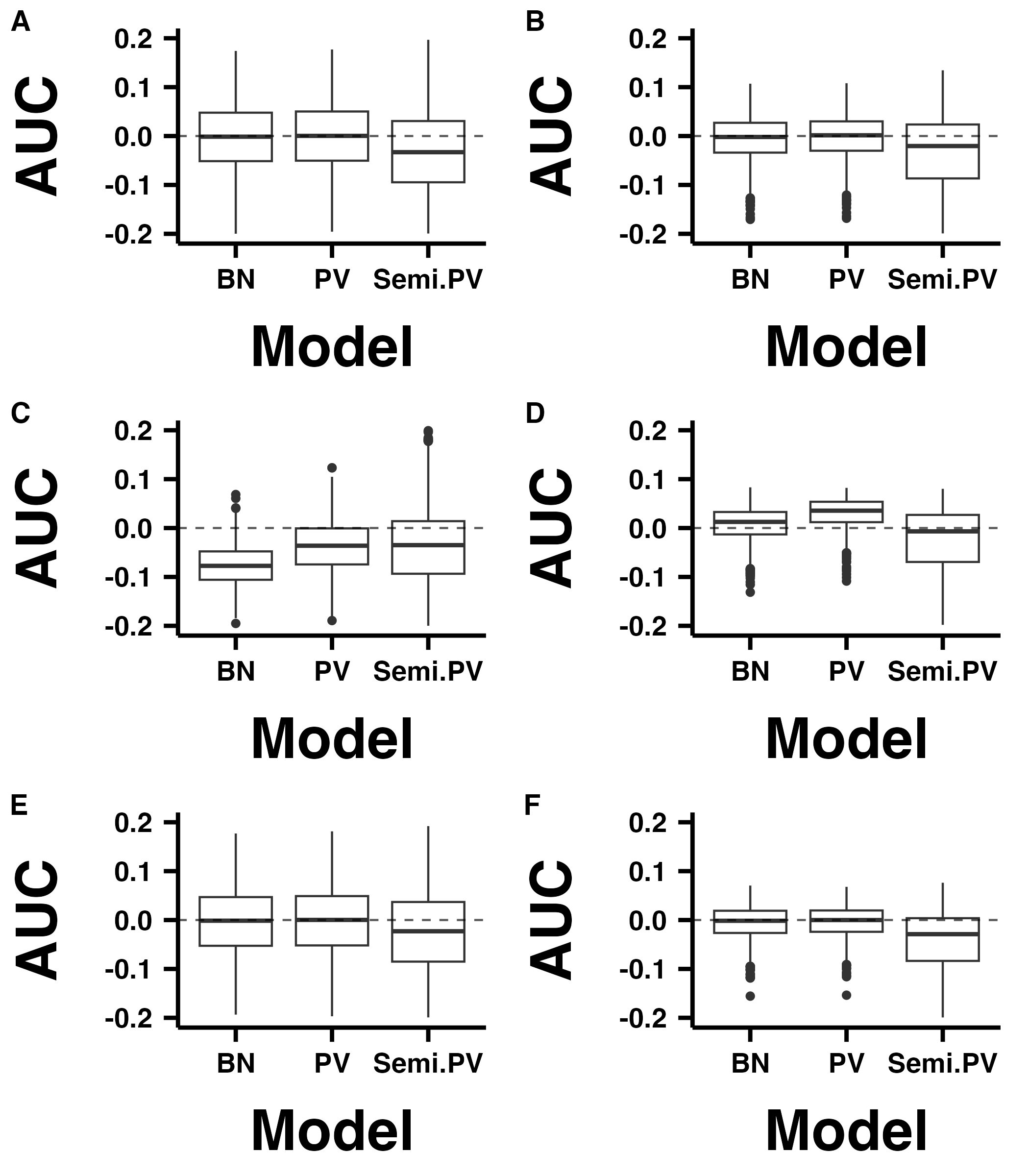}
    \caption{\label{Fig: Bias_lowSamp_cov_AUC}Bias in estimation of AUC for low sample size in the presence of covariates. The left panel (A, C, E) respectively corresponds to simulation scenarios: BN, Skewed, and Mixed for covariate value at 0. The right panel (B, D, F) corresponds to the covariate value at 1.}
\end{figure}

Figure~\ref{Fig: Bias_lowSamp_cov} shows the biases for different covariate levels, obtained from different ROC methodologies in estimating four different cutpoints under different data generating mechanism, when covariates impacts the data generating process and the sample size is low, i.e. $N=50$ for both healthy and diseased biomarker. Figure~\ref{Fig: Bias_lowSamp_cov_AUC} shows the corresponding biases for estimating AUCs.

Table~\ref{Tab: bias_cov_highSamp} presents the biases in estimating AUCs and different cutoffs from different ROC methodologies under different data generating mechanism and different covariate levels with high sample size.

\begin{table}[hbt!]
\begin{center}
\caption{\label{Tab: bias_cov_highSamp} Biases of estimating AUC and optimal thresholds for different fitting models at different covariate levels for high sample size.}
\begin{adjustbox}{width=0.95\textwidth}
\begin{tabular}{cccccccccccc}
\hline
\textbf{Data}           & \textbf{Fitting} & \multicolumn{5}{c}{\textbf{x = 0}}                                                                     & \multicolumn{5}{c}{\textbf{x = 1}}                                                                       \\ \cline{3-12} 
\textbf{generating}     & \textbf{model}   & \multicolumn{5}{c}{\textbf{Median $\pm$ IQR}}                                                          & \multicolumn{5}{c}{\textbf{Median $\pm$ IQR}}                                                            \\ \cline{3-12} 
\textbf{mechanism}      & \textbf{}        & AUC                & J                  & ER                 & CZ                 & IU                 & AUC                  & J                  & ER                 & CZ                 & IU                 \\ \hline
\multirow{3}{*}{BN}     & BN               & 0 $\pm$   0.029    & 0 $\pm$ 0.126      & -0.001 $\pm$ 0.065 & -0.002 $\pm$ 0.069 & 0.001 $\pm$ 0.114  & -0.001 $\pm$   0.019 & 0.001 $\pm$ 0.058  & 0 $\pm$ 0.056      & 0.001 $\pm$ 0.056  & 0.001 $\pm$ 0.058  \\
                        & PV               & 0 $\pm$ 0.029      & 0 $\pm$ 0.127      & -0.002 $\pm$ 0.065 & -0.002 $\pm$ 0.069 & 0 $\pm$ 0.127      & -0.001 $\pm$ 0.019   & 0.001 $\pm$ 0.057  & 0 $\pm$ 0.056      & 0.001 $\pm$ 0.056  & 0.071 $\pm$ 0.112  \\
                        & Semi.PV          & -0.02 $\pm$ 0.118  & 0.024 $\pm$ 0.341  & 0.005 $\pm$ 0.15   & 0.005 $\pm$ 0.161  & 0 $\pm$ 0.233      & -0.019 $\pm$ 0.142   & -0.017 $\pm$ 0.245 & -0.026 $\pm$ 0.271 & -0.026 $\pm$ 0.276 & 0.03 $\pm$ 0.443   \\ \hline
\multirow{3}{*}{Skewed} & BN               & -0.083 $\pm$ 0.018 & 5.178 $\pm$ 0.577  & 2.901 $\pm$ 0.387  & 4.023 $\pm$ 0.422  & -0.564 $\pm$ 0.344 & 0.005 $\pm$ 0.015    & 2.075 $\pm$ 0.609  & 1.59 $\pm$ 0.555   & 2.179 $\pm$ 0.581  & -1.209 $\pm$ 0.707 \\
                        & PV               & -0.043 $\pm$ 0.022 & 3.985 $\pm$ 0.585  & 2.211 $\pm$ 0.384  & 2.895 $\pm$ 0.405  & 1.813 $\pm$ 0.397  & 0.031 $\pm$ 0.013    & 1.353 $\pm$ 0.608  & 1.461 $\pm$ 0.559  & 1.564 $\pm$ 0.583  & -3.653 $\pm$ 0.469 \\
                        & Semi.PV          & -0.015 $\pm$ 0.16  & 1.386 $\pm$ 7.347  & 0.761 $\pm$ 3.196  & 0.753 $\pm$ 3.443  & 0.447 $\pm$ 4.154  & 0.001 $\pm$ 0.087    & 0.609 $\pm$ 4.827  & 1.274 $\pm$ 4.479  & 0.833 $\pm$ 4.69   & -2.168 $\pm$ 4.012 \\ \hline
\multirow{3}{*}{Mixed}  & BN               & 0 $\pm$ 0.032      & -0.001 $\pm$ 0.122 & -0.003 $\pm$ 0.075 & -0.002 $\pm$ 0.079 & -0.435 $\pm$ 0.08  & 0 $\pm$ 0.014        & 0.001 $\pm$ 0.068  & 0.001 $\pm$ 0.071  & 0.001 $\pm$ 0.067  & -0.189 $\pm$ 0.068 \\
                        & PV               & 0 $\pm$ 0.032      & -0.003 $\pm$ 0.122 & -0.003 $\pm$ 0.075 & -0.003 $\pm$ 0.08  & -0.001 $\pm$ 0.076 & -0.002 $\pm$ 0.014   & 0.001 $\pm$ 0.068  & 0.001 $\pm$ 0.071  & 0.001 $\pm$ 0.067  & -0.695 $\pm$ 0.119 \\
                        & Semi.PV          & -0.018 $\pm$ 0.118 & 0.022 $\pm$ 0.431  & -0.012 $\pm$ 0.187 & -0.02 $\pm$ 0.202  & -0.019 $\pm$ 0.278 & -0.019 $\pm$ 0.128   & -0.053 $\pm$ 0.249 & -0.048 $\pm$ 0.342 & -0.055 $\pm$ 0.33  & -0.741 $\pm$ 0.382 \\ \hline
\end{tabular}
\end{adjustbox}
\end{center}
\end{table}

\begin{figure}[hbt!]%
    \includegraphics[width=0.95\textwidth]{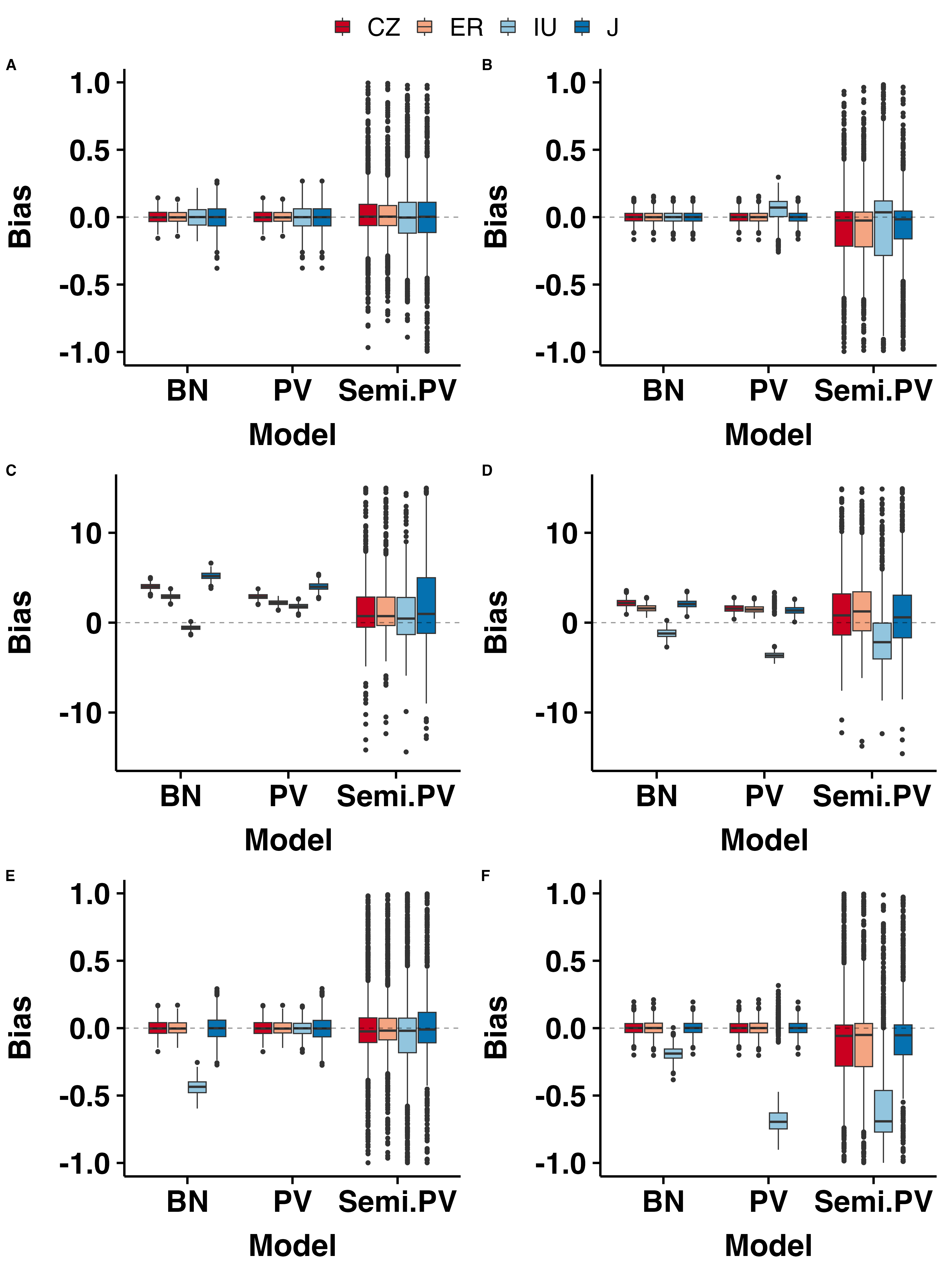}
    \caption{\label{Fig: Bias_highSamp_cov}Bias in estimation AUC and optimal cut-points for high sample size in the presence of covariates. The left panel (A, C, E) respectively corresponds to simulation scenarios: BN, Skewed, and Mixed for covariate value at 0. The right panel (B, D, F) corresponds to the covariate value at 1.}
\end{figure}

\begin{figure}[hbt!]%
    \includegraphics[width=\textwidth]{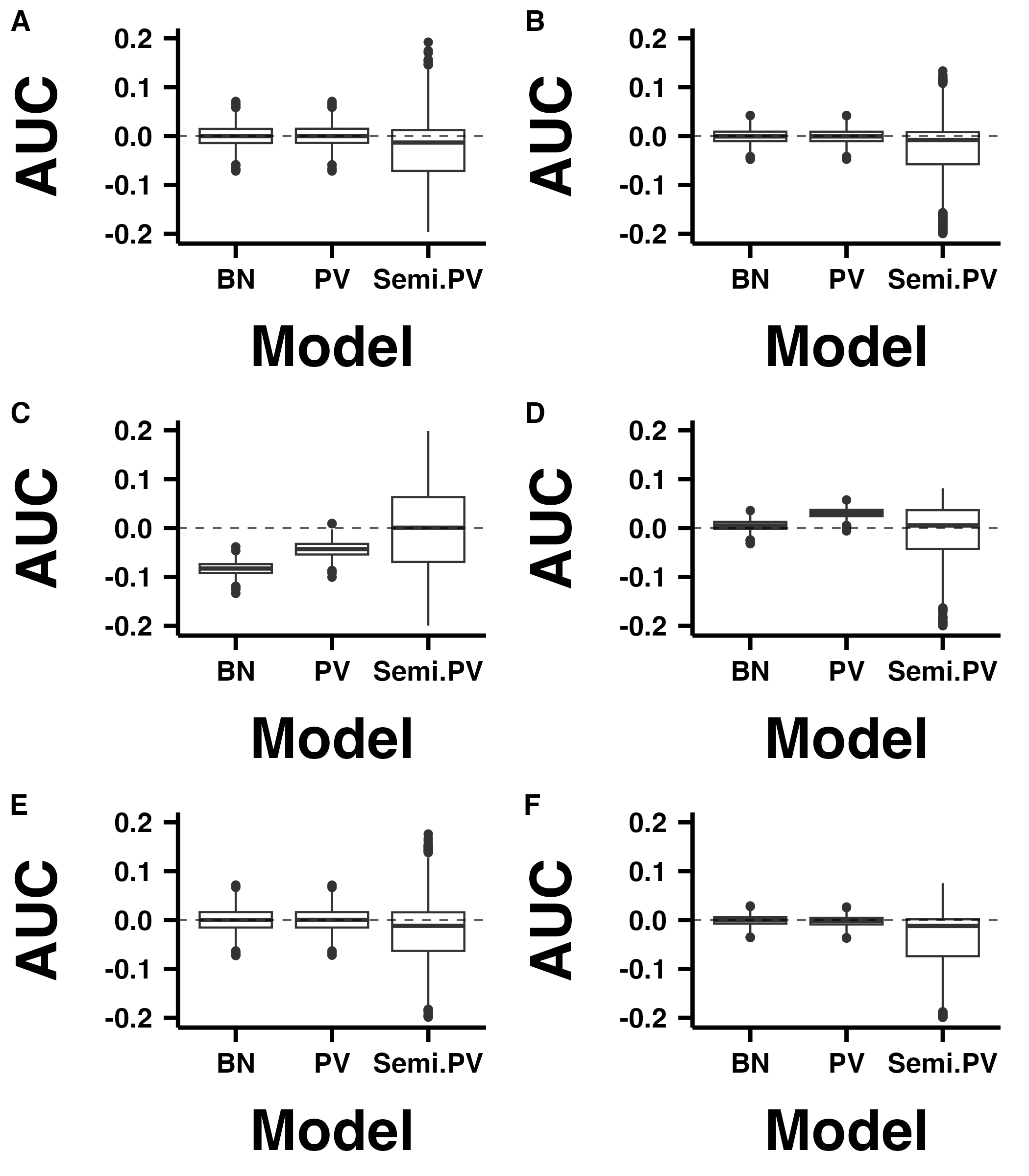}
    \caption{\label{Fig: Bias_highSamp_cov_AUC}Bias in estimation of AUC for high sample size in the presence of covariates. The left panel (A, C, E) respectively corresponds to simulation scenarios: BN, Skewed, and Mixed for covariate value at 0. The right panel (B, D, F) corresponds to the covariate value at 1.}
\end{figure}

Figure~\ref{Fig: Bias_highSamp_cov} shows the biases for different covariate levels, obtained from different ROC methodologies in estimating four different cutpoints under different data generating mechanism, when covariates impacts the data generating process and the sample size is high, i.e. $N=500$ for both healthy and diseased biomarker. Figure~\ref{Fig: Bias_highSamp_cov_AUC} shows the corresponding biases for estimating AUCs.

\section{Other details of data analysis} \label{AppSec: Other_data_analysis}

\begin{figure}[hbt!]%
    \begin{center}
    \includegraphics[width=\textwidth]{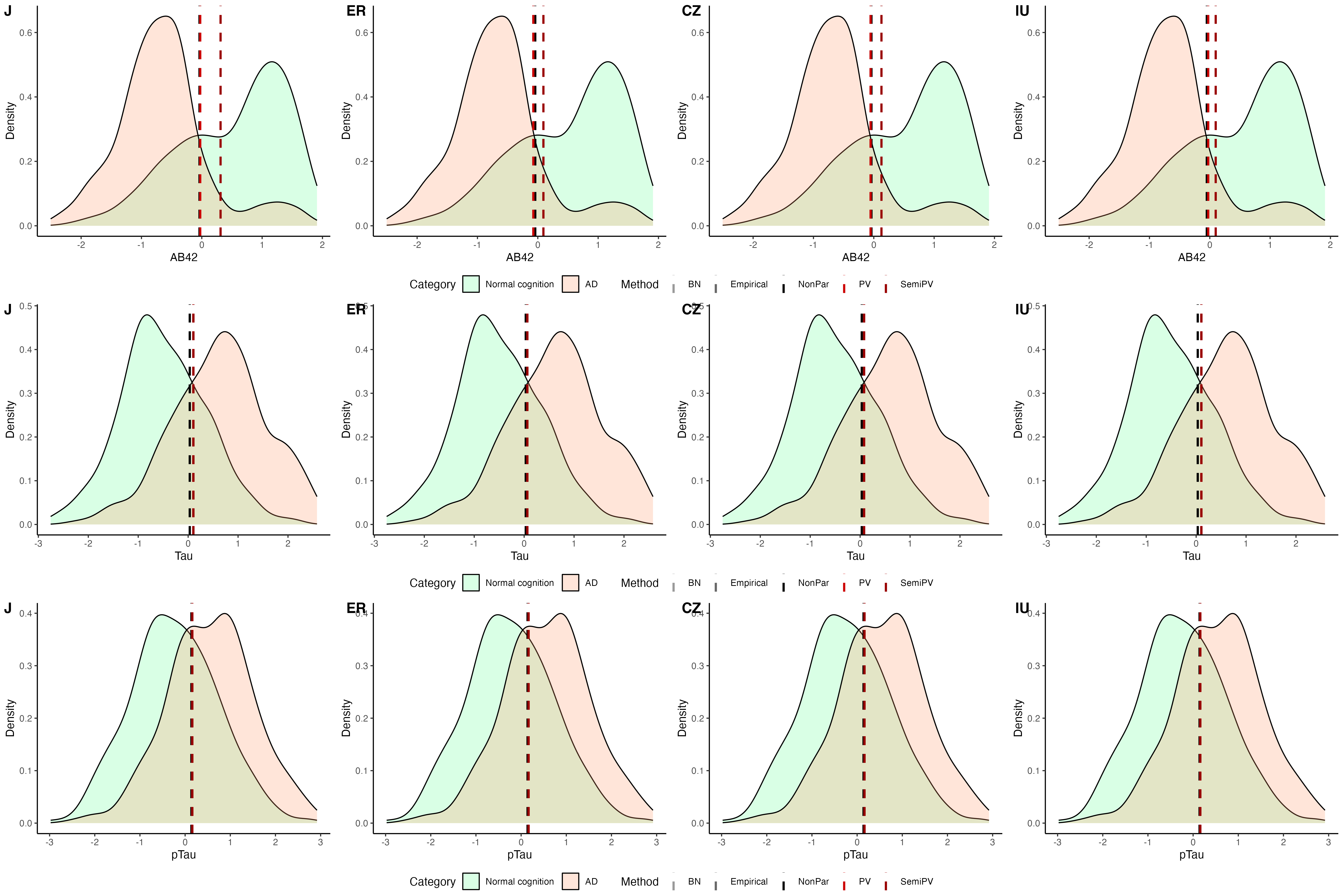}
    \caption{\label{Fig: data_analysis_nocov_density} Biomarker densities and estimated cutoffs of all biomarkers. The three rows respectively corresponds to the estimated cutoffs of all the optimal threshold estimators for the biomarkers $A\beta 42$, Tau, and pTau.}
    \end{center}
\end{figure}

\begin{figure}[hbt!]%
    \begin{center}
    \includegraphics[width=0.8\textwidth]{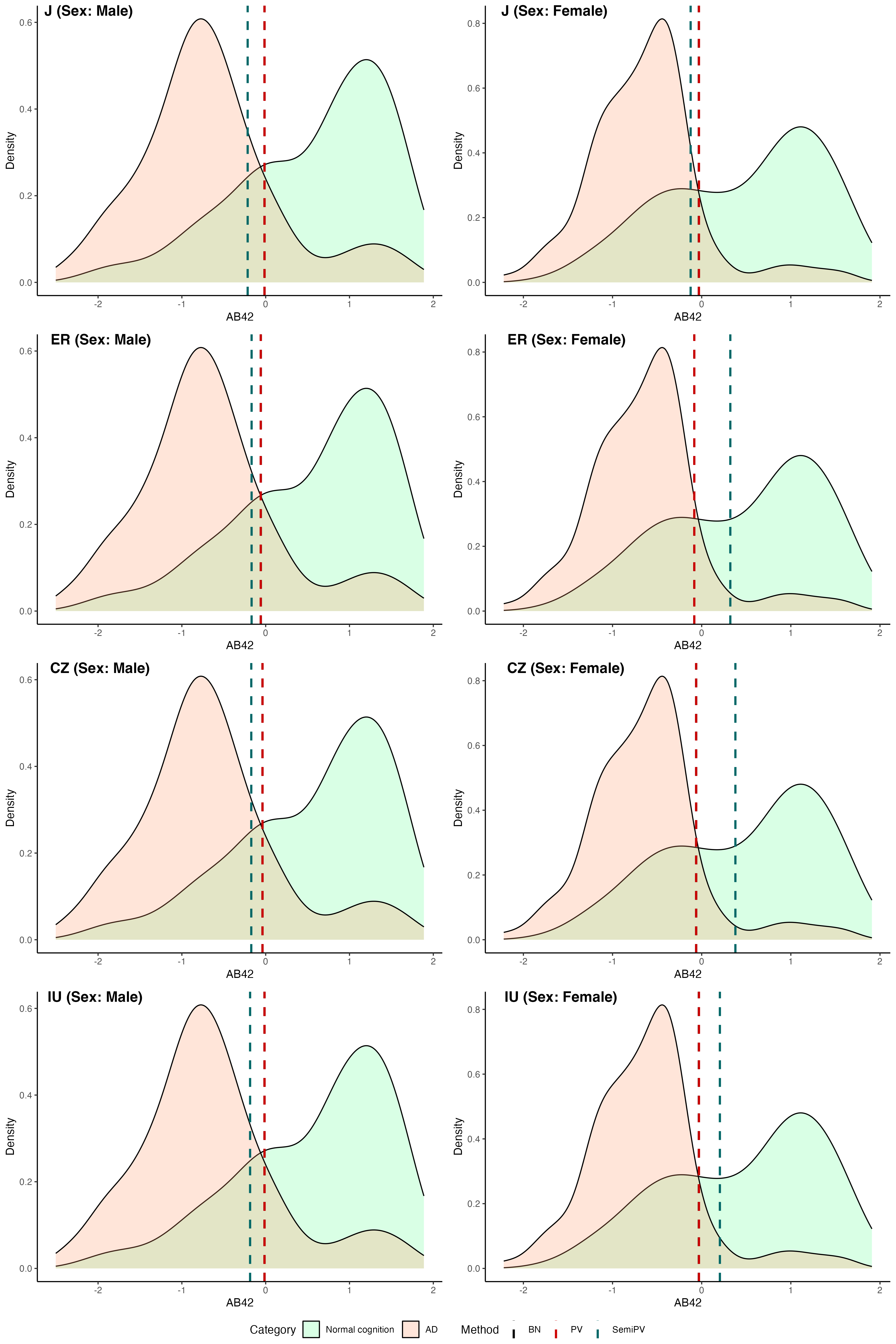}
    \caption{\label{Fig: data_analysis_cov_density_AB42} Sex-specific biomarker densities and estimated cutoffs of biomarker $A\beta 42$. The left row corresponds to the cutoff estimates of \enquote{Male} subjects and the right row corresponds to that of the \enquote{Female} subjects.}
    \end{center}
\end{figure}

\begin{figure}[hbt!]%
    \begin{center}
    \includegraphics[width=0.8\textwidth]{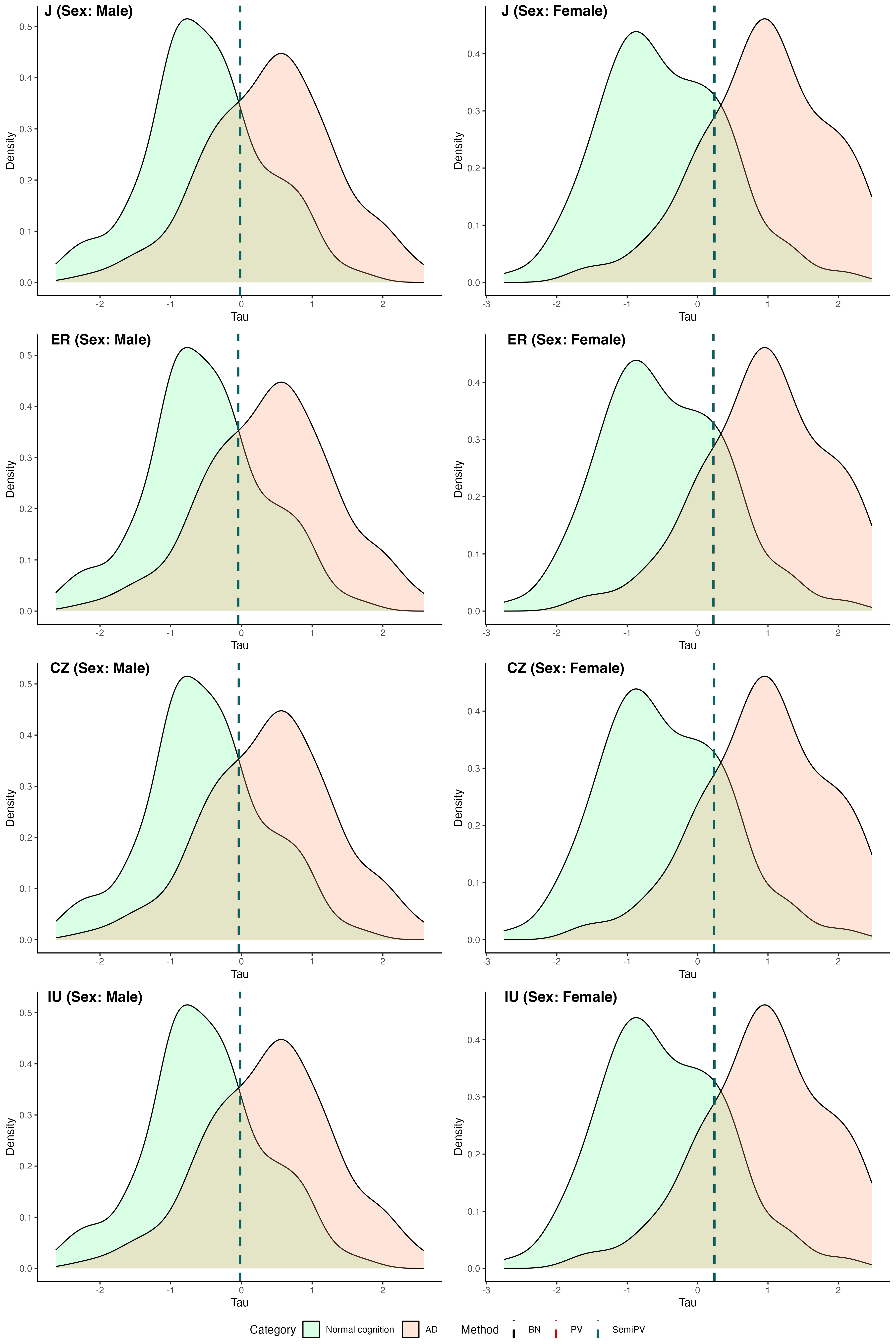}
    \caption{\label{Fig: data_analysis_cov_density_Tau} Sex-specific biomarker densities and estimated cutoffs of biomarker Tau. The left row corresponds to the cutoff estimates of \enquote{Male} subjects and the right row corresponds to that of the \enquote{Female} subjects.}
    \end{center}
\end{figure}

\begin{figure}[hbt!]%
    \begin{center}
    \includegraphics[width=0.8\textwidth]{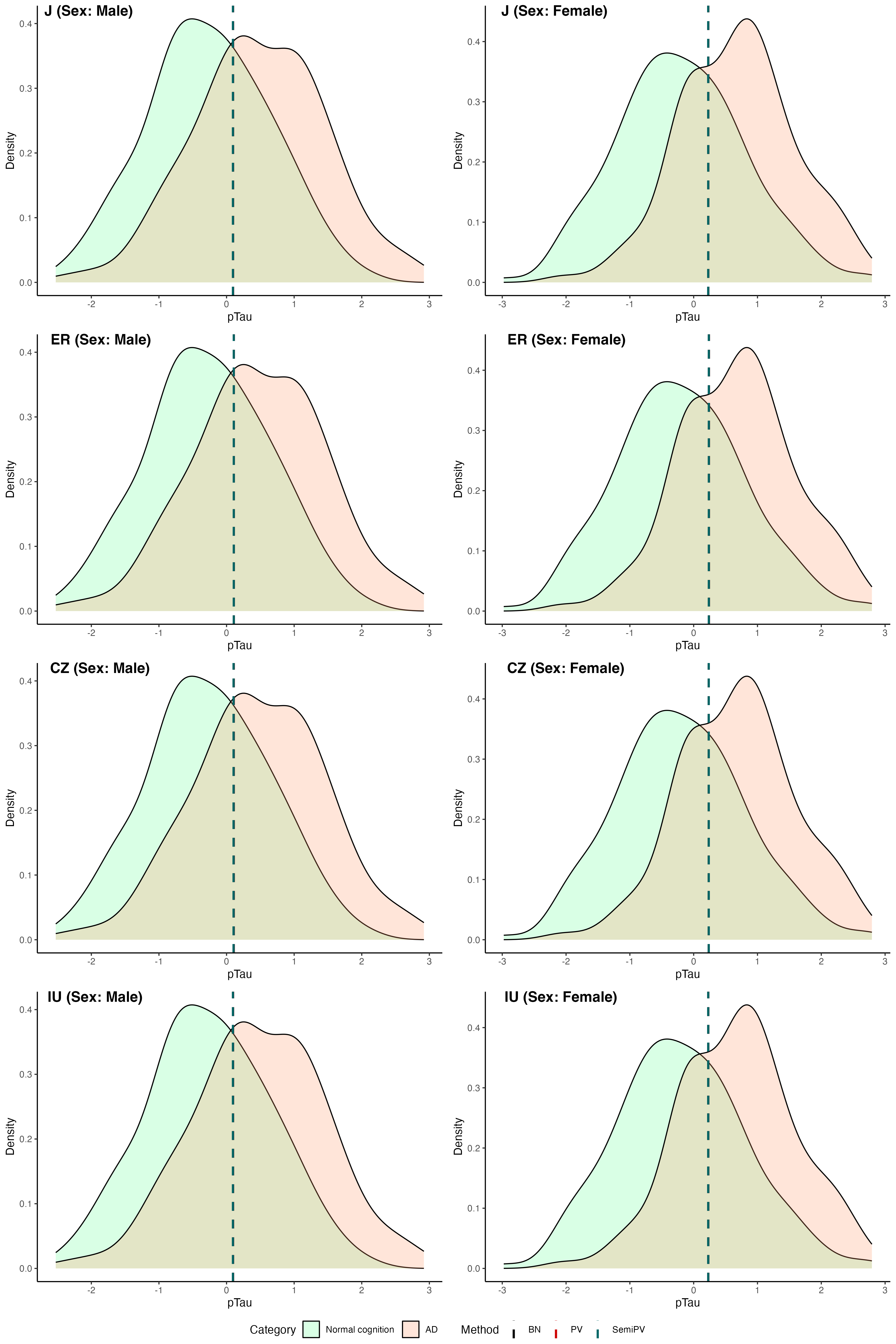}
    \caption{\label{Fig: data_analysis_cov_density_pTau} Sex-specific biomarker densities and estimated cutoffs of biomarker pTau. The left row corresponds to the cutoff estimates of \enquote{Male} subjects and the right row corresponds to that of the \enquote{Female} subjects.}
    \end{center}
\end{figure}

In this section, we primarily present density plots aligned with the data analysis performed in the main paper to illustrate the positions of the estimated optimal cutoffs on the biomarker distributions. Figure~\ref{Fig: data_analysis_nocov_density} displays the overall cutoff estimates of different biomarkers $A\beta 42$, Tau, and pTau, obtained by various ROC methodologies. Figures~\ref{Fig: data_analysis_cov_density_AB42} - \ref{Fig: data_analysis_cov_density_pTau}, respectively for biomarkers $A\beta 42$, Tau, and pTau, depict the same, but separately for males and females.

%\end{appendices}

%\bibliographystyle{apalike}
%\bibliography{biblio_cutoff}